\documentclass[twocolumn,aps,pra,superscriptaddress]{revtex4}
\usepackage{graphicx}
\usepackage{amsfonts}
\usepackage{pxfonts}
\usepackage{bm}

\begin{document}

\title{Topological states in two-dimensional optical lattices}
\author{Tudor D. Stanescu} 
\affiliation{Condensed Matter Theory Center and Joint Quantum Institute,
Department of Physics, University of Maryland, College Park, Maryland 20742-4111, USA}
\affiliation{Department of Physics, West Virginia University, Morgantown, West Virginia 26506, USA}
\author{Victor Galitski}
\affiliation{Condensed Matter Theory Center and Joint Quantum Institute,
Department of Physics, University of Maryland, College Park, Maryland 20742-4111, USA}
\author{S. Das Sarma}
\affiliation{Condensed Matter Theory Center and Joint Quantum Institute,
Department of Physics, University of Maryland, College Park, Maryland 20742-4111, USA}

\begin{abstract}
We present a general analysis of two-dimensional optical lattice models that give rise to topologically non-trivial  insulating states. We identify the main ingredients of the lattice models that are responsible for the non-trivial topological character and argue that such states can be realized within a large family of realistic optical lattice Hamiltonians with cold atoms. We focus our quantitative analysis on the properties of topological states with broken time-reversal symmetry specific to cold-atom settings. In particular, we analyze finite-size effects, multi-orbital phenomena that give rise to a variety of distinct topological states and transitions between them, the dependence on the trap geometry, and most importantly, the behavior of the edge states for different types of soft and hard boundaries. Furthermore, we demonstrate the possibility of experimentally detecting the topological states through light Bragg scattering of the edge and bulk states. 
\end{abstract}

\maketitle

\section{Introduction}\label{SecI}

It has been shown recently that band structures of non-interacting lattice models and quadratic mean-field Hamiltonians can be classified according to the topological character of the wave functions associated with the bands. The most complete classification of this type of Hamiltonians in the physically relevant two and three dimensions was recently presented by Kitaev~\cite{PTTI} and by Ryu {\it et al}.~\cite{Ludwig2} who identified all distinct topological classes, which differ sharply depending on the presence or absence of particle-hole symmetry and time-reversal symmetry. 
The general interacting case was addressed by Volovik~\cite{Volovik2009} using the Green function, rather than the Hamiltonian,  as the object for the topological classification~\cite{Volovik1989,Volovik2009a}.
With this understanding achieved, a question appears on how to realize various such topological states in physical systems.  Until now a few promising solid-state materials have been identified that are expected to host certain topological phases. However, in solid-state settings, we are bound to work with the existing compounds provided by nature and we have no choice but to rely on serendipity in our search for physical realizations of 
topological states, rather than on a controlled "engineering" of appropriate lattice Hamiltonians that are guaranteed to host these exotic phases. 

On the other hand, optical lattices populated with cold atoms offer a very promising alternative avenue to build topological insulating states. Cold-atom systems provide more control in constructing specific optical lattice Hamiltonians by allowing both tunable hoppings and interparticle interactions that can be adjusted as needed, hence opening the possibility of accessing interacting topological states such as topological Mott insulators. However, cold-atom settings bring in their own specific challenges associated with the trapping potential, the effective vector potential responsible for the nontrivial topological properties, the soft boundaries, and also with the fact that cold atom experiments involve neutral particles and therefore make any transport measurement irrelevant or very difficult, thus bringing up the question of how to probe experimentally  the topological character of these phases. Motivated by the opportunity of creating topological insulating states with cold atoms and by the aforementioned challenges, we discuss in this article a general prescription for building certain types of topological optical lattice models and analyze in detail the properties of the emergent states in the presence of trapping potentials with different geometries.

Until the discovery in the early 1980s of the quantum Hall effect~\cite{QHE1,QHE2}, the standard way of classifying quantum states of condensed matter systems was to consider the symmetries they break. The existence of extremely robust properties, such as the quantized Hall conductance, was found to be linked to the nontrivial topological structure of the quantum Hall states. These states do not break any symmetry, hence cannot be described by the Landau symmetry breaking theory~\cite{Landau}, but possess a more subtle organizational structure sometimes called topological order~\cite{Wen}.  In two-dimensional systems, such as the quantum Hall fluids, the nontrivial topological structure is intrinsically connected with the existence of robust gapless edge modes.  In the three-dimensional case it leads to robust gapless surface or interface modes, such as the interface midgap states in  heterojunctions composed of semiconductors with opposite band-edge symmetry~\cite{Korenman1986,Korenman1987}. 
In recent years a significant number of different models and solid state systems with topologically ordered ground states were found and studied both theoretically~\cite{Moessner,Balents,Misguich,Kitaev,Wen2003,KaneMele,KaneMele2005,FuKane,BernevigZ,Murakami,Wu,Moore,FuKaneMele} and experimentally~\cite{Konig,Hsieh2008,Hsieh2009,Hsieh2009a,Roushan}. While most of the efforts are concentrated on solid state systems, it was recently proposed to realize topological quantum states with cold atoms trapped in optical lattices~\cite{CWu2008,Wang,StanSarma}. The original focus was on the realization of a particular model that supports topological quantum states, the Haldane model~\cite{Haldane}. However, to take full advantage of the great flexibility in constructing an optical lattice and of the high possibility of parameter control offered by cold atom systems, a generalization scheme that can easily generate new models would come in handy. In this article we describe a very intuitive scheme to construct new families of models with nontrivial topological properties starting from the model introduced by Haldane~\cite{Haldane} in the late 1980s.

Before proceeding to the  main technical part, we first explain our choice of the model, which as we will show below gives rise to topological states within the same class as the lattice quantum Hall state described by the Haldane model. This quantum Hall-like state explicitly breaks time-reversal symmetry and therefore does not represent a time-reversal invariant topological insulator of the type that most recently has been of main interest in the solid-state context. This focus on time-reversal-invariant systems is understandable there, because the driving force that generates the non-trivial topological structure arises from the spin-orbit coupling, which in a sense is responsible for the "internal magnetic fields" associated with the spin-split bands. The absence of any required external field is of course a huge experimental simplification in solid-state experiments, which deal with given material compounds with predetermined properties. It is also a limitation restricting the topological insulator states that are practically accessible. In the cold-atom context, however, the time-reversal-invariant topological insulators and their equivalents (in the presence of a pseudospin variable) are not necessarily experimentally preferable.  A cold-atom Hamiltonian has to be build from scratch and typically there are no predetermined chiral hopping terms and spin- or pseudo-spin-orbit interactions or no relevant spin degree of freedom at all. It has been shown recently both theoretically and experimentally that one indeed can construct an analog of a spin-orbit-coupled system with cold atoms~\cite{SZG}. However, the corresponding schemes are by no means easier to realize than an analog of a magnetic field, dubbed a synthetic magnetic field, which may suffice to produce the topological insulating states with broken time-reversal symmetry. In fact, because the artificial magnetic fields do not involve any spin (or pseudospin) degree of freedom, the broken time-reversal symmetry lattice quantum Hall states are expected to be easier to realize than the  time-reversal-invariant topological insulators.  These later systems require additional optical setups to produce equivalents of the spin-orbit interaction and their realization will probably represent the second stage of building topological quantum states with cold atoms.  For this reason, we focus specifically our discussion on the two-dimensional lattice quantum Hall states, which as explained above, are of more direct experimental relevance. 

\subsection{Main results and open questions} 

\subsubsection{Model and implementation} We show that there are  infinitely many lattice models, descendants of the canonical Haldane model~\cite{Haldane}, which host the same type of lattice quantum Hall states. The topological character of such a state is associated with chiral hoppings, which usually are thought of in the context of a simple honeycomb lattice. We argue instead that one can start with a local model that includes  chiral hoppings as the main initial ingredient and then add other ordinary hopping terms to produce non-local dispersion on various lattices. We show that the nature of the latter is not germane to the topological nature of the state and that in particular, one can construct a square optical super-lattice, which will give rise to topological insulating behavior and which may be easier to realize by optical means with cold atoms. 

The main ingredient for realizing topological insulators (TIs) in cold atom systems is a periodic vector potential, which generates a Peierls phase for certain hopping matrix elements. Such vector potentials can be induced by the interaction of atoms with spatially modulated light fields~\cite{Dum,Dutta,Jaksch,oberg1,oberg2,obergRusec,Soren,Zhu,Osterloh,Spiel}.  We provide the functional spatial dependence of the artificial vector field consistent with the realization of topological quantum states. The description of the model and the proposed scheme for realizing it in optical lattices are presented in Sec.~\ref{SecII}.

{\it Open questions}: In this article we propose the realization of topological insulators with noninteracting spinless atoms. The analysis can be generalized to the case atoms with pseudo-spin degrees of freedom subjected
to synthetic $SU(2)$ gauge fields, which allow building time-reversal invariant topological insulators\cite{SpielII} similar to the quantum spin Hall state in $HgTe$ quantum wells~\cite{Konig}. More exotic topological phases could, in principle, be realized using various types of non-Abelian synthetic gauge fields~\cite{Rusec,Osterloh}. Some of these phases may have no realization in solid-state systems. However, the main direction that requires further study is considering the particle-particle interactions. Given the robustness of the topological states, weak interactions are not expected to modify significantly the present results. In fact, most of the topological states found in condensed matter systems (with the exception of fractional quantum Hall states) belong to various classes of noninteracting TIs. On the other hand, the study of strongly interacting topological insulators is only at the beginning and many fundamental questions remain to be answered. Nonetheless, considering the high capability of tuning the interaction, it is clear that ultra-cold atoms trapped in optical lattices represent the ideal platform for the potential realization of strongly-interacting TIs~\cite{TopoMott1,TopoMott2}.

\subsubsection{Edge states properties} We describe in detail the edge states of topological insulators  that can be realized using an optical lattice implementation of the square super-lattice model (Sec.~\ref{Edge1}). Using ideal boundaries, i.e., boundaries created by an infinitely steep potential wall, we demonstrate that the nature of the gapless edge state mode is independent of the geometry of the two-dimensional system. In particular, we discuss the properties of the edge states for systems with stripe (subsection~\ref{s_stripe}) and disk (subsection~\ref{s_disk}) geometries. This equivalence reflects the topological nature of the edge states and allows for a more convenient computational treatment of large systems. For example, the edge states within a small region near the boundary of a large disk are similar to the edge states within a comparable region near the boundary of a stripe. 

The broken time-reversal symmetry TIs belong to different classes labeled by an integer number ($Z$-type TIs). There is a direct connection between this integer and the number of gapless edge modes. We show explicitly that different types of topological insulators can be obtained by filling multiple bands (subsection~\ref{multi_band}). The number of characteristic edge modes is arbitrary, in contrast to the case of time reversal invariant topological insulators ($Z_2$-type TIs), which support only odd numbers of pairs of edge modes.
    
{\it Open questions}: The most natural and interesting manifestations of the nontrivial topological properties of an insulator take place at the boundary.  Exotic manifestations, such as Majorana fermions, require interfaces between a TI and a superconductor. Creating well defined boundaries in cold atom systems represents a significant challenge. The requirement of an infinitely sharp edge can be relaxed and the TI survives even in shallow traps (see below). However, a soft boundary determines the softening of the edge mode(s) and the proliferation of edge states, a process which modifies some of the physical properties of the system. 

\subsubsection{Phase transitions} We study transitions between topologically distinct band insulators (Sec.~\ref{transitions}). Being able to drive the system through these transitions is an important tool that allows identifying various topological states and distinguishing them from trivial insulating states. The transitions can be driven by an additional staggered potential  (subsection~\ref{tr_stag}) or, in the multi-band situation,  by simply tuning the system parameters (subsection~\ref{tr_multi}). The latter case turns out to be especially interesting as we find that band crossings controlled by optical lattice tunable parameters may "transfer" or "exchange" Chern numbers between different bands, while conserving the total Chern number of the bands.  

{\it Open questions}: A significant challenge for realizing TIs with cold atoms in optical lattices is controlling the filling. The number of atoms has to correspond to a certain number of filled bands and the chemical potential has to lie within a band gap. This issue is connected with the problem of realizing and controlling the boundary (see above).  

\subsubsection{Stability of the TI states} We address the very important experimental question regarding the stability of the edge states, which represent the hallmark of the lattice TI phase (Sec.~\ref{stability}). In particular we focus on the finite-size effects (subsection~\ref{finite_size}) and the effects of soft boundaries generated by a confining potential (subsection~\ref{confining}). We show that the finite size effects are a consequence of the overlap between different edge states. The amplitude of the edge states decreases exponentially away from the boundary with a certain characteristic length scale. If this length scale is much smaller than the system size, the gaps in the edge states spectrum scale as the inverse of the boundary length. 

We also find that a shallow confining potential determines a strong softening of the edge mode(s) and the proliferation of edge states, which acquire a quasi continuous spectrum. Nonetheless, the TI survives even in a shallow harmonic trap, but in this case the insulating core is surrounded by a non-homogeneous chiral metal. The fact that TIs with hard and soft boundaries have different physical characteristics, yet are topologically identical, illustrates vividly the nature of topological order.  For an arbitrary boundary potential characterized by a length scale $L$ associated with the width of the boundary, we find that the edge mode velocity is rescaled by a factor $a/L$, where $a$ is the lattice constant.

{\it Open questions}: If one considers the phenomenology of the boundary, TIs in optical lattices can be divided into two categories: TIs with edge states (in systems with well defined boundaries) and TIs with inhomogeneous (chiral) metallic clouds (in systems with shallow confinement). Realizing experimentally well defined boundaries is a serious challenge (see above). On the other hand, a more detailed analysis of the properties of the inhomogeneous chiral metal is an important direction for future study.  

\subsubsection{Detection of topological edge states}

We propose three different methods of probing topological quantum states (Sec. \ref{SecDetect}). First, we propose imaging the edge states with bosons. The procedure involves loading bosons into the edge states and  then imaging the atoms using a direct {\it in situ}
imaging technique~\cite{Nelson,gillen2008hst}. This technique does not involve the realization of an equilibrium topological insulating state, but rather a real space analysis of the properties of the single particle states. As the nontrivial topological properties of the system represent a feature of the single particle Hamiltonian, identifying the edge states is an effective way of "seeing" a topological phase. 

A very convenient way of identifying an insulator is to perform density profile measurements on fermionic atomic systems. The presence of an insulator generates a characteristic plateau in the density profile, hence the procedure can be used for studying metal-insulator transitions. However, as we show explicitly by performing a model calculation, this method cannot distinguish between a TI and a trivial insulator. 

The chiral edge states of a TI can be detected using optical Bragg spectroscopy. We calculate the dynamical structure factor for a TI model on a square super-lattice and show that the edge mode generates a characteristic low-frequency peak. The chiral nature of the edge mode can be probed by inverting the scattering wave vector (or, equivalently, the vector potential responsible for the nontrivial Peierls phases): the characteristic peak is present for one orientation and absent for the opposite. We also discuss the effect of softening the confinement of the system. 

{\it Open questions}: In the case of imaging the edge states using bosons, future theoretical studies are required for a quantitative estimate of the transfer probabilities and for determining the optimal parameters of the lasers.  
Density profile measurements could be supplemented by probes involving perturbations with opposite angular orientations. Calculations of the response of a TI with broken time reversal symmetry to such perturbations are not yet available. Finally, while for strong and moderate confinement the Bragg spectroscopy provides a direct way to observe the chiral edge states, probing  the inhomogeneous chiral metal requires further analysis.

\section{\bf Topological Insulators on a Square Superlattice: the Model}\label{SecII}

The goal of this section is twofold: (i) to show that there is an unlimited number of different families of topological insulator models and describe a simple method of constructing such models (this flexibility in building quantum states with nontrivial topological properties is particularly relevant in view of their possible realization in cold atom systems) and  (ii) to introduce a particular two-dimensional model of a topological insulator with broken time-reversal symmetry on a square superlattice. The properties of this model will be studied in detail in the subsequent sections.

\subsection{A recipe for constructing topological insulator models}

The Haldane model~\cite{Haldane} is a tight-binding representation of motion on a hexagonal lattice having as key feature a direction-dependent complex next-nearest -neighbor hopping. A periodic vector potential ${\mathbf A}({\mathbf r})$ that generates a magnetic field with zero total flux trough the unit cell is responsible for the imaginary components of the hopping matrix elements. The vanishing of the magnetic flux through each unit cell ensures that the nearest-neighbor hoppings remain unaffected by the vector potential. The quantization of the Hall conductance in integer quantum Hall systems can be intuitively linked to the formation of Landau levels in a uniform magnetic field. However, using the simple tight-binding model Haldane showed that quantum Hall-like states may result from breaking time-reversal symmetry in the presence of a periodic vector potential without having a net magnetic flux, i.e., without Landau levels. In both cases it is the non-trivial topology of the ground state that ensures the quantization of the Hall conductance, which can be interpreted as the topological Chern number of the $U(1)$ bundle over the Brillouin zone of the bulk states~\cite{Thouless}. While the value of the Chern number for a given occupied band is far from obvious without an explicit calculation, a more direct and intuitive  signature of the non-trivial topological properties
of a system is the existence of chiral gapless edge (in two dimensions) or surface (in three dimensions) states robust against disorder effects and interactions. The basic features of these states are intrinsically linked to the topological properties of the system, but their detailed structure is dictated by the boundary. As the bulk of the system is an insulator, it is the edge or surface states that participate in transport. The quantization on the transverse Hall conductance can understood within this edge states picture~\cite{Halperin} using Laughlin's gauge invariance argument~\cite{Laughlin}.
\begin{figure}[tbp]
\begin{center}
\includegraphics[width=0.4\textwidth]{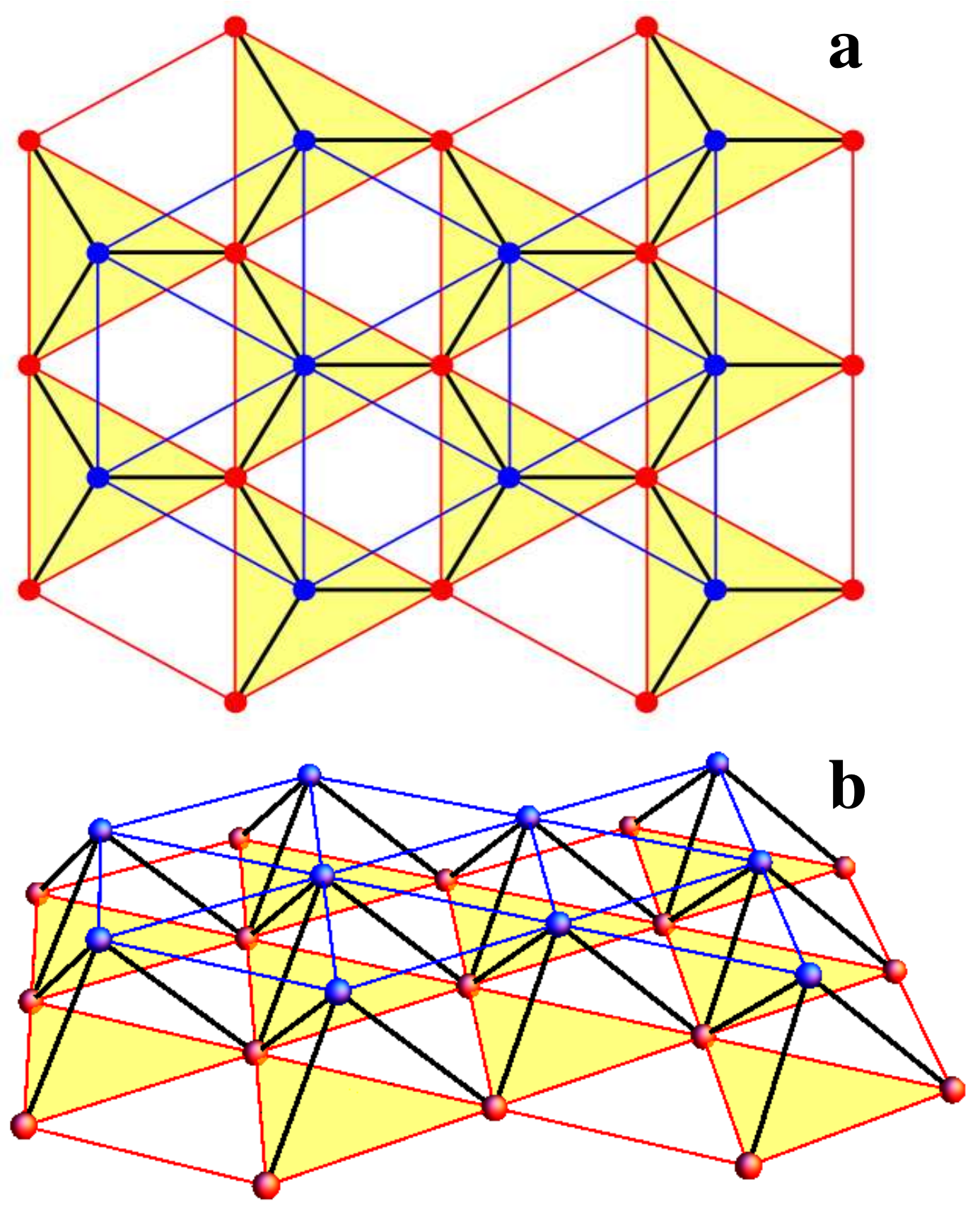}
\end{center}
\caption{(Color online) (a) Two-dimensional (2D) hexagonal lattice for the Haldane tight-binding model, consisting of real nearest-neighbor hoppings (black lines) and complex direction-dependent next-nearest-neighbor hoppings (red and blue  or gray lines). The imaginary components of the hopping matrix elements are generated by an effective vector potential that produces a ``magnetic'' field with zero total magnetic flux through the unit cell [i.e., the magnetic fluxes through the white and yellow (light gray) triangles have equal magnitudes and opposite signs]. (b) A three-dimensional (3D) realization of the model obtained by translating one sub-lattice (the blue spheres) along the direction perpendicular to the plane ($z$-direction). Staking such layers in the $z$-direction with the $A$ sub-lattice sites on top of $B$ sites generates a 3D generalization of the Haldane model on a diamond lattice. Alternatively, we can treat the model  as a quasi-2D lattice of triangular pyramids.  Neglecting the hopping between the apex sites (blue lines) does not change the topological properties of the model. Pyramids with a different base will generate similar models with non-trivial topological properties.}
\label{Fig1}
\end{figure}

The hexagonal (honeycomb) lattice for the Haldane model is shown in Fig. \ref{Fig1}a. It consists of two interpenetrating triangular sublattices $A$  and $B$. The nearest neighbor hoppings between $A$-type and $B$-type sites (black lines) are real, while the next-nearest-neighbor hoppings (red and blue/gray lines) contain imaginary components due to the presence of a periodic vector potential ${\mathbf A}({\mathbf r})$. The total magnetic flux generated by ${\mathbf A}({\mathbf r})$ through each hexagonal unit cell vanishes, but the magnetic fluxes through the white and yellow (light gray) triangles are nonzero and have equal magnitudes and opposite signs. It is crucial that, in the presence of the vector potential, $A$-type and $B$-type sites are not equivalent. Consequently, after changing the sign of ${\mathbf A}({\mathbf r})$ (i.e., exchanging the white and yellow triangles) the original configuration cannot be restored by any translation or rotation operation. By contrast, if for example  we remove the  sublattice $B$ altogether we obtain a triangular lattice in a staggered magnetic field. The nearest-neighbor hoppings are complex. However, in this case the original configuration can be recovered after a time reversal operation by a $\pi/3$ rotation.

Next, we modify the model while preserving the crucial ingredients that ensure the breaking of time reversal symmetry, as discussed above. For example, we can view the two-dimensional (2D) lattice shown in Fig. \ref{Fig1}(a) as a projection of the three-dimensional (3D) model shown in Fig. \ref{Fig1}(b). If we use the same tight-binding parameters, the two geometries will generate identical results. However, the 3D version suggests a direct way of generalizing the Haldane model to three dimensions. For example, staking layers as the one shown in Fig. \ref{Fig1}(b) on top of each other with the $A$ sublattice sites directly above the $B$ sites generates a family of models that represents the 3D generalization of the Haldane model on a diamond lattice. Different members of this family may be obtained by making further choices for the vector potential. If only the original in-plane components of ${\mathbf A}({\mathbf r})$ are considered, there are no anomalous interplane hoppings. However, complex inter-plane hopping matrix elements can be generated by including a field component in the $z$-direction. Alternatively, we can simplify the structure shown in Fig. \ref{Fig1}(b) and reduce it to the bare essentials. For example, we can ignore the hopping between the $B$-type sites (the blue lines) and treat the model  as a quasi-2D lattice of triangular pyramids with complex direction-dependent hoppings between the base sites. Note that the system represents a triangular lattice with a two-point basis. Within a single band tight-binding model we cannot eliminate the apex sites without restoring the equivalence between the white and yellow (light gray) triangles. However, this elimination  is possible within a multiband model. Intuitively one can easily understand this property if we notice that hopping between $s$-orbitals is isotropic, while $p$-orbitals generate direction-dependent hopping matrix elements that carry the information about the nonequivalence of white and yellow (light gray) triangles.

\subsection{Topological insulator model on a square super-lattice} \label{SQSLM}

\begin{figure}[tbp]
\begin{center}
\includegraphics[width=0.35\textwidth]{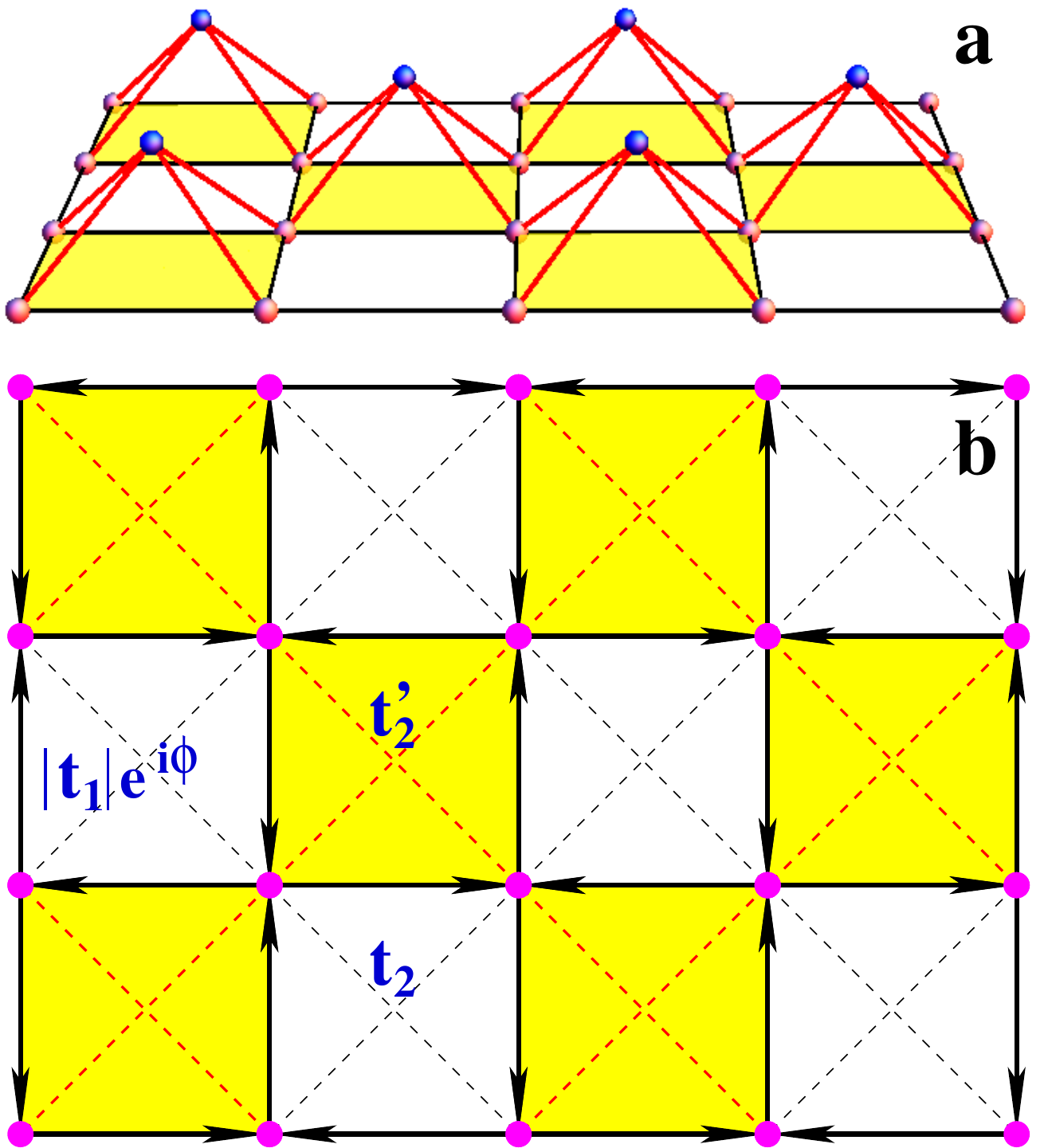}
\end{center}
\caption{(Color online) (a) Quasi-2D model of a topological insulator with broken time-reversal symmetry on a lattice of square pyramids. Expanding the structure in the $z$ direction will generate 3D models of topological insulators. Alternatively, one can move the apex site into the base plane and generate a 2D square superlattice model. (b) Topological insulator model on a 2D square super-lattice. Instead of extra apex sites, as in  (a), we consider different next-nearest-neighbor hoppings $t_2^{\prime}\neq t_2$. A vector potential  $\mathbf{A}(\mathbf{r})$ produces an effective "magnetic" field with opposite flux through the yellow (light gray) and white squares and generates a complex direction-dependent nearest-neighbor hopping $t_1$. Hoppings with a given sign of the phase are marked by arrows.}
\label{Fig2}
\end{figure}

The fact that the pyramids in the quasi-2D model described above are triangular does not have any particular significance and does not determine the topological properties of the model. One can imagine for example a similar system of square pyramids, as shown in Fig. \ref{Fig2}(a). Again, we can stack such structures in the $z$-direction and generate a family of topological insulators on a cubic lattice.  Alternatively, we can project the structure onto the base plane and generate a 2D square superlattice model~\cite{VY1,VY2,VY3}. As before, a periodic  vector potential  $\mathbf{A}(\mathbf{r})$ generates a staggered magnetic field with opposite flux through the yellow (light gray) and white squares. The unit cell consisting of one  yellow (light gray) square and one white square contains three sites. Unlike the triangular lattice case discussed previously, we can now remove the former apex sites and replace them with an effective next-nearest-neighbor hopping within a single-band model without restoring time reversal symmetry. The unit cell of the simplified model contains two-sites and we can view the lattice as consisting of two rectangular sublattices $A$ and $B$. The resulting 2D square superlattice effective model is shown in  Fig. \ref{Fig2}(b). The next-nearest-neighbor hoppings $t_2$ and $t_2^{\prime}$ are real and have different values.
The nearest-neighbor hopping $t_1$ is complex and has a direction-dependent phase. If we choose a coordinate system with the axes along the next-nearest-neighbor directions and set the nearest-neighbor distance $a = 1/\sqrt{2}$, the tight-binding model can be expressed analytically by the Hamiltonian
\begin{equation}
{\cal{H}} = \sum_{\mathbf k} \left(c_{A{\mathbf k}}^{\dagger} ~c_{B{\mathbf k}}^{\dagger} \right)
\left(\begin{array}{cc}
\tilde{t}_2(k_x, k_y) &\left[\tilde{t}_1({\mathbf k})\right]^* \\
\tilde{t}_1({\mathbf k}) & \tilde{t}_2(k_y, k_x)
\end{array}\right)
\left(\begin{array}{c}
c_{A{\mathbf k}} \\
c_{B{\mathbf k}}
\end{array}\right),  \label{modelH}
\end{equation}
with
\begin{eqnarray}
&&\tilde{t}_1({\mathbf k}) = \vert t_1 \vert \left[e^{-i \phi}(1+e^{i(k_x+k_y)}) + e^{i \phi}(e^{i k_x} + e^{i k_y})\right], \nonumber \\
&&\tilde{t}_2(k_x, k_y) = 2t_2\cos k_x + 2t_2^{\prime}\cos k_y.
\end{eqnarray}
In Eq. (\ref{modelH}) the operators $c_{A{\mathbf k}}^{\dagger}$ and $c_{B{\mathbf k}}^{\dagger}$ create a particle with wave vector ${\mathbf k}$ on the sublattices $A$ and $B$, respectively.

So far we did not mention the possible role of the spin (or pseudospin) degree of freedom in generating nontrivial topological quantum states. All the topological insulator models generated according the scheme described above can be easily generalized to include spin, similar to the construction used by Kane and Mele who proposed a tight-binding Hamiltonian for graphene~\cite{KaneMele} that generalizes Haldane's model to include spin with time-reversal invariant spin-orbit interactions. Basically, for spin $1/2$ particles the models should include a spin-dependent vector potential $\mathbf{A}_{\sigma}(\mathbf{r})$ that has opposite orientations for the two spin components, $\mathbf{A}_{\uparrow}(\mathbf{r})=-\mathbf{A}_{\downarrow}(\mathbf{r})$. While each spin component breaks time-reversal symmetry, the system as a whole is time-reversal invariant. These systems form new classes of topological insulators~\cite{Ludwig} that cannot be classified using Chern numbers. For example, in two-dimensions one obtains a quantum spin Hall state~\cite{KaneMele,KaneMele2005,BernevigZ,Konig2008}, which carries no net charge current along the system edges. If a U(1) part of the SU(2) spin-rotation symmetry is preserved, particles with opposite spin will propagate along a given edge in opposite directions giving rise to a quantized spin Hall conductance~\cite{KaneMele,KaneMele2005,BernevigZ}. However, the system remains topologically ordered even in the presence of small perturbations that break the full spin-rotation symmetry, when the spin Hall conductance is no longer quantized. To classify these time-reversal invariant topological states, Kane and Mele introduced a $Z_2$ topological invariant~\cite{KaneMele2005}, which can be interpreted in terms of doublets of edge modes.
In three dimensions  the $Z_2$ topological invariant is associated with the number of Kramers degenerate points (Dirac points) in the spectrum of the surface states. In both two and three dimensions, the existence of an odd number of Kramers degenerate points ensures the stability of the edge or surface states against disorder and interactions~\cite{Wu,XuMoore,Ostrovsky,Obuse,Bardarson}. We note that spin plays a crucial role in solid state topological insulators as the band gap itself is opened by strong spin-orbit interactions~\cite{Konig,Hsieh2008,Hsieh2009,Hsieh2009a,Roushan}. On the other hand, in cold atom systems an effective spin-orbit interaction can be generated using certain spin-dependent vector potentials~\cite{SZG}. These artificial light-induced vector potentials can be realized in a system of multi-level atoms interacting with a spatially modulated laser field~\cite{Dum,Dutta,Jaksch,oberg1,oberg2,obergRusec,Soren,Zhu,Osterloh}. However, as a first step in the realization of topological insulator with cold atoms a spin-independent vector potential~\cite{Spiel} is probably easier to implement. Therefore in this article we ignore spin and focus on the relatively simpler case of topological insulators with broken time-reversal symmetry.

\subsection{Cold atom realization of the square super-lattice model}

The relatively simple geometrical structure of the square super-lattice model described by Eq. (\ref{modelH}) and Fig. \ref{Fig2}(b) is particularly appealing if we address the problem of constructing topological insulators with cold atoms. The crucial ingredients for constructing topological quantum states with cold atoms are~\cite{StanSarma}: (i) the optical lattice (in the present case the optical super-lattice), obtained as a superposition of co-planar standing waves with properly chosen wave-vectors; (ii) the additional confining potential that determines the properties of the boundary; and (iii) the effective vector potential. The general form of the effective single-particle Hamiltonian describing the atoms trapped in the optical lattice moving in the presence of the light-induced vector potential is
\begin{equation}
H = \frac{1}{2m} \left[{\mathbf p} - \mathbf{A}({\mathbf r})\right]^2 + V_{\rm latt}({\mathbf r}) + V_c({\mathbf r}), \label{H1}
\end{equation}
where $m$ is the atom mass, ${\mathbf p}=-i \hbar {\mathbf \nabla}$ the momentum, $\mathbf{A}(\mathbf{r})$ the effective vector potential, $V_{\rm latt}({\mathbf r})$ the optical lattice potential and $V_c({\mathbf r})$ the extra confining potential. The role of $V_c({\mathbf r})$, in addition to preventing the atoms from escaping the optical lattice, is to create appropriate boundaries for the system and thus make possible the formation and observation of the characteristic topological edge states~\cite{StanSarma}. We start by assuming an infinitely sharp confining potential, and then in Sec. V we discuss explicitly the case of smooth confining.
\begin{figure}[tbp]
\begin{center}
\includegraphics[width=0.45\textwidth]{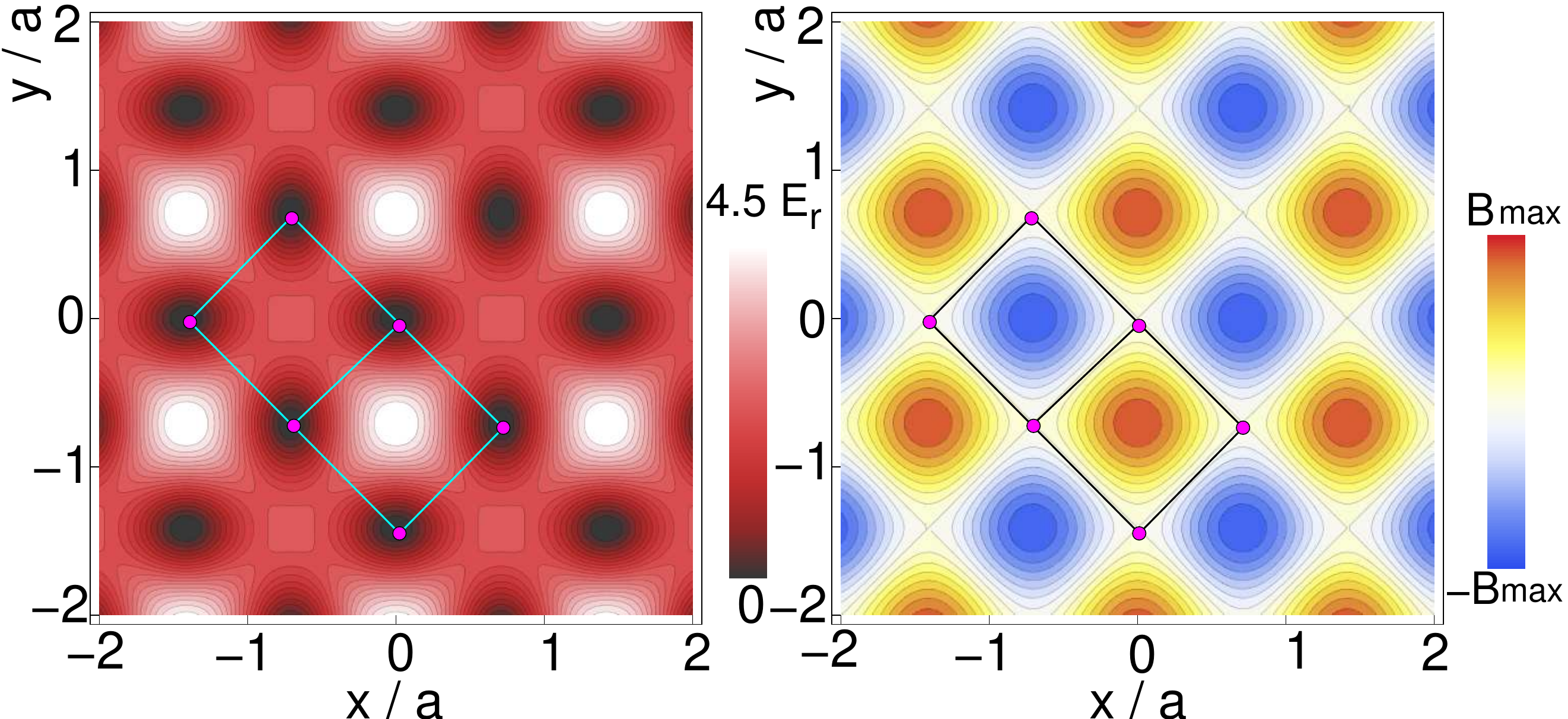}
\end{center}
\caption{(Color online) (Left) Optical super-lattice potential corresponding to $V_1=3.4 E_r$, $V_2=1.7 E_r$ and $\alpha=2\hbar/a$ (see main text). A unit cell consisting of two squares with side length $a$ is marked with light blue (light gray) lines. Notice the $\pi/2$ rotation of the axes relative to lattice in Fig. \ref{Fig2}(b). (Right) Effective "magnetic" field generated by the vector potential $\mathbf{A}(\mathbf{r})$. The total flux through the unit cell is zero.}
\label{Fig3}
\end{figure}
A crucial ingredient is the light-induced vector potential ${\mathbf A}(\mathbf{r})$ that generates the effective ''magnetic'' field with zero total flux through the unit cell. The construction of synthetic Abelian and non-Abelian gauge potentials coupled to neutral atoms is an emerging theme in the field of cold atom systems, which has been investigated theoretically in some detail but is just beginning to receive experimental attention~\cite{Dum,Dutta,Jaksch,oberg1,oberg2,obergRusec,Soren,Zhu,Osterloh,Spiel,Clark}. In order to realize the square superlattice model (\ref{modelH}) we propose a vector potential of the form ${\mathbf A}(\mathbf{r}) = \alpha{\bf \cal{A}}(\mathbf{r})$, where $\alpha$ is a parameter that measures the strength of the potential and
\begin{equation}
{\bf \cal{A}}(\mathbf{r}) = \left(\sin\left[\frac{\sqrt{2}\pi y}{a}\right], \sin\left[\frac{\sqrt{2}\pi x}{a}\right]\right), \label{calA}
\end{equation}
with $a$ being the nearest-neighbor distance of the lattice. The two-dimensional optical super-lattice, generated as a  superposition of co-planar standing waves with properly chosen wave-vectors~\cite{Grynberg,Ritt,Rey,Trotzky,Cheinet}, is characterized by the effective potential
\begin{eqnarray}
V_{\rm latt} &=& V_1\left(1-\frac12\cos^2\left[\frac{\pi(x+y)}{\sqrt{2}a}\right] - \frac12\cos^2\left[\frac{\pi(x-y)}{\sqrt{2}a}\right] \right) \nonumber \\
&+& V_2\left(\cos^2\left[\frac{\pi x}{\sqrt{2}a}\right] + \sin^2\left[\frac{\pi y}{\sqrt{2}a}\right] -1\right), \label{Vlatt}
\end{eqnarray}
where the amplitude $V_1$ controls the overall depth of the optical lattice while  $V_2$ generates the super-lattice structure. The case $V_2=0$ corresponds to a simple square lattice with lattice constant $a$, while $V_2\neq 0$ produces the doubling of the unit cell. Note that the first term in Eq. (\ref{H1}) contains a quadratic contribution in the vector potential, ${\mathbf A}^2/2m$, which renormalizes the effective optical lattice potential. One potential challenge in realizing a topological quantum state with cold atoms is the precise matching of the light wavelengths for the laser generating the optical lattice and those generating the artificial vector potential. We note here that  a mismatch $\Delta\lambda$ between the two periods leads to a pseudo-random potential with a strength that cannot be made arbitrarily small. Basically, the strength of the pseudo-random potential is controlled by the amplitude of the effective vector potential, which also controls the magnitude of the insulating band gap. Consequently, in systems with a linear size larger than $\lambda^2/\Delta\lambda$ the pseudo-random potential leads to the closing of the insulating gap and the destruction of topological quantum states.
The structure of the optical superlattice potential, including the contributions from the ${\mathbf A}^2/2m$ term, are shown in Fig. \ref{Fig3} (left panel).

Throughout the article we will use the recoil energy $E_r = (\hbar\pi/a)^2/2m$ as the energy unit. Also, the parameter $\alpha$ which measures the strength of the vector potential is expressed in units of $\hbar/a$. The positions of the nodes of the square lattice generated by the potential in Fig. \ref{Fig3} are given by
the minima of the effective potential, $V_{\rm latt}^{(eff)}({\mathbf r}_i) \equiv V_{\rm latt}({\mathbf r}_i) + {\mathbf A}^2({\mathbf r}_i)/2m = 0$. In addition to renormalizing the optical lattice potential, ${\mathbf A}({\mathbf r})$ generates an effective "magnetic" field with zero total flux through the unit cell. The position dependence of the "magnetic" field is shown in the right panel of Fig. \ref{Fig3}. If $(\delta x, \delta y)$ represents a small deviation
away from one of the minima of the effective optical lattice potential, we have
\begin{eqnarray}
&&V_{\rm latt}^{(eff)}(x_i+\delta x, y_i+\delta_y) \label{harmonicV} \\
&\approx&\frac{\pi^2}{a^2}\left(\frac{V_1\mp V_2}{2} + \frac{\alpha^2}{m} \right)\delta x^2
+ \frac{\pi^2}{a^2}\left(\frac{V_1\pm V_2}{2} + \frac{\alpha^2}{m} \right)\delta y^2 \nonumber \\
&=& \frac{m}{2} \left(\omega_{1(2)}^2\delta x^2 + \omega_{2(1)}^2\delta y^2  \right), \nonumber
\end{eqnarray}
i.e., near a minimum the effective optical potential can be approximated by a two-dimensional anisotropic harmonic oscillator potential with characteristic frequencies
\begin{equation}
\omega_{1(2)} = \pi \sqrt{\frac{V_1\mp V_2}{m} + \frac{2\alpha^2}{m^2}}. \label{omgi}
\end{equation}
Consequently, the harmonic oscillator eigenfunctions represent a natural basis for a tight-binding treatment of the quantum problem described by the Hamiltonian (\ref{H1}).

At this point we note that the experimental observability of topological quantum states in cold-atom systems depends on the energy separation between edge states and bulk states, i.e., on the size of the bulk gap. If the gap for bulk states is not large compared to the lowest temperatures that are accessible experimentally, the standard signature of a topological insulator cannot be observed in any type of transport measurement, because of the significant contribution from thermally excited bulk states. The scheme proposing the direct mapping of the edge states~\cite{StanSarma} is equally inapplicable, because the lack of energy resolution does not allow loading a significant fraction of particles into specific edge states. Other schemes are also likely to fail. Hence systems with large values of the bulk gap are desirable. As the gap scales with the hopping parameters and, in turn, these hopping matrix elements depend on the depth of the lattice potential, we conclude that rather shallow optical lattices may be required for observing topological quantum states. To capture, at least qualitatively, this regime when solving the quantum problem (\ref{H1}) within the tight-binding approximation one has to consider not only the orbital associated with the ground state of the harmonic oscillator (\ref{harmonicV}), but also higher energy states. In our calculations we include the ground state $\psi_{0,0}$ and the first two excited states $\psi_{1,0}$ and $\psi_{0,1}$ with energies $(\omega_1+\omega_2)/2$,  $(3\omega_1+\omega_2)/2$ and  $(\omega_1+3\omega_2)/2$, respectively. As the square super-lattice model is defined on a lattice with a two-point basis, the three orbitals that we consider will generate six bands. Note that the orbitals $\psi_{n,m}$ for the sublattice $B$ are rotated with $\pi/2$ relative to those of the sublattice $A$. Explicitly,
\begin{equation}
\psi_{n,m}^{({\mathbf r}_0)}({\mathbf r}) = \left\{
\begin{array}{c}
\varphi_n^{(\omega_1)}(x-x_0) \varphi_m^{(\omega_2)}(y-y_0), ~~{\mathbf r}_0\in A \\
\varphi_n^{(\omega_1)}(y-y_0) \varphi_m^{(\omega_2)}(x-x_0),  ~~{\mathbf r}_0\in B
\end{array}\right. \label{psinm}
\end{equation}
where ${\mathbf r}_0=(x_0, y_0)$ is the position of a certain lattice site (i.e., minimum of $V_{\rm latt}^{(eff)}$), and  $\varphi_n^{(\omega_j)}(\xi)$ are eigenstates of the one-dimensional quantum harmonic oscillator with angular frequency $\omega_j$. We calculate the hopping parameters for the effective tight-binding model, $t_{ij}^{(n,m)(n^{\prime},m^{\prime})} = \langle\psi_{n,m}^{({\mathbf r}_i)}\vert H \vert \psi_{n^{\prime},m^{\prime}}^{({\mathbf r}_j)}\rangle$, and include nearest-neighbor and next-nearest-neighbor contributions, which add up to a total of 22 different hopping parameters. We have determined analytic expressions for all these hopping matrix elements as functions of the fundamental parameters of the model, $V_1$, $V_2$ and $\alpha$. The key contributions coming from the vector potential, $\langle\psi_{n,m}^{({\mathbf r}_i)}\vert {\mathbf p}\cdot{\mathbf A} \vert \psi_{n^{\prime},m^{\prime}}^{({\mathbf r}_j)}\rangle$, are complex with an imaginary component that is maximal for nearest-neighbor hopping. As the values of the hopping parameters decrease rapidly with the inter-site distance, having anomalous nearest-neighbor components represents a potential advantage of this model over the honeycomb geometry of the original Haldane model, where the anomalous hopping responsible for the non-trivial topological properties occurs between next-nearest-neighbors. Finally, we note that the orbitals used as a basis for the tight-binding approximation are not orthogonal, so the corresponding overlap matrix $\langle\psi_{n,m}^{({\mathbf r}_i)}\vert \psi_{n^{\prime},m^{\prime}}^{({\mathbf r}_j)}\rangle$ has to be calculated and used in the diagonalization procedure. To summarize, we solve the single-particle quantum problem
\begin{equation}
H\Phi_q({\mathbf r}) = \epsilon_q\Phi_q({\mathbf r}), \label{HPhi}
\end{equation}
where $H$ is the Hamiltonian given by Eq. (\ref{H1}) and $q$ is a set of quantum numbers that label the single-particle states. Within the tight-binding approximation, we look for solutions of the form
\begin{equation}
\Phi_q({\mathbf r}) =\sum_j \sum_{(n, m)} \Psi_q^{(n,m)}({\mathbf r}_j) \psi_{n,m}^{({\mathbf r}_j)}({\mathbf r}),  \label{Phi}
\end{equation}
where the sum over $j$ runs over all the sites of the lattice, i.e., the locations of the minima of the effective potential $V_{\rm latt}^{(eff)}({\bf r}) =
V_{\rm latt}({\bf r)) + \mathbf A}^2({\bf r})/2m$ and the orbitals $\psi_{n,m}^{({\mathbf r}_j)}$ are harmonic oscillators wavefunctions given by Eq. (\ref{psinm}). In the calculations we include the components $(n,m) \in \{(0,0), (1,0), (0,1)\}$. Within the subspace spanned by the orbital basis, equation (\ref{HPhi}) reduces to
\begin{eqnarray}
&&\sum_j\sum_{(n^{\prime}, m^{\prime})} t_{ij}^{(n,m)(n^{\prime},m^{\prime})} \Psi_q^{(n^{\prime},m^{\prime})}({\mathbf r}_j) \label{tijchi} \\
&=& \epsilon_q  \sum_j\sum_{(n^{\prime}, m^{\prime})} s_{ij}^{(n,m)(n^{\prime},m^{\prime})} \Psi _q^{(n^{\prime},m^{\prime})}({\mathbf r}_j), \nonumber
\end{eqnarray}
with the hopping matrix $t_{ij}^{(n,m)(n^{\prime},m^{\prime})}$ and the overlap matrix $s_{ij}^{(n,m)(n^{\prime},m^{\prime})}$ defined above. For a system with translational symmetry, the problem can be diagonalized with respect to the position indices by a Fourier transform and the relevant quantum numbers are $q=(\lambda, {\mathbf k})$, where $\lambda$ is a band index and ${\mathbf k}$ is a wave vector in the reduced Brillouin zone associated with the super-lattice structure. In the case of a finite system, Eq. (\ref{tijchi}) is solved numerically for the full size matrices.
\begin{figure}[t]
\begin{center}
\includegraphics[width=0.47\textwidth]{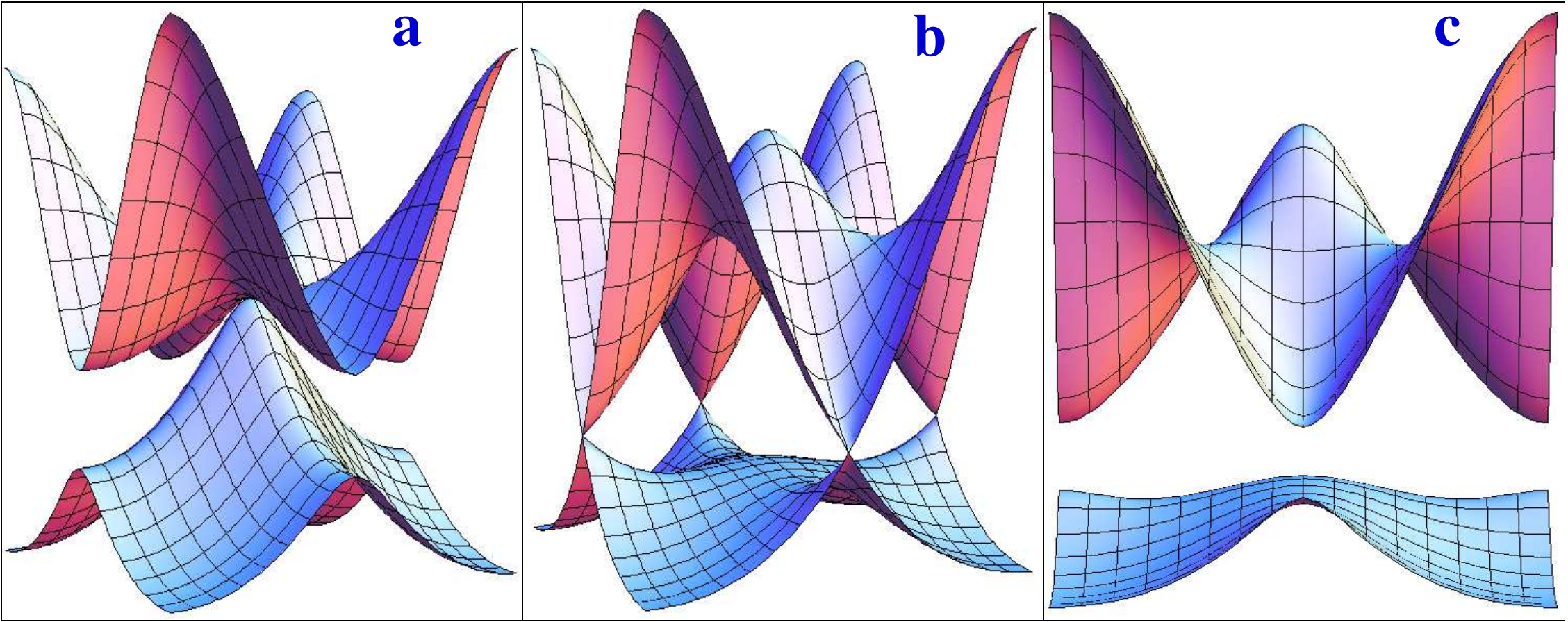}
\end{center}
\caption{(Color online) Spectrum of the square super-lattice model with periodic boundary conditions (no boundaries). Only the lowest two bands are shown, although four other bands were considered in the calculation. The wave-vector takes values in the first Brillouin zone, $0\leq k_x \leq 2\pi/\sqrt{2} a$, $0\leq k_y \leq 2\pi/\sqrt{2} a$, and we make the choice of length units $a=1/\sqrt{2}$. (a) If $V_2=0$ (no super-lattice structure), the two bands are degenerate at ${\bf k} =(\pi, \pi)$. (b) If $\alpha=0$ (no vector potential), the gap closes at two Dirac points $(0, \pi)$ and $(\pi,0)$. (c) For $V_1=3.4 E_r$, $V_2=1.7 E_r$ and $\alpha=2\hbar/a$ a full gap opens. The bands are shown along the $k_x$ direction (with $k_y$ out of the plane).}
\label{Fig4}
\end{figure}

We emphasize that using the harmonic oscillator basis imposes no restriction on the accuracy of the numerical analysis. By including more wave functions in the basis one can attain any desired accuracy. The results presented below have quantitative relevance for the lowest band and give a qualitative picture of the higher-energy bands. To estimate the accuracy of the approximation, we determine the component of the effective lowest band hopping due to virtual transitions to  higher bands, $\delta t_{ij}(n,m) = t_{ij}^{(0,0)(n,m)}t_{ij}^{(n,m),(0,0)}/[\epsilon_{(n,m)}- \epsilon_{(0,0)}]$. For the range of parameters used in this study, the $\delta t_{ij}(1,0)$ and $\delta t_{ij}(0,1)$ represent up to $15\%$ of the bare value $t_{ij}^{(0,0),(0,0)}$. Consequently, one expects a strong renormalization of the spectrum due to the hybridization with these bands. Higher-energy bands generate corrections smaller that $5\%$ and we neglect them. Another potential source of errors comes from neglecting longer   
range hoppings. Second neighbor hoppings are typically less that $10\%$ of the nearest neighbor hoppings (up to $25\%$ in a few cases) and have a crucial role in opening the insulating gap. Consequently, they have to be included. However, longer range hoppings have values that are always less that $3\%$ of the nearest neighbor hoppings  and are neglected. The estimates presented here are valid for deep enough optical lattices with $V_1 > 3 E_r$ and $V_2 < 0.65 V_1$. Note that higher values of $V_2$ will generate strongly anisotropic lattice minima with large hopping matrix elements along certain directions.

\begin{figure}[tbp]
\begin{center}
\includegraphics[width=0.47\textwidth]{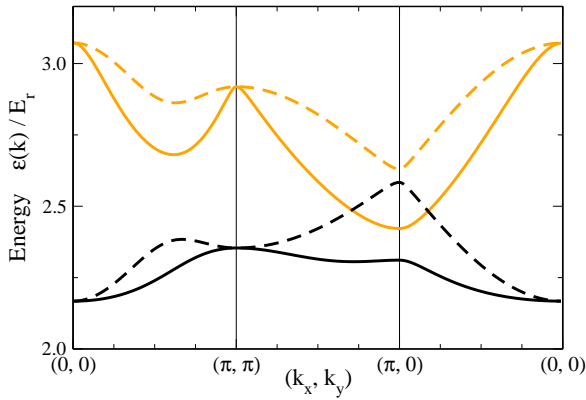}
\end{center}
\caption{(Color online) Energy dispersion for the first two bands of the square super-lattice model along the $(0,0)\rightarrow (\pi,\pi)\rightarrow (\pi,0)\rightarrow (0,0)$ path in the Brillouin zone. The full lines correspond to the parameters from Fig. \ref{Fig4}(c). The dashed lines show the energy dispersion obtained if we neglect the hybridization with higher-energy bands.}
\label{Fig5}
\end{figure}

\subsection{Bulk properties of the square superlattice model}

Before studying the properties of the edge states for the square superlattice model, let us convince ourselves that the system has nontrivial topological properties and therefore can support robust chiral edge states if a boundary is present. Figure \ref{Fig4} shows the spectrum obtained by solving Eq. (\ref{tijchi}) for an infinite lattice (or by imposing periodic boundary conditions). The vertical axis represents the energy and the horizontal axes the wave vector ${\mathbf k}$ taking values within the first Brillouin zone. The hopping and overlap matrix elements correspond to  different sets of original parameters $(V_1, V_2, \alpha)$ for the optical lattice: (a)~ $(3.4, 0, 2)$, (b)~ $(3.4, 1.7, 0)$,  and (c)~ $(3.4, 1.7, 2)$, where $V_i$ are measured in units of recoil energy, $E_r$, and $\alpha$ in units of $\hbar/a$. Notice that a full gap opens only if both the vector potential and the component of the optical lattice potential responsible for the supper-lattice structure, i.e., $\alpha$ and $V_2$, are nonzero. Moreover, for a given strength $\alpha$ of the vector potential there is a critical $V_{2}^*(\alpha)$ above which the full gap opens. For $V_2 < V_2^*(\alpha)$, a negative indirect gap will exist between the top of the first band at $(\pi, \pi)$ and the bottom of the second band at $(\pi, 0)$ or $(0, \pi)$. We note that four other bands, although not shown in Fig. \ref{Fig4}, were included in the diagonalization procedure.

Including the higher-energy bands is crucial for obtaining quantitatively relevant results. As mentioned above, the energy scale in the problem is set by the values of the hopping parameters, which in turn depend strongly on the depth of the optical lattice potential. For example, the nearest-neighbor hoppings contain  exponential factors of the form $\exp\left(-\frac{\pi^2}{8}\frac{\hbar\omega_i}{E_r}\right)$, where $\omega_i$ is given by Eq. (\ref{omgi}). Consequently, to have a large gap compared with the temperatures attainable experimentally, one has to use a lattice potential that is not very deep. In turn, this will determine strong inter-orbital hybridization. This property is exemplified by the results shown in Fig. \ref{Fig5}. The energy dispersion for the two lowest bands along a certain path in k-space was calculated, first including the mixing with higher-energy bands (full lines) and then neglecting it (dashed lines). The two sets of curves, although qualitatively similar, in the sense that both correspond to energy bands separated by a gap, show significant quantitative differences.
\begin{figure}[tbp]
\begin{center}
\includegraphics[width=0.47\textwidth]{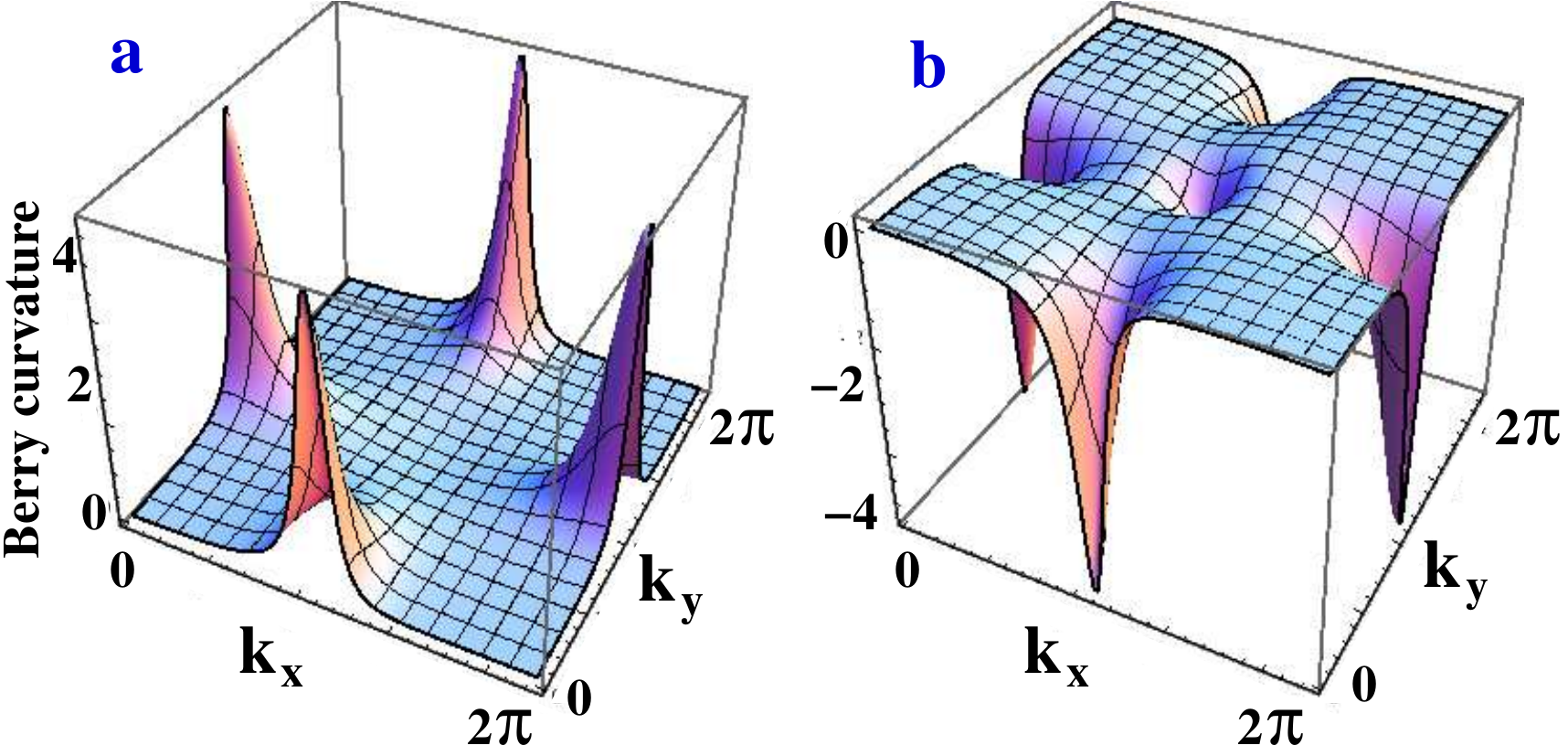}
\end{center}
\caption{(Color online) Momentum dependence of the Berry curvature of the lowest energy bands for the same parameters as in Fig. \ref{Fig4}(c). The integral of the Berry curvature over the first Brillouin zone (i.e., the total flux) is $2\pi$ for the lowest energy band (a) and $-2\pi$ for the second band (b), corresponding to the Chern numbers $1$ and $-1$, respectively. The nonvanishing Chern numbers reveal the nontrivial topological properties of the system.}
\label{Fig6}
\end{figure}

To unveil the topological properties of the band structure described above, we calculate the Berry curvature associated with the momentum space gauge field (or Berry connection) defined for a given band $\lambda$ as $\vec{\it A}_{\lambda}(\vec{k}) = i\langle \Phi_{\lambda \vec{k}}\vert \nabla_{\vec{k}}\vert\Phi_{\lambda \vec{k}}\rangle$~\cite{Thouless,Kohmoto}. The Berry curvature is the effective "magnetic field" generated by this momentum-space gauge field, $F_{\lambda}(\vec{k}) = \partial_{k_x}{\it A}_y(\vec{k}) - \partial_{k_y}{\it A}_x(\vec{k})$. The momentum-space gauge field $\vec{\it A}_{\lambda}({\vec k})$, which is a property of the single-particle wave functions, should not be confused with the real space vector potential ${\mathbf A}({\mathbf r})$, which is an externally applied field.
The distribution of Berry curvature over the Brillouin zone for the two lowest energy bands is shown in Fig. \ref{Fig6}. Note the large values of $F_{\lambda}$ in the vicinity of $(k_x, k_y) = (\pi, 0)$ and $(k_x, k_y) = (0, \pi)$. These are the points in momentum space where the gap closes when the strength of the vector potential approaches zero, $\alpha\rightarrow 0$, leading to the Dirac cone structure shown in Fig. \ref{Fig4}(b). In this limit the Berry curvature diverges at the location of the Dirac points. Similarly, if $V_2\rightarrow 0$, the band gap closes at $(\pi,\pi)$ and the Berry curvature diverges at that point in k-space. The total flux of berry curvature over the first Brillouin zone is an integer multiple of $2\pi$ and defines the Chern number $C_{\lambda} = \frac{1}{2\pi} \int d^2k~F_{\lambda}(\vec{k})$~\cite{Thouless,Kohmoto}. A nonzero value of the Chern number is a signature of the nontrivial topological properties of the system. In the case shown in Fig. \ref{Fig6} the Chern numbers for the first two bands are quantized to $C_1=1$ and $C_2=-1$, respectively. Changing the external parameters $(V_1, V_2, \alpha)$ can significantly modify the shape of the two bands and the distribution of Berry curvature in k-space without altering the Chern numbers. For example, $C_1$ can be modified only by passing through a critical point $(V_1^*, V_2^*, \alpha^*)$ where the band gap closes at least in one point in k-space. This type of transition will be addressed in Sec. IV. Having established that the square superlattice model supports quantum states with a non-trivial topology, we consider now systems with boundaries and study in detail the properties of the states localized in the vicinity of those boundaries.
\begin{figure}[tbp]
\begin{center}
\includegraphics[width=0.47\textwidth]{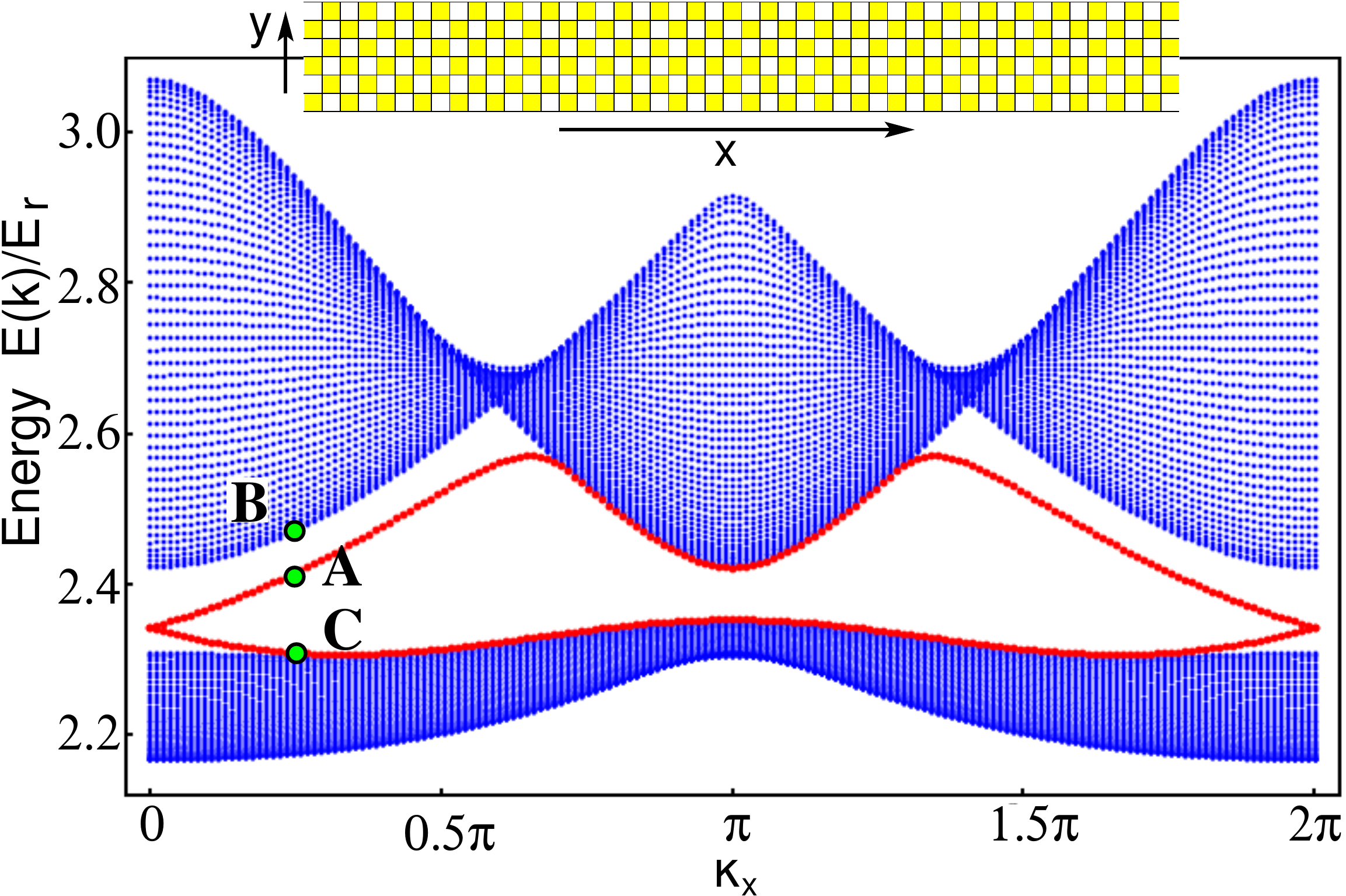}
\end{center}
\caption{(Color online) The band structure of Eq (\ref{H1}) in the stripe geometry. The corresponding lattice is schematically shown in the inset. Notice the edge state modes that populate the gap and merge with the bulk states. The edge modes cross the gap connecting the lower and upper bands and intersect at ${k_x}=0$ due to Kramers degeneracy. Small perturbations can modify the dispersion of the edge modes and the location of the  Kramers degeneracy points, but cannot open a gap for the edge states. Notice that the bulk bands can be obtained by projecting the two-dimensional spectrum shown in Fig. \ref{Fig4}(c) on a plane perpendicular to the $y$ axis. The same set of parameters as in Fig. \ref{Fig4}(c) was used.}
\label{Fig7}
\end{figure}

\section{Edge states: Properties and characterization}\label{Edge1}

In this section we consider a two-dimensional system described by the square super-lattice model in the presence of ideal boundaries, i.e., boundaries created by an infinitely steep potential wall. We discuss the properties of the edge states for systems with either stripe or disk geometry. First we concentrate on the edge states that populate the gap between the lowest energy bands, then we discuss higher-energy edge states.

\subsection{The s-bands edge states}

One defining characteristic of topological insulators is the existence of gapless edge states that are robust against disorder and interactions. While the characterization of topological insulators without boundaries using Berry curvatures and Chern numbers is mathematically elegant, the corresponding experimental manifestations are not straightforward. By contrast, the existence of gapless edge states should be much easier to address experimentally even in cold-atom systems~\cite{Wang,Umuca,StanSarma}, as proved by the experiments on solid state topological insulators~\cite{Konig,Hsieh2008,Hsieh2009,Hsieh2009a,Roushan}. A boundary can be formally introduced by turning on the extra confining potential $V_c({\bf r})$ in Eq. (\ref{H1}). We start with an idealized potential that vanishes in a certain region ${\cal S}$ and is infinite outside. The problems concerning realistic confining potentials will be addressed in Sec. V. We note, however, that the crucial assumption here is not the infinite value of $V_c$ outside ${\cal S}$, as any finite value  $V_c^{max}$ of the order of the total relevant bandwidth or larger produces similar consequences. The key assumption is that the transition between the region with $V_c=0$ and the region with $V_c=V_c^{max}$ is characterized by a length scale of the order of the lattice constant or smaller.

Without translation symmetry, the numerical complexity of the problem increases significantly. Therefore, it is convenient to address the problem of characterizing the edge states in two stages: (i) First, we consider a stripe geometry, in which  ${\cal S}$ is finite along one direction ($y$ in our calculations) but infinite along the orthogonal direction ($x$), and we characterize the edge states that form near the boundaries. (ii) Second, we consider a disk geometry and show that the basic properties of the edge states remain the same while pointing out the properties that depend on the system geometry. We start our analysis by focusing on the edge modes that populate the gap between the first two energy bands, i.e., the bands having the main contributions from $s$-type orbitals $\psi_{0,0}^{({\mathbf r}_j)}({\mathbf r})$. We call these bands "$s$ bands" but remind the reader that significant contributions from higher energy orbitals due to strong hybridization are included.
\begin{figure}[tbp]
\begin{center}
\includegraphics[width=0.47\textwidth]{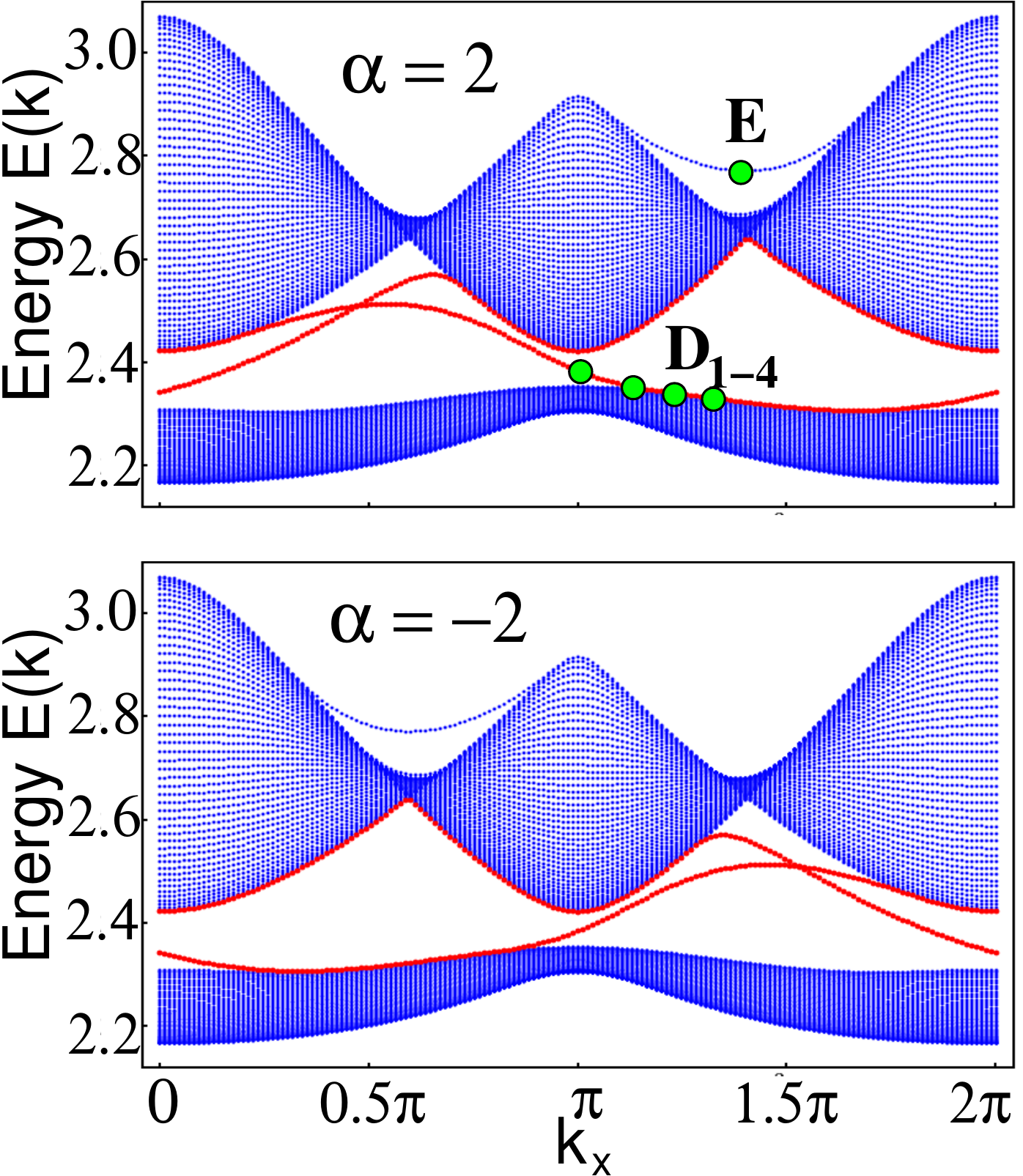}
\end{center}
\caption{(Color online) Band structure for a stripe with inequivalent edges (see main text). All the sites on one boundary belong to the $A$ sublattice, while all the sites on the other boundary are $B$-type. For the top panel the parameters are the same as in Fig. \ref{Fig7}, the only difference being an extra line of lattice sites at one of the boundaries. Notice the completely different dispersion of the edge modes and the different location of the degeneracy point. The symmetry between left-moving and right-moving states is broken. A mirror image of the dispersion lines can be obtained by adding the extra line of lattice sites to the opposite edge, or by reversing the direction of the vector potential, $\alpha\rightarrow -\alpha$ (lower panel). A topologically trivial edge mode develops at the top of the second band (state E belongs to this mode).}
\label{Fig8}
\end{figure}

\subsubsection{Stripe geometry}\label{s_stripe}

Let us consider an optical lattice generated by the potential $V_{latt}$ given by Eq. (\ref{Vlatt}) and having a real space profile as shown in Fig. \ref{Fig3}(a). In the stripe geometry, we consider the lattice as infinite in the $x$ direction and finite in the $y$ direction, as shown schematically in the inset of Fig. \ref{Fig7}. As translation invariance is preserved along the $x$ direction, so $k_x$ is still a good quantum number. For a given value of $k_x$ each band is expected to contain a number of states equal to the number of unit cells along the transverse direction of the stripe. The calculated spectrum corresponding to the first two bands is shown in Fig. \ref{Fig7}. In a stripe geometry, the contribution coming from bulk states can be inferred by projecting the two-dimensional spectrum (see Fig. \ref{Fig4}) on a plane perpendicular to the transverse direction of the stripe. In Fig. \ref{Fig4}(c) the view angle was chosen to visually facilitate this projection. In addition to the bulk contributions, the spectrum in Fig. \ref{Fig7} contains edge modes that populate the bulk gap. These modes cross the gap connecting the lower and upper bands and intersect at $k_x=0$ due to Kramers degeneracy. As we will show below, each of the states having the energy inside the bulk gap is spatially localized near one of the two edges of the system. Small perturbations, like disorder and interactions, or changing the boundary conditions will modify the edge mode dispersion, but the edge states will remain gapless. Considering the Fermi energy somewhere inside the bulk gap, it will always intersect each of the the edge modes containing states localized either near the lower boundary or near the upper boundary an odd number of times, i.e., these edge modes will necessarily connect the lower and upper bands.

\begin{figure}[tbp]
\begin{center}
\includegraphics[width=0.49\textwidth]{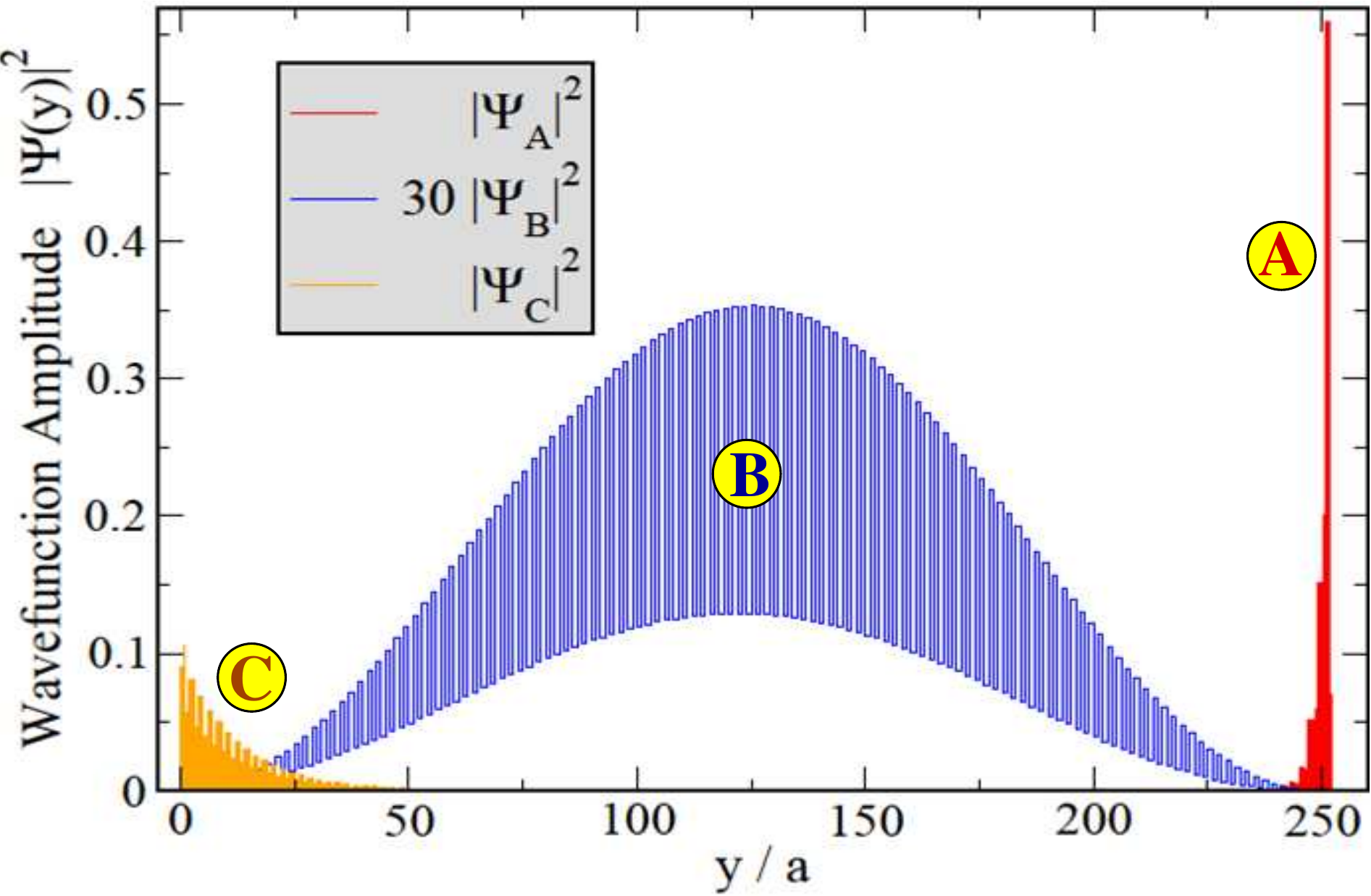}
\end{center}
\caption{(Color online) Spatial dependence of the "amplitude" function $\vert\Psi_{k_x,\nu}\vert^2 = \sum_{(n, m)}\vert \widetilde{\Psi}_{k_x,\nu}^{(n,m)}(y_j)\vert^2$ for the states marked by the letters A, B and C in Fig. \ref{Fig7}. The stripe has a width $d=252$ (in units of $a=1/\sqrt{2}$). State A, which is well inside the bulk gap, is localized near the upper edge and decays exponentially with a characteristic length scale of a few lattice constants. State B, which is at the gap edge, is a bulk state with a smoothly varying envelope function. Notice the multiplication factor of 30 introduced to make the function visible on the same scale as the edge states. State C belongs to the edge mode but is very close to the gap edge. It is localized near the bottom edge and decays exponentially but has a length scale significantly larger than state A.}
\label{Fig9}
\end{figure}

A simple way to exemplify the properties described above is to modify the boundary conditions for the stripe. As we discussed in the previous section when we described the square super-lattice model (see subsection \ref{SQSLM}), the structure of the lattice can viewed as consisting of two inter-penetrating sublattices $A$ and $B$. For an edge along the x-direction all the boundary sites will be of the same type, $A$ or $B$. Consequently, we can construct stripes with edges of the same type and stripes with edges of different types. The example shown in Fig. \ref{Fig7} belongs to the first category. We can modify one of the boundaries by adding (or removing) one line of points and we obtain a stripe that belongs to the second category. The corresponding spectrum is shown in Fig. \ref{Fig8}.
The dispersion of the edge modes is significantly modified as compared to Fig. \ref{Fig7}, as well as the location of the degeneracy point. However, the main property of the edge modes, namely that they connect the lower and upper bands, is not affected. This property is a signature of the topological nature of these edge states. Figure \ref{Fig8} also offers a counterexample, i.e., a topologically trivial edge mode. This mode, which develops at the top of the second band, does not connect two different bands and is not robust, as it can be absorbed into the bulk continuum in the presence of small perturbations.
\begin{figure}[tbp]
\begin{center}
\includegraphics[width=0.49\textwidth]{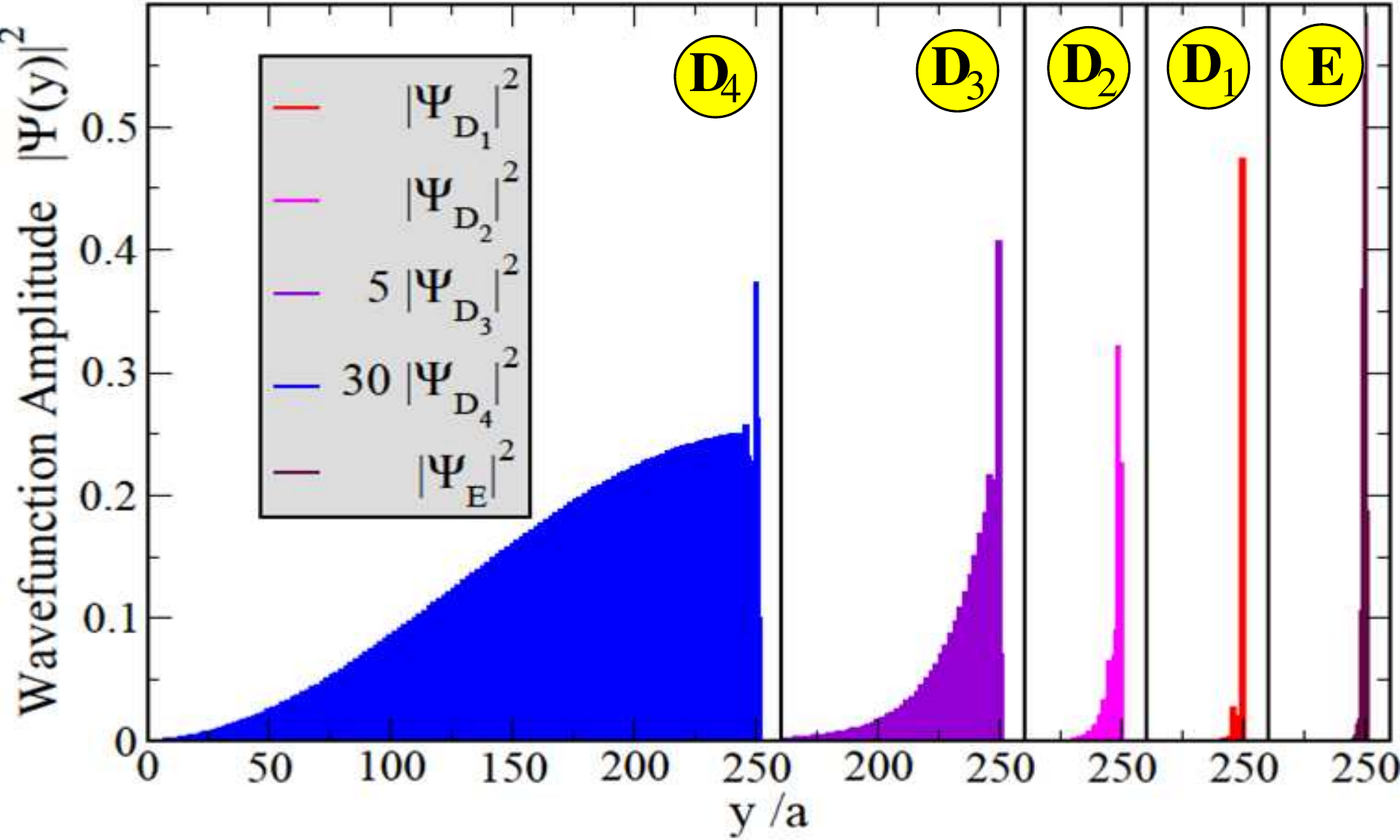}
\end{center}
\caption{(Color online) Spatial dependence of the "amplitude" function $\vert\Psi_{k_x,\nu}\vert^2 = \sum_{(n, m)}\vert \widetilde{\Psi}_{k_x,\nu}^{(n,m)}(y_j)\vert^2$ for state E and for the sequence of states $D_1 \rightarrow D_4$ from Fig. \ref{Fig8}. $D_1$ is positioned in the middle of the gap, while  $D_4$ is at the gap edge and has bulk character. During this transition the characteristic length scale for the exponential decay of the edge states increases continuously from a value of a few lattice sites to a value comparable to the width of the stripe. The state E has a very pronounced edge character, but is not topologically protected. Note that the horizontal axis was translated for clarity.}
\label{Fig10}
\end{figure}

So far we referred to the in-gap states as edge states without showing explicitly that they are indeed localized near the boundary of the system. If for a given wave vector $k_x$ we order the single-particle states according to their energy so $\Phi_{1,k_x}$ is the lowest energy state, then the spatial properties of a generic state are given by the norm $\vert\Phi_{\nu,k_x} ({\mathbf r})\vert^2$, where Eq. (\ref{Phi}) is used with the amplitudes $\Psi_{\nu,k_x}^{(n,m)}$ being solutions of Eq. (\ref{tijchi}). However, such a detailed description of the spatial dependence of the wave function is  not necessary for our purpose and instead we focus on the dependence of the envelope function, which does not contain the details of the orbital structure,  on the transverse coordinate $y$. More precisely, for a state $(\nu, k_x)$ we define the "density" or "amplitude" function $\vert\Psi_{\nu,k_x}\vert^2 = \sum_{(n, m)}\vert \widetilde{\Psi}_{\nu, k_x}^{(n,m)}(y_j)\vert^2$, where $\widetilde{\Psi}_{\nu,k_x}^{(n,m)}(y_j)$ is the Fourier transform of ${\Psi}_{\nu,k_x}^{(n,m)}(x_j, y_j)$ with respect to $k_x$. Note that the "density" is normalized, $\sum_j\vert\Psi_{\nu,k_x}\vert^2(y_j) = 1$.
The spatial dependence of the "amplitude" function for the states marked by the letters A, B and C in Fig. \ref{Fig7} is shown in Fig. \ref{Fig9}. The figure shows clearly that the states within the gap (A and C) are indeed localized in the vicinity of one of the two edges of the system and decay exponentially away from the boundary. For states well inside the gap the characteristic length scale is of the order of the lattice constant. This length scale increases as the edge mode merges into the bulk states.
\begin{figure}[tbp]
\begin{center}
\includegraphics[width=0.49\textwidth]{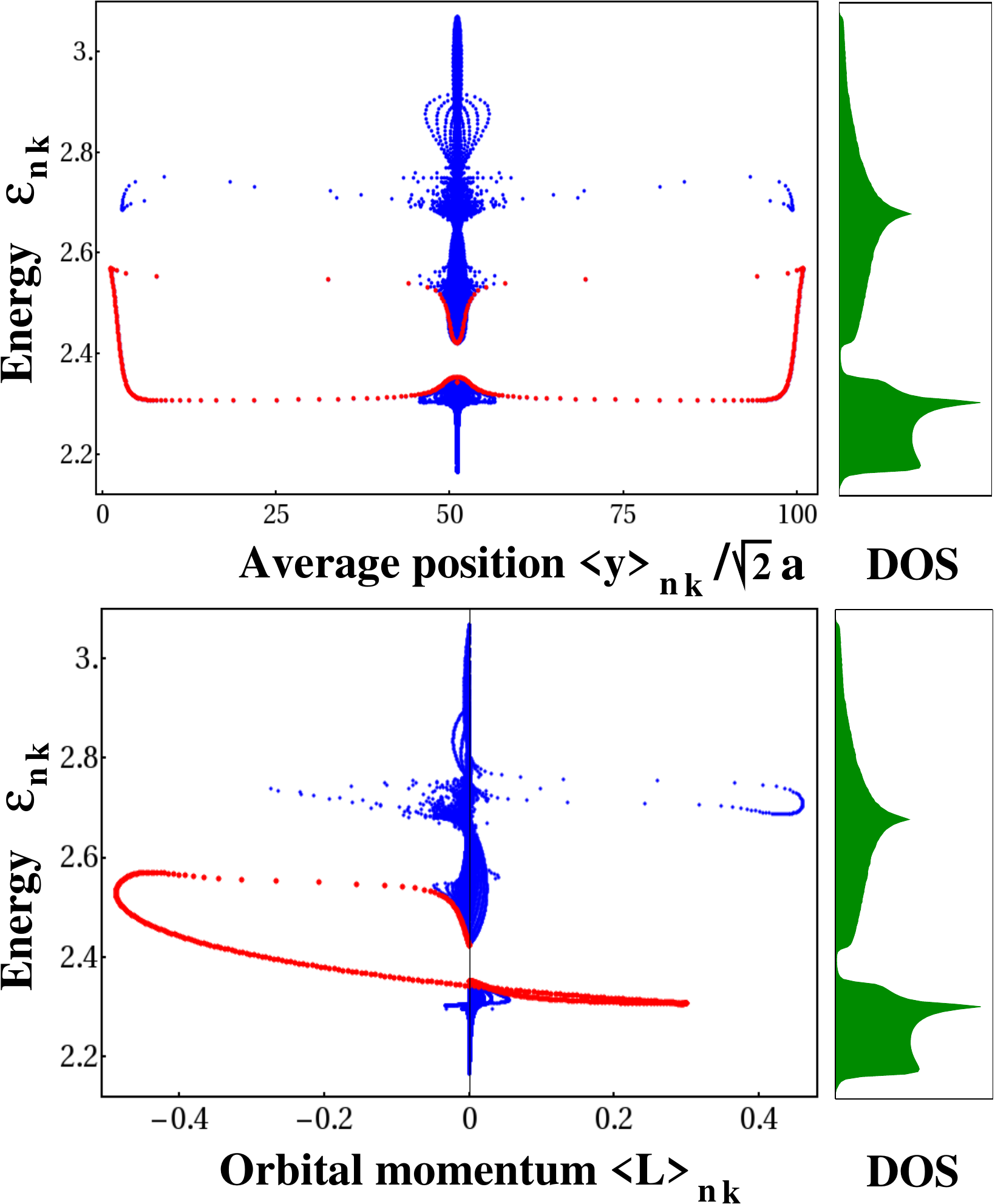}
\end{center}
\caption{(Color online) (Upper panel) Average position versus energy for the single particle states that are solutions of Eq. (\ref{H1}) in the stripe geometry (with equivalent edges). The system is characterized by the parameters $V_1 = 3.4 E_r$, $V_2 = 1.7 E_r$ and $\alpha=2~\hbar/a$ and has a width $d = 100 (\sqrt{2}a)$. The bulk states are characterized by $\langle y\rangle_{n k_x} \approx d/2$, while the edge states have $\langle y\rangle_{n k_x} \approx 0$ or $\langle y\rangle_{n k_x} \approx d$, corresponding to the positions of the two edges. (Lower panel) Average orbital momentum (in arbitrary units) versus energy for the same system. Notice the chiral nature of the topological edge states and the extra edge modes located in the upper band. The spectrum for this system is shown in Fig. \ref{Fig7} and the density of states (DOS) is shown in the right panels.}
\label{Fig11}
\end{figure}

To examine further the transition from edge to bulk states we show in Fig. \ref{Fig10} the "amplitude" function for the sequence of states $D_1 \rightarrow D_4$ from Fig. \ref{Fig8}. The state $D_1$ is positioned in the middle of the gap and in real space it decays exponentially away from the top edge with a length scale of a few lattice constants. As the edge states  mode approaches the gap edge (states $D_2$ and $D_3$) the characteristic length scale increases and eventually becomes comparable to the size of the system. $D_4$ is a bulk state with a very small amplitude near the top boundary. Also shown in
Fig. \ref{Fig10} is the state E from Fig. \ref{Fig8}. This is a state with a very pronounced edge character, but which is not topologically protected, as discussed above.

In the spectra shown in Figs. \ref{Fig7} and \ref{Fig8} the topologically protected edge modes are very well defined, yet the edge mode at the top of the second band (state E) is manifest only for the inequivalent edge stripe.
\begin{figure}[tbp]
\begin{center}
\includegraphics[width=0.42\textwidth]{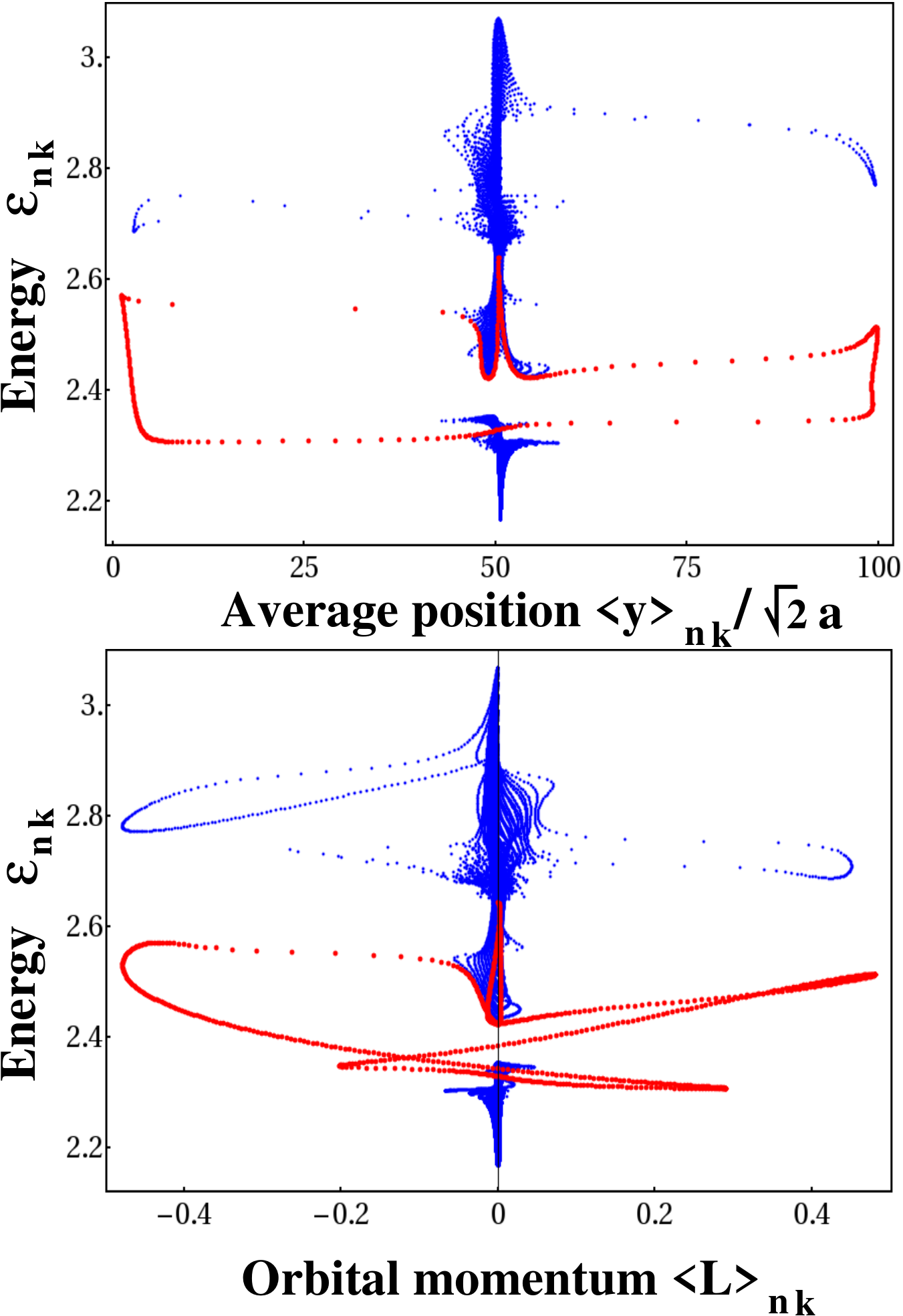}
\end{center}
\caption{(Color online) Average position (upper panel) and average orbital momentum (lower panel) versus energy for the single-particle states that are solutions of Eq. (\ref{H1}) in the stripe geometry (inequivalent edges). The system is characterized by  $V_1 = 3.4 E_r$, $V_2 = 1.7 E_r$ and $\alpha=2~\hbar/a$ and has a width $d = 101 (\sqrt{2}a)$. The spectrum for this system is shown in Fig. \ref{Fig8}. \vspace{-3mm}}
\label{Fig12}
\end{figure}
However, a detailed analysis reveals the existence of edge states at the top of the second band even for equivalent edge stripes. This raises the more general question of distinguishing between bulk and edge states and representing this difference. Of course, determining the "amplitude" of each single-particle state will provide the answer, but this is a rather cumbersome process and a more global characterization would be desirable instead. Two possible quantities that offer such a characterization are the average position in the transverse direction, $\langle y \rangle_{\nu, k_x} = \langle \Phi_{\nu,k_x}\vert y \vert \Phi_{\nu,k_x}\rangle$, and the average orbital momentum, $\langle L \rangle_{\nu, k_x} = \langle \Phi_{\nu,k_x}\vert L \vert \Phi_{\nu,k_x}\rangle$.
Diagrams of these average quantities versus the energy are shown in Fig. \ref{Fig11} for a stripe with equivalent edges and Fig. \ref{Fig12} for a stripe with inequivalent edges. The corresponding spectra were already shown in Figs. \ref{Fig7} and \ref{Fig8}, respectively. Each point in these diagrams corresponds to a single-particle state, solution of Eq. (\ref{H1}). The edge-type states are characterized by average positions corresponding to the location of the two boundaries and relatively large orbital momenta. These can be easily distinguished from the bulk like states, which are characterized by average positions close to the middle of the stripe and which carry small orbital momenta.  The topologically unprotected edge modes that can be partially seen in Fig. \ref{Fig8} but are totally obscured by bulk states in Fig. \ref{Fig7} can now be easily identified. If we focus on the topological edge states within the gap ($\epsilon_{n k_x} \approx 2.4$), note that for a stripe with equivalent edges (Fig. \ref{Fig11}) the two modes localized on the opposite boundaries carry the same orbital momentum, thus revealing their chiral nature.
In other words, for each energy within the gap there is a pair of counter propagating edge states localized on opposite edges. For a stripe with inequivalent edges (Fig. \ref{Fig12}) this symmetry is broken, and for some energy values within the gap it possible to find states localized on opposite edges, yet propagating in the same direction. However, the negative orbital momentum of one state is always larger (in absolute value) than the positive orbital momentum of its pair.

\begin{figure}[tbp]
\begin{center}
\includegraphics[width=0.43\textwidth]{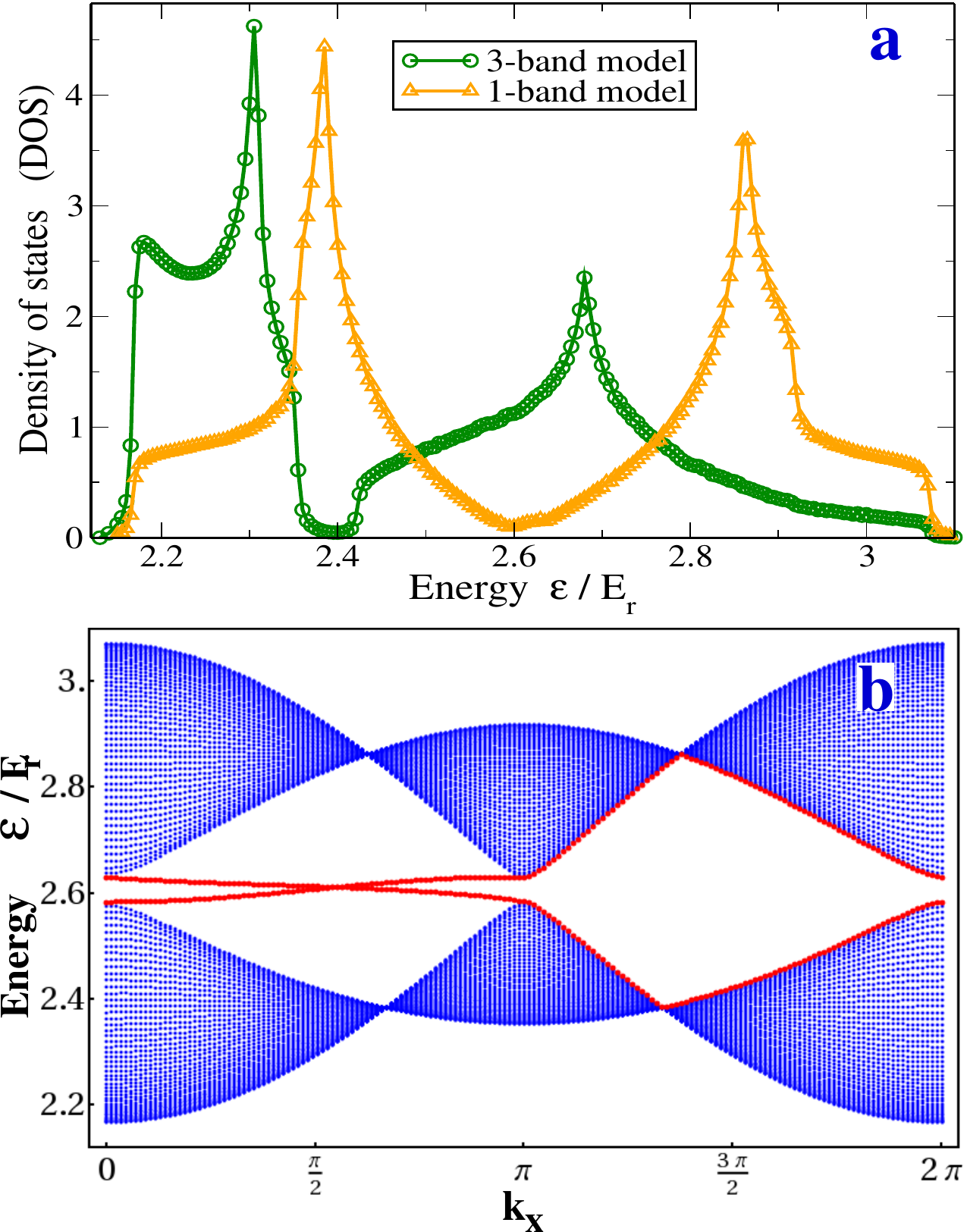}
\end{center}
\caption{(Color online) Density of states (a) and spectrum (b) showing the $s$ bands of a stripe with inequivalent edges calculated within a simplified tight-binding approximation that neglects hybridization with higher bands. For comparison, the density of states calculated within a three-orbital tight-binding approximation is also shown. The corresponding spectrum is given in Fig. \ref{Fig8} (top panel). Note that the topology of the edge modes is not altered, in spite of a significant redistribution in the density of states.\vspace{-3mm}}
\label{Fig13}
\end{figure}

The usefulness of these diagrams showing the average position (orbital momentum) versus energy is even greater for geometries without any translation symmetry, when the standard energy versus momentum spectra cannot be constructed. Before we switch to a different geometry, let us note that the fundamental properties of the topologically protected edge states are not affected by approximations used in the calculation as long as the bulk gap is preserved. Shown in Fig. \ref{Fig13} are the density of states and the spectrum of a stripe with inequivalent edges calculated for the $s$ bands within a simplified tight-binding approximation that neglects the hybridization with higher-energy bands. The density of states for the same system calculated within a three-orbital approximation is also shown for comparison, while the corresponding spectrum is presented in Fig. \ref{Fig8} (top). We can say that the edge states are protected against approximations, as long as these approximations do not affect the gap structure of the (bulk) spectrum. This is not surprising, as approximations can be viewed as effective perturbations applied to the Hamiltonian.

\subsubsection{Disk geometry}\label{s_disk}

So far we have discussed the properties of the edge states in systems with translation symmetry in one direction (stripes). Because momentum along one direction is a good quantum number, spectra showing the energy dispersion as a function of momentum are a very effective way of characterizing the system and, in the case of condensed matter systems, have a direct connection with experimentally measurable quantities. By contrast, cold-atom systems may contain a relatively small number of sites, so the explicit treatment of a finite system may be required, and have a circle or an ellipse as the most natural shape for the boundary. This raises two questions: (i) What is the impact of the boundary geometry on the edge states? and (ii) How important are the finite size effects for the stability of the edge states? We start by addressing the first question, while the second  will be discussed in Sec. V.

\begin{figure}[tbp]
\begin{center}
\includegraphics[width=0.47\textwidth]{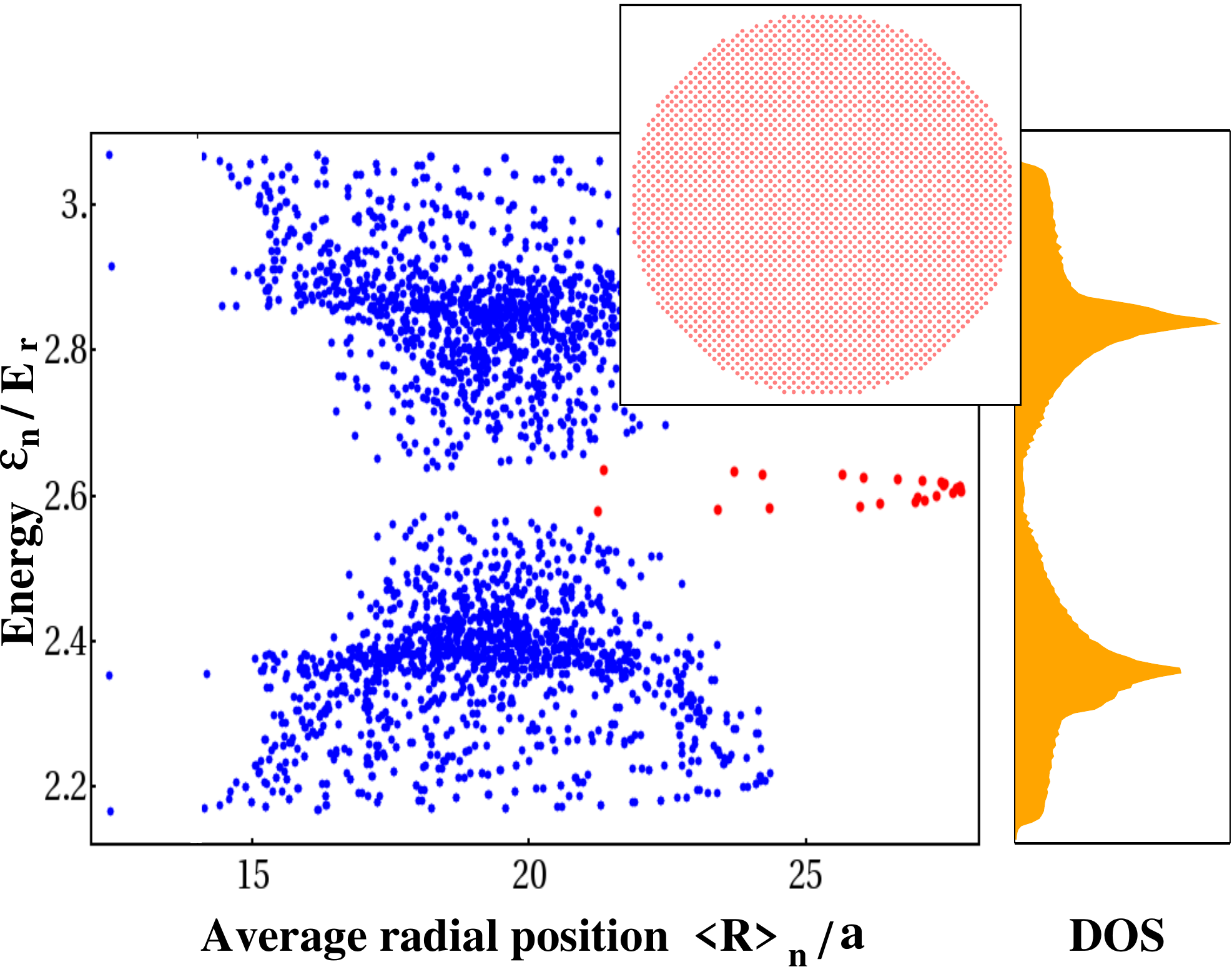}
\end{center}
\caption{(Color online) Diagram showing the average radial position versus energy for the single-particle states of a system described by Eq. (\ref{H1}) in a confining potential (\ref{VcR}). The parameters of the system are $V_1 = 3.4 E_r$, $V_2 = 1.7 E_r$, and $\alpha=2~\hbar/a$. The underlying lattice (i.e., the minima of the effective optical lattice potential) is shown in the inset. The edge states are characterized by values of $\langle r \rangle_n$ comparable with $R_0=29a$, the disk radius, as a result of their localization in the vicinity of the boundary. The clearly defined  edge mode crosses the bulk gap and connects the lower and upper bands, thus revealing its topological nature. The density of states (right) is practically identical with that shown in Fig. \ref{Fig13}(a) for a similar system with stripe geometry.\vspace{-3mm}}
\label{Fig14}
\end{figure}

Let us consider the single particle quantum problem described by Eq. (\ref{H1}) with an extra confining potential given by
\begin{equation}
V_c({\mathbf r}) = \left\{
\begin{array}{c}
0 ~~~~~~ \mbox{if}~~\vert {\mathbf r}\vert \leq R_0, \\
\infty ~~~~~ \mbox{if}~~\vert {\mathbf r}\vert > R_0.
\end{array}\right.    \label{VcR}
\end{equation}
The system consists of a disk-shaped piece of the square super-lattice with a boundary that contains sites from both sublattices, $A$ and $B$ with a distribution that depends on the radius $R_0$. For simplicity, we solve the problem within the single-orbital tight-binding approximation, as the full size matrices (i.e., $N\times N$ matrices, with $N$ the number of lattice sites inside the disk) have to be used in Eq. (\ref{tijchi}).
To describe globally the system we use the type of diagrams introduced in the previous section. More precisely, we represent the average radial position for a given single-particle state, $\langle r \rangle_n = \langle \Phi_{n}\vert r \vert \Phi_{n}\rangle$, versus the state energy, $\epsilon_n$. The results for a disk with radius $R=29 a$ are shown in Fig. \ref{Fig14}.
The main conclusion suggested by the data is that the edge mode is robust against deformations of the boundary. The states with energies near the middle of the gap are localized within a few lattice spacings from the boundary, while this characteristic length increases as one approaches the gap edge. The number of edge states is proportional to the length of the boundary (i.e., $R_0$), while the number of bulk states scales with the area of the system (i.e.,  $R_0^2$). Finally, we note that the topologically unprotected edge states  that were present in the stripe geometry (see Figs. \ref{Fig8}, \ref{Fig11}, and \ref{Fig12}) do not survive in the absence of translational symmetry.

\begin{figure}[tbp]
\begin{center}
\includegraphics[width=0.47\textwidth]{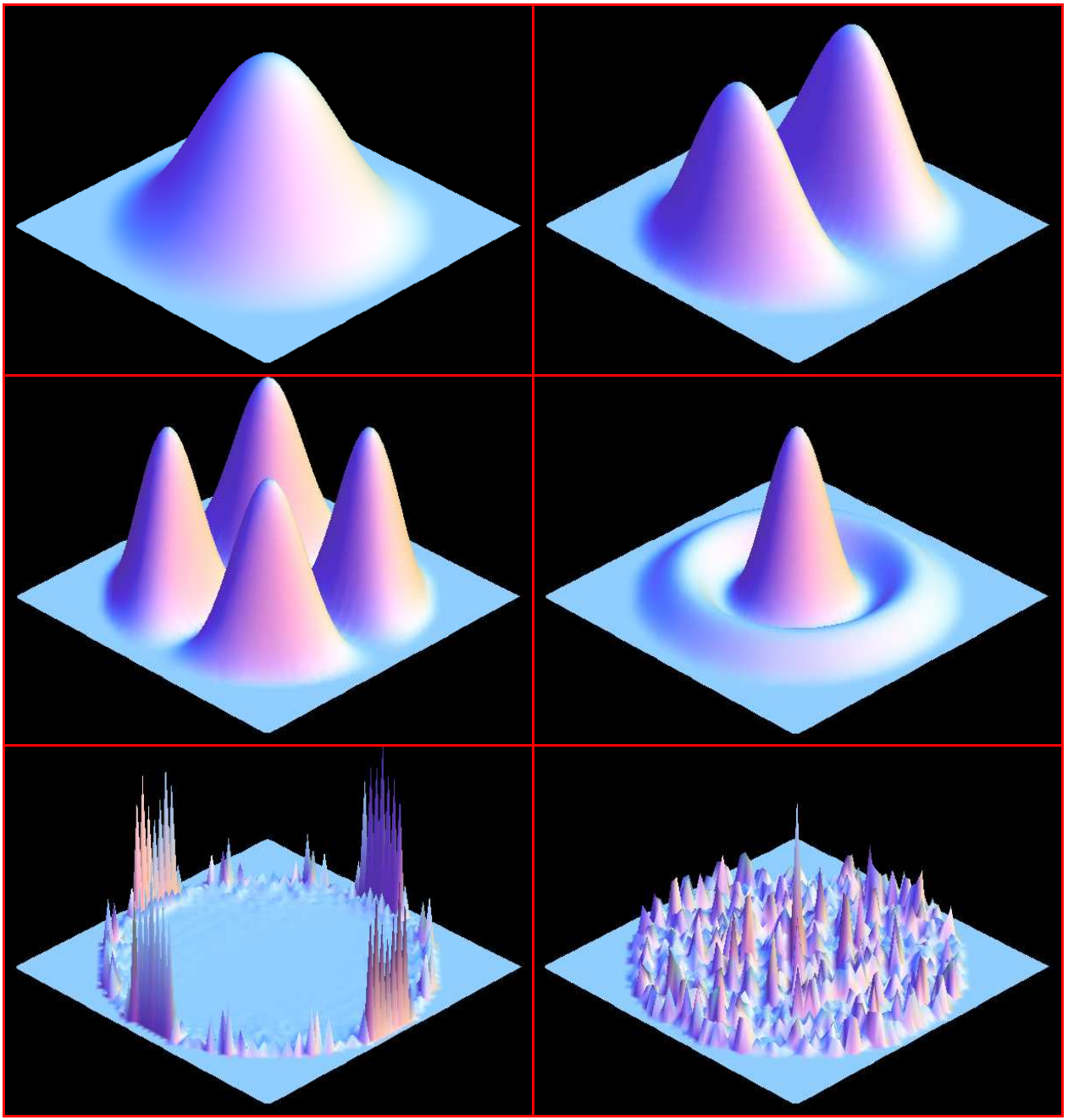}
\end{center}
\caption{(Color online) "Amplitude" functions for single-particle states in the disk geometry. The top and middle panels show "densities" corresponding to the first four energy levels starting with the ground state (top left corner). We note that the actual density is obtained by multiplying these envelope functions with a sum of $s$-type orbitals centered at each lattice site. The lower panels show a typical edge state (left) and a typical bulk state (right). The edge state decays exponentially away from the boundary, with a characteristic length scale of a few lattice constants. \vspace{-3mm}}
\label{Fig15}
\end{figure}

To visualize the spatial dependence of the single particle states in the disk geometry, we show in Fig. \ref{Fig15} the "amplitude" function $\vert\Psi_{n}\vert^2 = \vert {\Psi}_{n}^{(0,0)}({\mathbf r}_j)\vert^2$ for several states. The edge state shown in the bottom-left panel has an energy near the middle of the bulk gap. The amplitude of the edge state decays exponentially away from the boundary, with a characteristic length scale of the order of the lattice constant. This length scale increases for states with energies closer to the gap edge and eventually becomes comparable to the system size (i.e., to $R_0$) as the edge mode merges with the bulk bands. This behavior, as well as the characteristic length scales, are the same as those observed in the stripe geometry and are independent of the boundary geometry. What depends on the details of the boundary is the actual distribution of the "density" {\it along} the boundary. For example, the edge state shown in Fig. \ref{Fig15} has four regions with higher amplitude. These regions correspond to sections of the boundary that contain only sites that belong to one of the sublattices and are locally similar to the boundary in the stripe geometry. Modifying those regions determines a "density"  redistribution along the boundary, but the transverse properties (e.g., the characteristic length scale for the exponential decay) are not affected. In conclusion, the fundamental properties of the edge states do not depend on the geometry of the system and, therefore, can be studied using the most convenient geometry. However, if we are interested in the detailed behavior of the edge states along the boundary, a precise characterization of this boundary is required and has to be explicitly included in the calculations.

\subsection{Edge states in a multi-band system}
\label{multi_band}

So far we have discussed the properties of the edge states that populate the gap between the lowest energy bands. From a practical point of view, in cold atom systems these may be the most relevant states for two reasons: (i) the extra confining potential necessary for defining a boundary~\cite{StanSarma} for higher energy bands has to be stronger  and may be harder to realize and (ii) for a relatively shallow optical  lattice, which is the optimal condition for observing topological edge states, the higher bands may strongly overlap, thus filling any possible gap. Nonetheless, in some cases the p-type bands may offer some advantages, most notable the possibility of having larger bulk gaps and the direction dependence of the $p$-orbitals (see Sec. II), which can be critical in certain models. In addition, from a theoretical standpoint it is interesting to investigate if there is any major difference between various types edge states that may be present in a multi-band system. We note that all the results presented in this section for the $p$-type bands are qualitative. Quantitative results would require taking into account contributions from several higher energy orbitals, as they hybridize strongly with the $p$ orbitals.
\begin{figure}[tbp]
\vspace{-7mm}
\begin{center}
\includegraphics[width=0.47\textwidth]{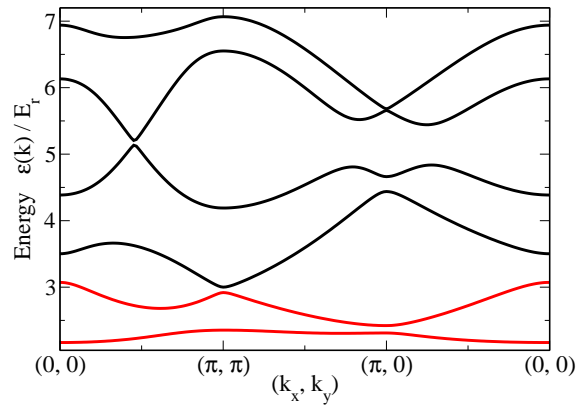}
\end{center}
\caption{(Color online) Energy dispersion for the first six bands
of the square super-lattice model along the $(0, 0) \rightarrow (\pi, \pi) \rightarrow (\pi, 0) \rightarrow (0, 0)$ path in the Brillouin zone. The parameters used in the calculation are $V_1=3.4 E_r$, $V_2=1.7 E_r$, and $\alpha=2\hbar/a$. The $s$ bands (red lines) are identical with those shown in Fig. \ref{Fig5}. \vspace{-3mm}}
\label{Fig16}
\end{figure}

Shown in Fig. \ref{Fig16} is the spectrum of single-particle Hamiltonian (\ref{H1}) with $V_c({\mathbf r}) = 0$ (no boundaries) obtained in a three-orbital tight-binding approximation. The parameters for this calculation are $V_1=3.4 E_r$, $V_2=1.7 E_r$, and $\alpha=2\hbar/a$, i.e., the parameters used to derive the majority of the results in the previous subsection. For these parameters most of the $p$ bands overlap and no gap opens except between the fourth and the fifth bands (at energies around $5.2 E_r$). The first question that we address is whether there are any other topological edge states except those located in the gap between the first two bands, which we studied in the previous subsection. To answer this question, we can either consider a system with boundaries and identify the edge modes or remain within a bulk description and calculate the Chern numbers $C_n$ for each band.
\begin{figure}[tbp]
\begin{center}
\includegraphics[width=0.49\textwidth]{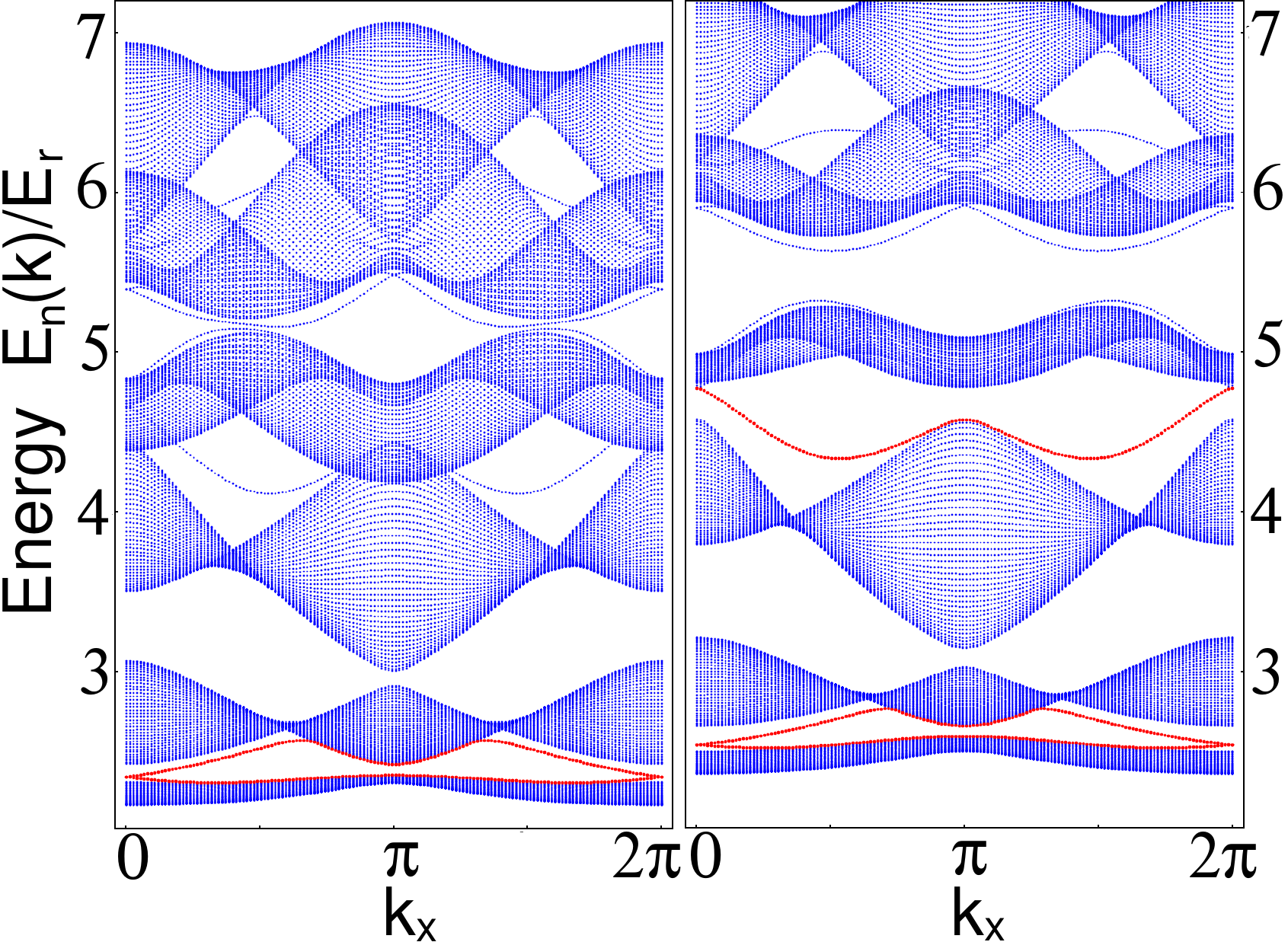}
\end{center}
\caption{(Color online) The band structure of the Hamiltonian given by Eq. (\ref{H1}) in the stripe geometry. The left panel corresponds to the parameters  $V_1=3.4 E_r$, $V_2=1.7 E_r$ and $\alpha=2\hbar/a$, while the right panel corresponds to a slightly deeper lattice with $V_1=3.7 E_r$. Note that there are no edge states between the second and the third bands, but an edge mode is clearly visible between the third and the fourth bands. Increasing  $V_1$ opens a gap in the spectrum (right panel) and reveals the topological nature of that edge mode. By contrast, the edge states that populate the gap between the fourth and the fifth bands are topologically trivial, as they do not connect the two bands. \vspace{-3mm}}
\label{Fig17}
\end{figure}
We start with the second approach, as it is much easier to implement numerically. The Chern numbers for the first four bands are $C_1=1$, $C_2=-1$, $C_3=1$, and $C_4=-1$. The last two bands are degenerate at ${\mathbf k} = (0, \pi)$ and  ${\mathbf k} = (\pi, 0)$. As the Berry curvature diverges in the vicinity of the degeneracy points, the Chern numbers for those bands are not defined. In principle, topological edge states exist in a gap if the {\it total} curvature (i.e., the sum of the Chern numbers) of all the bands below that gap is nonzero. The first band has non-zero curvature and topological edge states exist inside the gap above it, as we have seen above. The first two bands, {\it as a whole}, have zero {\it total} curvature and, consequently, no topological edge states should be present on the top of the second band. This is consistent with our previous observation of the topologically unprotected edge states in the stripe geometry. Similarly, we expect topological edge states to exist between the third and the fourth bands (if a full gap is opened), but not between the fourth and the fifth.

To confirm that the structure inferred from the values of the Chern numbers is indeed realized, we calculate the band structure of a system with boundaries in the stripe geometry. The results are shown in Fig. \ref{Fig17} and offer a picture that is consistent with the above analysis. Two features are worth mentioning. First, note that the topological edge mode that develops inside the gap between the third and fourth bands (Fig. \ref{Fig17}, right panel)  partially survives the gap collapse (Fig. \ref{Fig17}, left pane). These edge states are protected by the translation symmetry and are robust against perturbations that conserve this symmetry. However, they will be destroyed by the presence of disorder. Second, we notice some very well defined edge modes inside the gap between the fourth and fifth bands. These modes are good example of topologically trivial edge states: they do not cross the gap connecting two different bands, but rather start from and return to the same band. Consequently, at any given energy there will be an even number of such edge states near a given boundary and any small perturbation will make them to become localized. By contrast, in the case of topological edge states there is an odd number of states near a given boundary at any given energy within the gap, so even in the presence of perturbations a dispersive mode consisting of de-localized states is preserved.

\begin{figure}[tbp]
\vspace{-7mm}
\begin{center}
\includegraphics[width=0.47\textwidth]{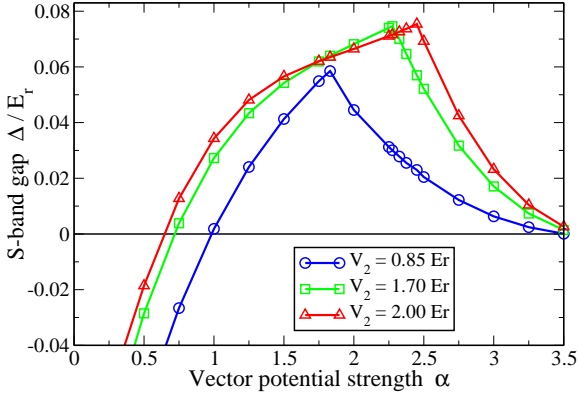}
\end{center}
\caption{(Color online)  The $s$ band gap as function of the vector potential strength for several values of the super-lattice generating component $V_2$. The main amplitude of the optical lattice potential is fixed, $V_1=3.4 E_r$. Notice that each curve has a singularity point and that the slope to the left (right) of the singularity point is positive (negative). The positive slope correspond to an indirect gap between the maximum of the lower band at $(\pi, \pi)$ and the minima of the upper band at $(0,\pi)/(\pi, 0$, while the negative slope corresponds to a direct gap at $\pi, \pi)$. Note that for weak vector potentials with the strength $\alpha$ below a critical value $\alpha^*(V_2)$ the indirect gap becomes negative. \vspace{-3mm}}
\label{Fig18}
\end{figure}

As we mentioned at the beginning of this subsection, including higher energy orbitals is expected to modify quantitatively our picture of the $p$ bands. However, we expect minor changes as far as the $s$ bands are concerned. Therefore it is relevant and useful to have a quantitative estimate of the $s$ band gap and a general idea on how it depends on the parameters of our model. Shown in Fig. \ref{Fig18} is the gap dependence on the strength of the vector potential for three different values of $V_2$. Depending on the parameters, the system has either a direct gap at $\pi, \pi)$ or an indirect gap between the maximum of the lower band at $(\pi, \pi)$ and the minima of the upper band at $(0,\pi)/(\pi, 0$. Note that the indirect gap becomes negative for $\alpha<\alpha^*(V_2)$, i.e., vector potentials with a strength lower than a certain critical value.

\section{Transitions between Topologically Distinct Quantum States}\label{transitions}

In this section we give several examples of phase transitions between quantum states with distinct topological properties. These transitions can be induced by applying certain perturbations, i.e., adding some extra terms to the Hamiltonian, or, in the case of a multi-band topological insulator, by simply varying the parameters that characterize the system.

The topological properties of a quantum state are not modified by any perturbation of the Hamiltonian that does not close the bulk gap. Nonetheless, when such a perturbation determines the closing of the bulk gap, the system undergoes a phase transition to either a metallic state or an insulating state with possibly different topological properties. The topological-insulator-to-metal transition can be exemplified by the band structure shown in Fig. \ref{Fig17}. Let us assume that the first three bands in the right panel are completely filled, so the only gapless excitations are provided by the edge mode that crosses the gap between the third and fourth bands. By simply reducing the depth of the optical lattice, this bulk gap collapses (see left panel) and the system becomes metallic. More interesting are the transitions between two different insulating states. We discuss two possible ways of inducing such transitions: (i) by adding an extra-term to the Hamiltonian that opens a (topologically trivial) gap and (ii) by tuning the parameters $V_1$, $V_2$, and $\alpha$ that characterize the system.
\begin{figure}[tbp]
\begin{center}
\includegraphics[width=0.47\textwidth]{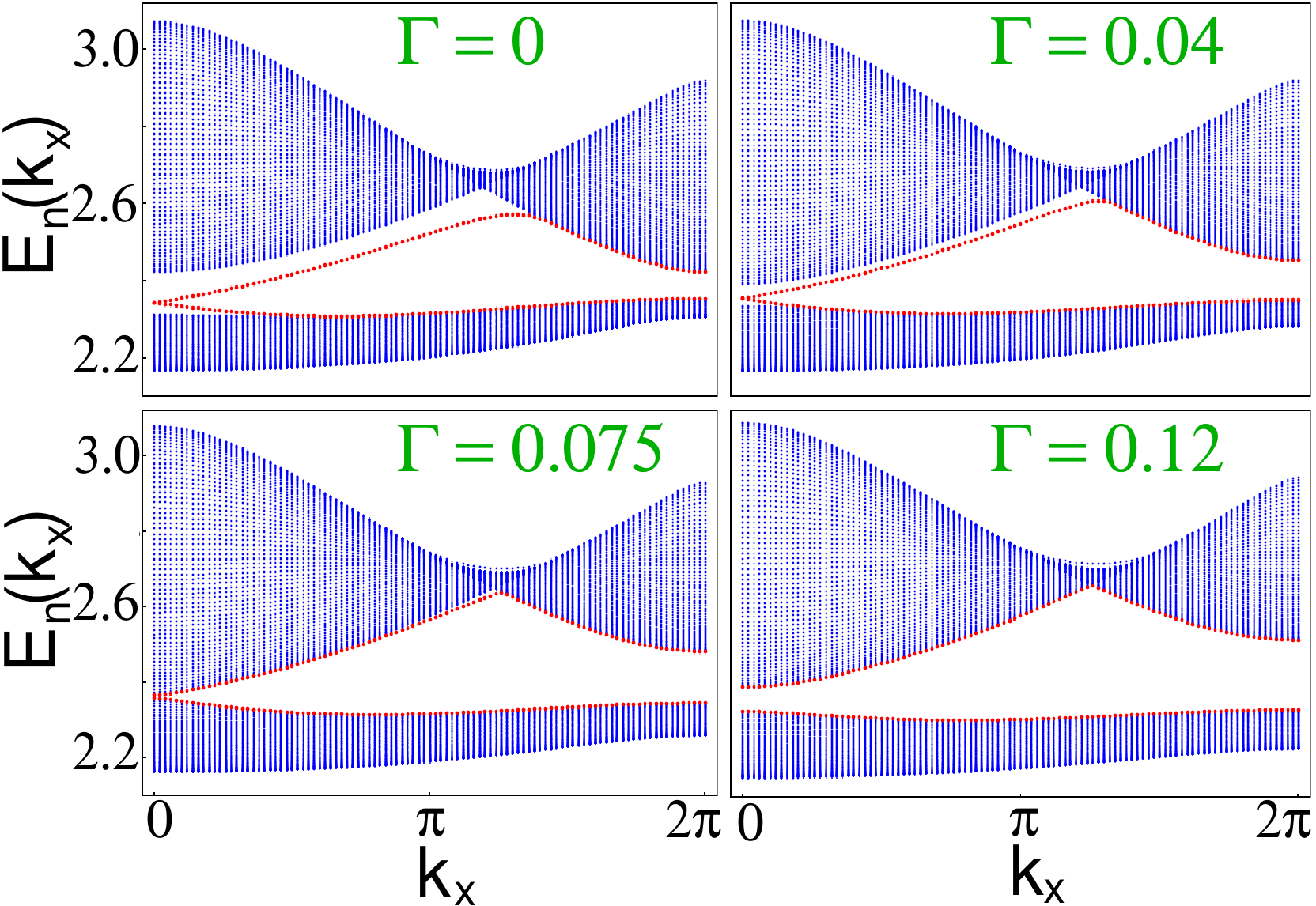}
\end{center}
\caption{(Color online)  Band structure for a system with stripe geometry described by the parameters $V_1=3.4 E_r$, $V_2=1.7 E_r$, $\alpha=2\hbar/a$
and different values of the staggered potential  $\Gamma$ (given in units of $E_r$). $\Gamma=0$ corresponds to the case shown in Fig. \ref{Fig7}. Applying a  small staggered potential reduces the gap in the vicinity of $k_x=0$ (top right panel). At the critical value $\Gamma_c \approx 0.075 E_r$ the gap closes at $k_x=0$, while for  $\Gamma>\Gamma_c$ a full gap opens again. At large values of $\Gamma$ the system is a standard band insulator with no edge modes inside the gap. Notice that the spectra are shown for half of the one-dimensional Brillouin zone while the other half can be obtained by mirror symmetry with respect to $k_x=\pi$.  \vspace{-3mm}}
\label{Fig19}
\end{figure}

\subsection{Transitions driven by a staggered potential}\label{tr_stag}

The tight-binding model described by Eq. (\ref{modelH}) is defined on a square
super lattice consisting of two inter penetrating sublattices $A$ and $B$, as discussed in Sec. I. If the second neighbor hopping is anisotropic, $t_2^{\prime} \neq t_2$, and the nearest-neighbor hopping is complex, $\phi \neq 0$ and $\phi \neq \pi$, a full gap $\Delta({\mathbf k})$ opens in the spectrum, with an explicit wave vector dependence given by
\begin{eqnarray}
&&\Delta^2(k_x, k_y) = 4(t_2-t_2^{\prime})^2(\cos k_x - \cos k_y)^2 \label{gap}\\
&+& 16\vert t_1\vert^2\left[1+\cos k_x \cos k_y + \cos(2\phi)(\cos k_x+\cos k_y) \right]. \nonumber
\end{eqnarray}
For the cold-atom realization of this model described by Eq. (\ref{H1}), the condition for anisotropic second-neighbor hopping becomes $V_2 \neq 0$, while the imaginary components of the nearest-neighbor hopping are generated by the effective vector potential, so the second condition becomes $\alpha \neq 0$.  Spectra for the Hamiltonian  (\ref{H1}) corresponding to three different sets of parameters are shown in Fig. \ref{Fig4}. Another possibility to open a gap in a square lattice tight-binding model is to simply add a staggered potential, i.e., a potential that generates on-site energies $\Gamma$ and $-\Gamma$ for the sublattices $A$ and $B$, respectively. This amounts to adding a term of the form
\begin{equation}
V_{\mbox{stagg}} = \Gamma \sum_{\mathbf k} \left(c_{A{\mathbf k}}^{\dagger}c_{A{\mathbf k}} - c_{B{\mathbf k}}^{\dagger}c_{B{\mathbf k}} \right) \label{Vstagg}
\end{equation}
to the Hamiltonian. In the presence of such a term, a simple tight binding  model with $t_2^{\prime} = t_2$ and $\exp(i\phi)=\pm 1$ is characterized by a full gap with a minimum value $\Delta_{min} = 2\vert\Gamma\vert$. If the lowest band is completely filled, the system represents a standard band insulator with trivial topological properties. The question that we want to address concerns the evolution of the spectrum and the fate of the edge modes as the transition between a topological insulator and a standard band insulator is induced by tuning the staggered field strength.

In the cold-atom realization of the model described by Eq. (\ref{H1}) the staggered potential can be introduced as an extra term. However, we want to point out the possibility that such a component be generated in the process of constructing the superlattice itself. This represents a potential problem for realizing topological quantum states. For example, if the components $V_1$ and $V_2$ of the optical lattice potential are produced by different lasers, a misalignment corresponding to $x\rightarrow x+\delta x$ and $y\rightarrow y+\delta y$ in the $V_2$ term will effectively generate a staggered potential. In the calculations we neglect the detailed effects of such a misalignment on the hopping matrix elements and consider only the on site staggered contributions parameterized by $\Gamma$. We start with a system in the stripe geometry with parameters $V_1=3.4 E_r$, $V_2=1.7 E_r$, $\alpha=2\hbar/a$
and $\Gamma=0$, then we turn on $\Gamma$ while keeping the other parameters fixed. The corresponding spectra are shown in Fig. \ref{Fig19}. For $\Gamma=0$ the system is a topological insulator and the corresponding spectrum is characterized by an edge mode that crosses the bulk gap. Applying a small staggered potential reduces the gap in the vicinity of $k_x=0$ and, eventually, at the critical value $\Gamma_c \approx 0.075 E_r$ the gap closes at $k_x=0$. For larger values of $\Gamma$ the gap opens again but no edge states are present inside the gap. We conclude that the system is a topological insulator for $\Gamma<\Gamma_c(V_1, V_2, \alpha)$ and a standard band insulator for $\Gamma>\Gamma_c(V_1, V_2, \alpha)$.
\begin{figure}[tbp]
\begin{center}
\includegraphics[width=0.47\textwidth]{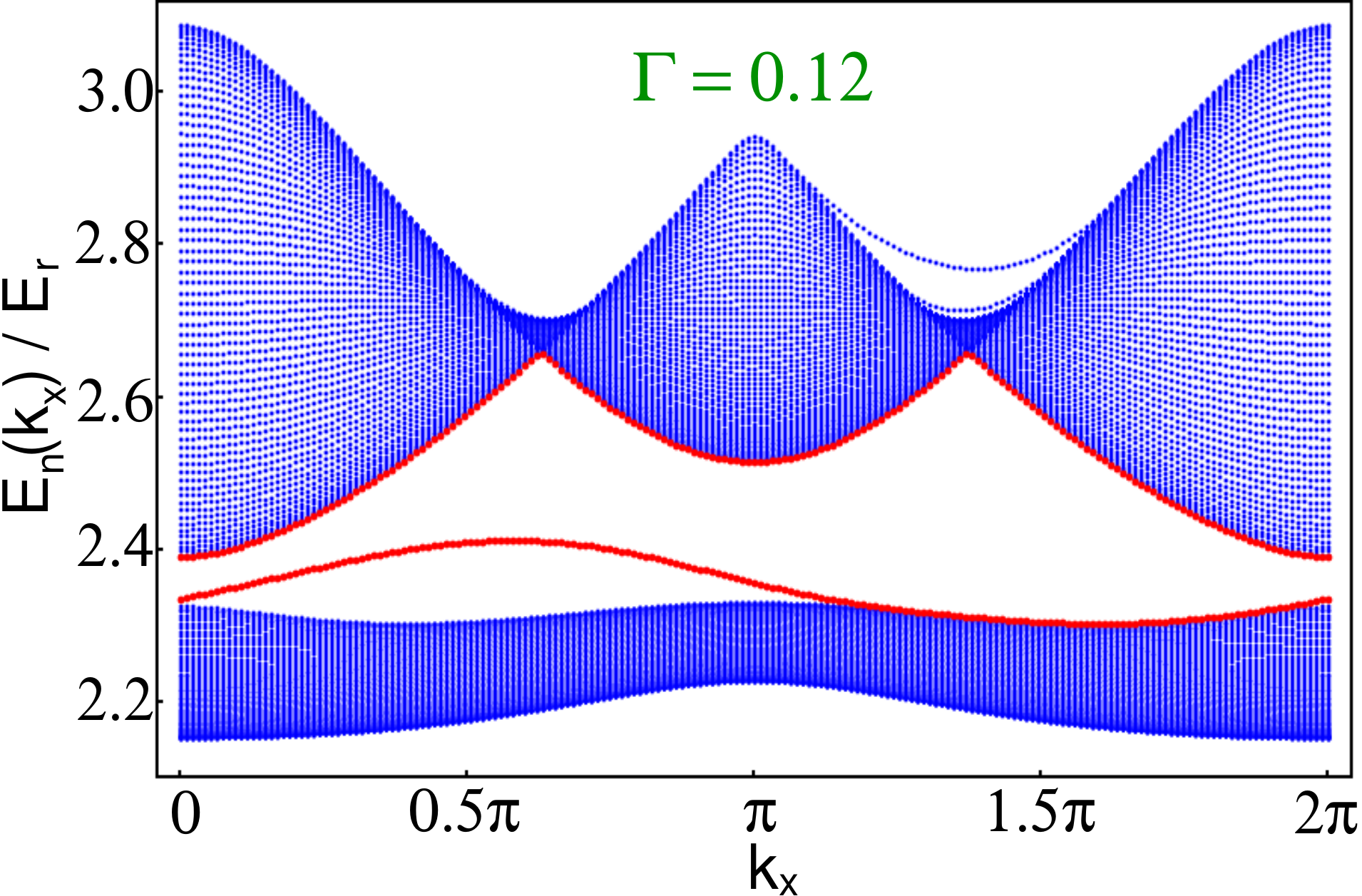}
\end{center}
\caption{(Color online) Band structure for a stripe with inequivalent edges and with the same parameters as in Fig. \ref{Fig19}. For $\Gamma=0.12 E_r$ the system is a standard band insulator but, in contrast with the equivalent edge case (see Fig. \ref{Fig19}), a well-defined edge mode populates the gap. A clean insulator with the chemical potential inside the bulk gap supports gapless edge excitations. However, any small perturbation (such as disorder or interactions) will open a gap in the edge mode.\vspace{-3mm}}
\label{Fig20}
\end{figure}

The mechanism described above is quite general. Any perturbation capable of opening a gap in the spectrum will have similar effects and will induce a transition at a particular critical strength. This critical strength is independent of the boundary geometry. However, the details that characterize the edge modes in either side of the transition depend on the properties of the boundaries. For example, if instead of the stripe geometry with equivalent edges considered in Fig. \ref{Fig19} we study a system with inequivalent edges (see Fig. \ref{Fig8}), we observe a transition (i.e., the closing of the gap) at the same critical value $\Gamma_c \approx 0.075 E_r$. However, the band insulator with $\Gamma>\Gamma_c$ has now a clearly defined edge mode, as shown in Fig. \ref{Fig20}. In particular, this edge mode ensures the existence of gapless excitations for any value of the chemical potential inside the gap. Nonetheless, this mode is topologically trivial, as revealed by the fact that it does not connect the two bands, and consequently it is not protected against disorder and interactions, in the sense that any weak perturbation will open a gap in the edge mode.
\begin{figure}[tbp]
\begin{center}
\includegraphics[width=0.47\textwidth]{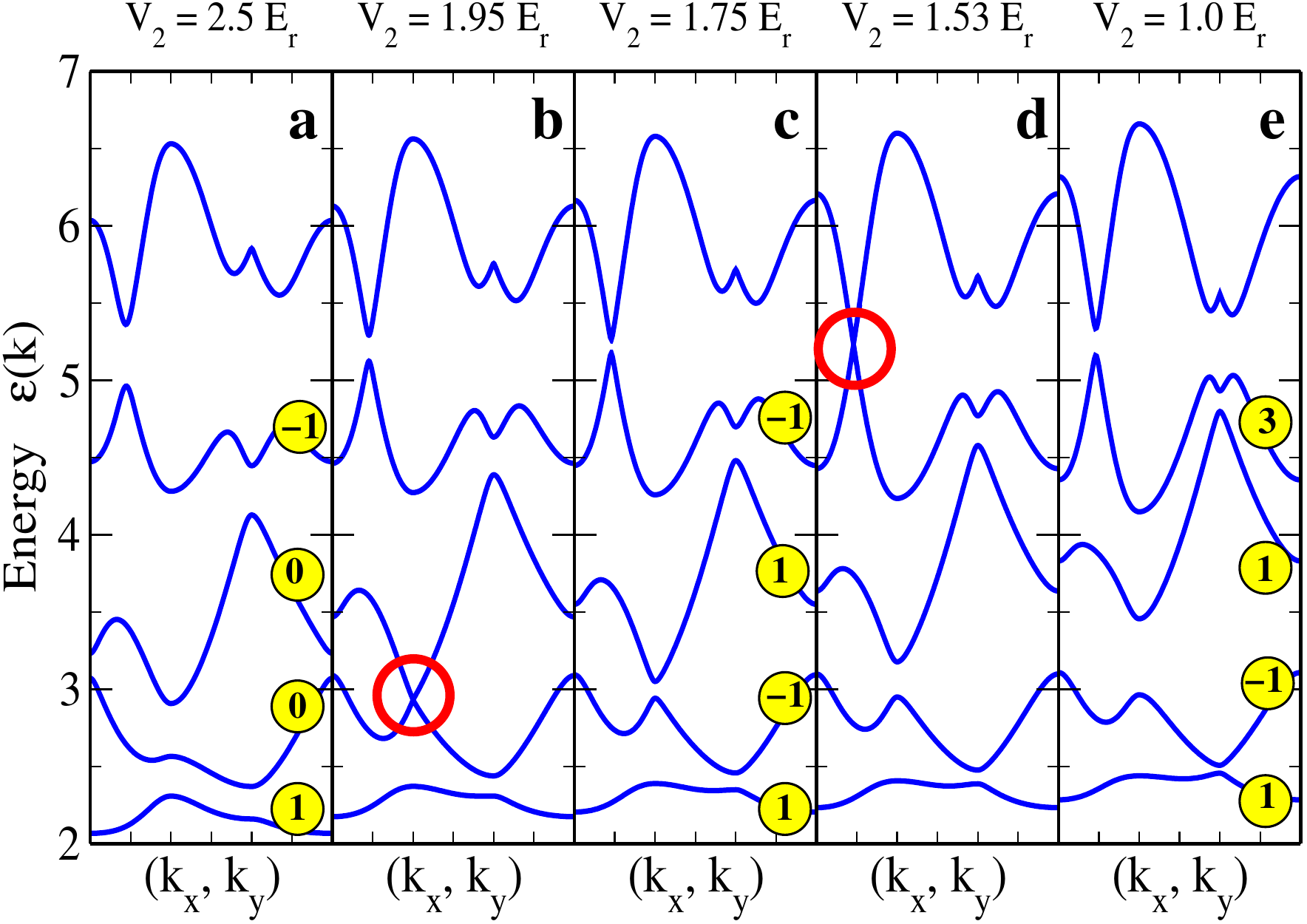}
\end{center}
\caption{(Color online) Energy dispersion curves for a system with $V_1=3.7 E_r$, $\alpha=2\hbar/a$, and different values of $V_2$ along the $(0,0)\rightarrow (\pi,\pi)\rightarrow (\pi,0)\rightarrow (0,0)$ path in the Brillouin zone. The numbers inside the yellow circles represent the Chern numbers of the bands. Note the closing of the direct gap at certain critical values of $V_2$ (panels b and d) and the corresponding change of the Chern numbers. The spectra shown in panels a, c, and e are consistent with different insulating states: (a) topological insulator (with the first three bands filled) and conventional insulator (four bands filled); (c) topological insulator (one band filled) and conventional insulator (four bands filled); and (e) topological insulator I (one band filled), conventional insulator (two bands filled) and topological insulator II (four bands filled).
 \vspace{-3mm}}
\label{Fig21}
\end{figure}

\subsection{Transitions in a multi-band system}\label{tr_multi}

The transition studied in the previous section was driven by the competition between the contributions to the Hamiltonian that generate its nontrivial topological properties and terms like $V_{\rm stagg}$ that tend to open a conventional band gap. However, the model described by Eq. (\ref{H1}) contains  three independent parameters, $V_1$, $V_2$, and $\alpha$, and only the vector potential (i.e., $\alpha$) is directly responsible for the nontrivial topology of the bands. Therefore, we expect several topologically distinct states to exist in various regions of the parameter space. We have already seen that small variations of the parameters can lead to the opening/closing of the gaps corresponding to a topological insulator to metal transition, as shown in Fig. \ref{Fig17}. The question that we address now is whether insulator to insulator transitions can be induced by tuning the parameters of the model. Our purpose is not to determine the full phase diagram of the model, but rather to give an example showing that such transitions are possible. In general, the parameter space can be characterized by two independent quantities, for example $V_2/V_1$ and $\alpha/\sqrt{2m V_1}$. However, we fix $V_1=3.7 E_r$ and $\alpha=2\hbar/a$  and vary $V_2$, which corresponds to a certain cut in the parameter space. The resulting bulk spectra are shown in Fig. \ref{Fig21} for several values of $V_2$. The total flux of Berry curvature (i.e., the Chern number) for a given band remains unchanged as long as the direct gaps separating that band from the neighboring bands do not vanish. When the direct gap between two bands vanishes (see panels b and d), the total  flux of Berry curvature is redistributed between the two bands and the topological properties of the system change accordingly.
\begin{figure}[tbp]
\begin{center}
\includegraphics[width=0.47\textwidth]{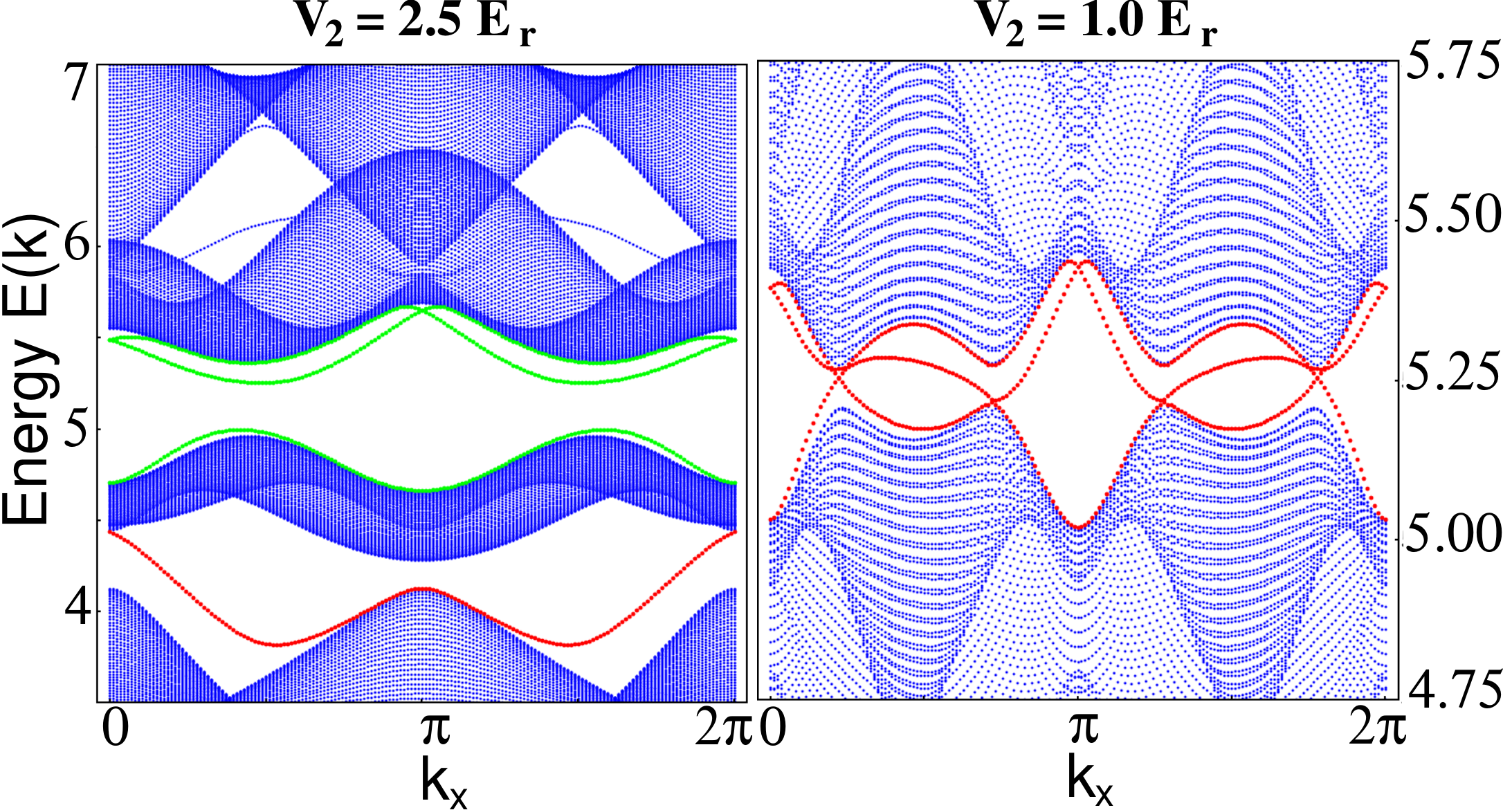}
\end{center}
\caption{(Color online) Detail of the band structure for a system in the stripe geometry with the same parameters as in Fig. \ref{Fig21}(a) (left) and \ref{Fig21}(e) (right). In the left panel notice one pair of topological edge modes inside the gap above the third band and the topologically trivial edge states above the fourth band, consistent with the Chern numbers shown in Fig. \ref{Fig21}(a). On the right we observe four pairs of edge modes inside the gap above the fourth band. This number of pairs of edge modes is equal to the total Chern number of the bands below the gap [see  Fig. \ref{Fig21}(e)].
 \vspace{-3mm}}
\label{Fig22}
\end{figure}

As we mentioned above, the topological character of an insulating state, and hence the existence of robust edge states, is determined by the sum of the Chern numbers of the occupied bands. An interesting case revealed by the results shown in Fig. \ref{Fig21} is when the first four bands are completely filled. Then, for $V_2<1.53 E_r$ the system is a conventional band insulator, while for
$V_2>1.53 E_r$ it is a topological insulator with the sum of the Chern numbers of the occupied bands equal to four. To identify the structure of the corresponding edge modes we consider a system with boundaries in the stripe geometry and determine the spectra for $V_2=1.0 E_r$ and $V_2=2.5 E_r$. The results are shown in Fig. \ref{Fig22}. As evident from these results, the number of pairs of topological edge states that populate a gap equals the total Chern number of the bands that are below that gap. In the case shown in right panel of
 Fig. \ref{Fig22} four pairs of topological edge modes populate the gap between the fourth and the fifth bands. Hence, for any given energy inside the gap at least four different edge states will exist on each of the two boundaries. The stability of these edge states is a natural question that we address next.
\begin{figure}[tbp]
\begin{center}
\includegraphics[width=0.47\textwidth]{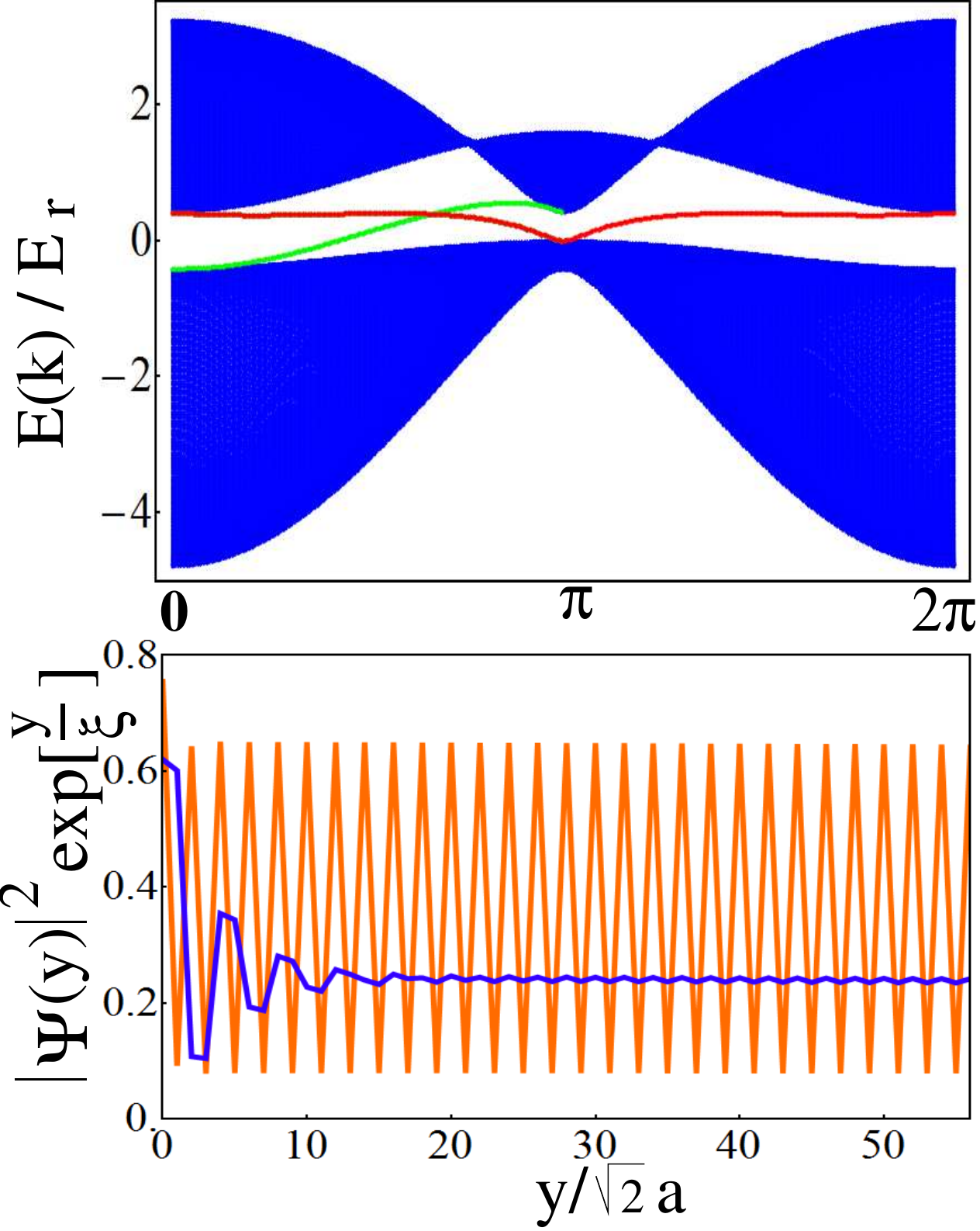}
\end{center}
\caption{(Color online) (Upper panel) Spectrum for the square superlattice model (\ref{modelH}) with parameters $t_1 = (1+0.2i)E_r$, $t_2=-0.3E_r$, and $t_2^{\prime}=-0.06E_r$ in the stripe geometry. The system is finite in the $y$ direction and infinite in the $x$ direction, with the axes are oriented along the next-nearest neighbor directions (see Fig. \ref{Fig3}). For an odd number of layers (i.e., equivalent edges) the edge states (red/dark gray lines) with $k_x<\pi$ are localized near $y=0$, while the edge states with $k_x>\pi$ are localized near the opposite boundary. For an even number of layers (i.e., inequivalent edges) the mode localized near $y=0$ remains unchanged, while the other mode is now characterized by $k_x<\pi$ (green/light gray line). (Lower panel) Edge state amplitudes multiplied by exponential factors to reveal the oscillatory behavior. Blue (dark gray): edge state corresponding to $k_x=0.97\pi/\sqrt{2}a$ and $\xi = 1.441 \sqrt{2} a$. Orange (light gray): edge state with $k_x=0.7\pi/\sqrt{2}a$ and  $\xi = 1.351 \sqrt{2} a$.
 \vspace{-3mm}}
\label{Fig23}
\end{figure}

\section{Stability of the edge states}\label{stability}

The physics that emerges from the nontrivial topological properties of a system can be directly related to the behavior of the edge states. The topological features are present as long as the spectrum is gaped for all bulk excitations and the bulk gap is populated by gapless edge modes. Weak local perturbations cannot induce a phase transition to a state with different topological properties and, consequently, the edge or surface states are robust against disorder and interactions~\cite{Wu,XuMoore,Ostrovsky,Obuse,Bardarson}. These perturbations modify the bulk properties of the system and generate extra terms in the Hamiltonian that describes its quantum mechanical properties. However, the solution of a quantum mechanical problem is determined not only by the  Hamiltonian but also by the boundary conditions. In this section we address the question of how changing these boundary conditions  impact the properties of the edge states. In particular, we discuss the stability of the edge states against finite size effects and their dependence on the confining potential that defines the boundaries of the system. The answers to these questions are particularly relevant for cold atom systems, but they can shed meaningful light on the physics of solid state topological insulators, for example, in the case of topological insulator thin films~\cite{KeXue2009,ZhangWu,Linder,LuShen,LiuZhang}. or in the case of topological insulator heterostructures.
\begin{figure}[tbp]
\begin{center}
\includegraphics[width=0.47\textwidth]{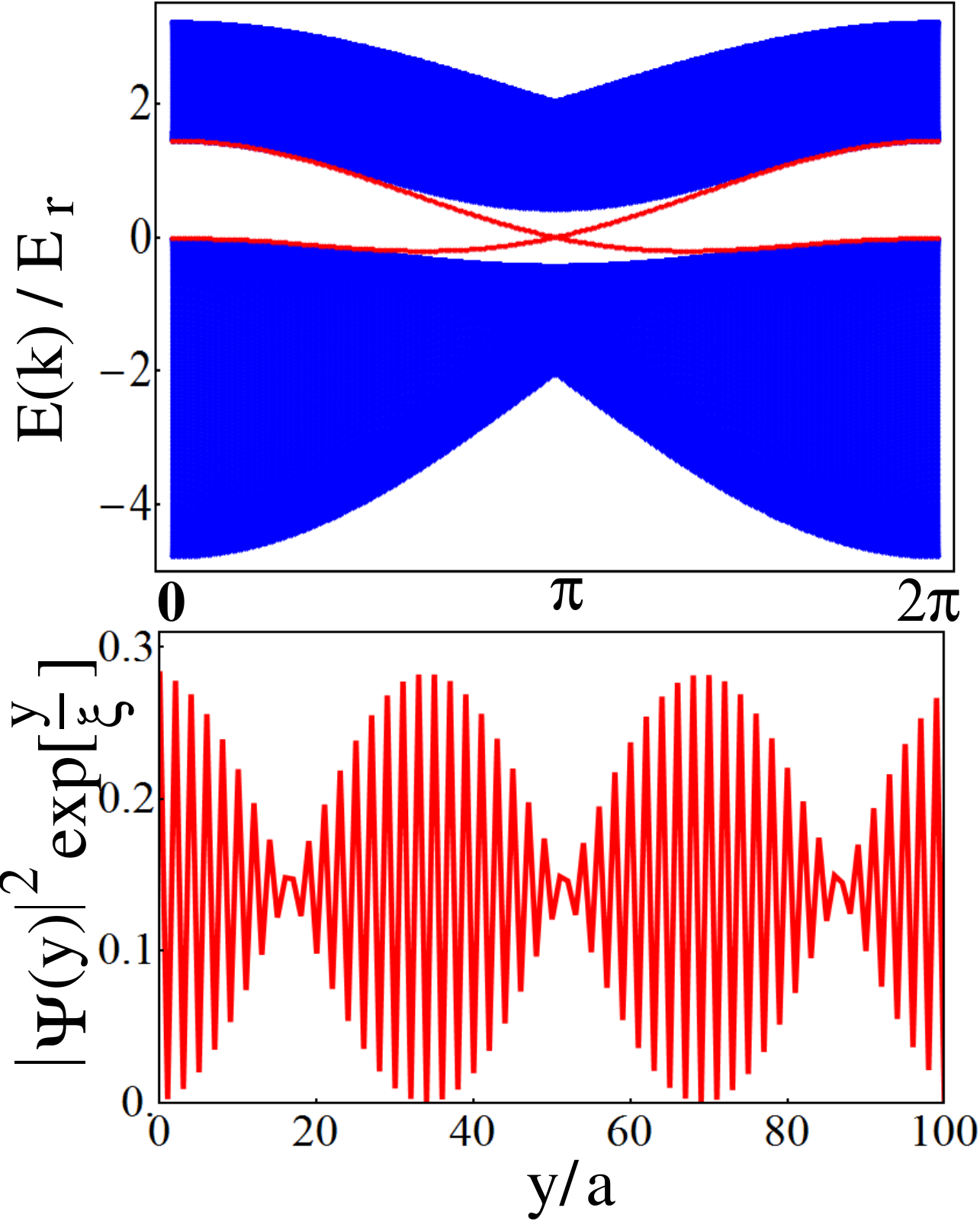}
\end{center}
\caption{(Color online)  (Upper panel) Spectrum for the square super-lattice model (\ref{modelH}) in the stripe geometry for the same parameters as in Fig. \ref{Fig23} but with the symmetry axes oriented along the nearest neighbor directions. The dispersion of the edge modes is independent of the parity of the number of layers. (Lower panel) Edge state amplitude multiplied by an exponential factor with $\xi = 2.396 a$ for the state with $k_x=0.97\pi/a$. Note the even-odd oscillations and the k-dependent long wavelength oscillatory component. At $k_x=\pi/a$ the extra oscillatory component is absent.
 \vspace{-3mm}}
\label{Fig24}
\end{figure}

\subsection{Finite size effects}\label{finite_size}

The chiral edge states robustness against disorder and interactions can be linked to the absence of backscattering. In a large system, counter propagating edge modes are localized near boundaries that are well separated spatially and, consequently, have a vanishing overlap. However, as the size of the system in a certain direction is reduced, edge states propagating along opposite edges may acquire a finite overlap. Consequently, a gap opens in the edge states spectrum.
The dependence of this gap on the size of the system depends on the spatial behavior of the edge states, in particular on how fast they decay away from the boundary. As suggested by the profiles shown in Figs. \ref{Fig9} and \ref{Fig10}, the edge state amplitude decreases exponentially with the distance from the edge. The characteristic length scale for this exponential decay, $\xi$, depends on the size of the bulk gap, roughly scaling as the bandwidth over the gap size, but also on the location of the edge state within the bulk gap. In addition, the exponential decay is generally non monotonic and includes one or more oscillatory components.

To illustrate this general behavior, we show in Fig. \ref{Fig23} the spectrum
for a system with a stripe geometry described by the square super-lattice model given by Eq. (\ref{modelH}) with the parameters $t_1 = (1+0.2i)E_r$, $t_2=-0.3E_r$, and $t_2^{\prime}=-0.06E_r$. The edges of the stripe are chosen along one of the next nearest neighbor directions and, as discussed above, may contain sites from the same sublattice (what we call equivalent edges) or from different sublattices  (inequivalent edges). As shown in the upper panel of Fig. \ref{Fig23}, the dispersion of the edge modes is extremely sensitive to changes in the boundary conditions (see also Figs. \ref{Fig7} and \ref{Fig8}). In addition, different states from a given edge mode have different asymptotic behaviors, as shown in the lower panel of Fig. \ref{Fig23}. As the parameters describing the asymptotic behavior of an edge state deep inside the system depend on bulk properties but also on the position of the state within the gap, they can be modified by changing the boundary conditions.

If the edges of the stripe are chosen along one of the nearest neighbor directions, the corresponding spectrum does not exhibit any even-odd variation with the number of layers in the system. The band structure for such a stripe is shown in Fig. \ref{Fig24} for the same model parameters as in Fig. \ref{Fig23}. The amplitude of an edge state from the vicinity of the Dirac point is also shown in the lower panel. Although the bulk parameters are the same as in Fig. \ref{Fig23}, the asymptotic behavior of the edge states is generally different, as a consequence of the new boundary conditions. Note that the edge state shown in Fig. \ref{Fig24} is characterized by multiple oscillatory components, in addition to the exponential decay. The relative amplitude of those components depend on $k_x$, the wave vector component parallel to the edge, for example at
$k_x=\pi/a$ only the short wavelength oscillatory component is present.
\begin{figure}[tbp]
\begin{center}
\includegraphics[width=0.47\textwidth]{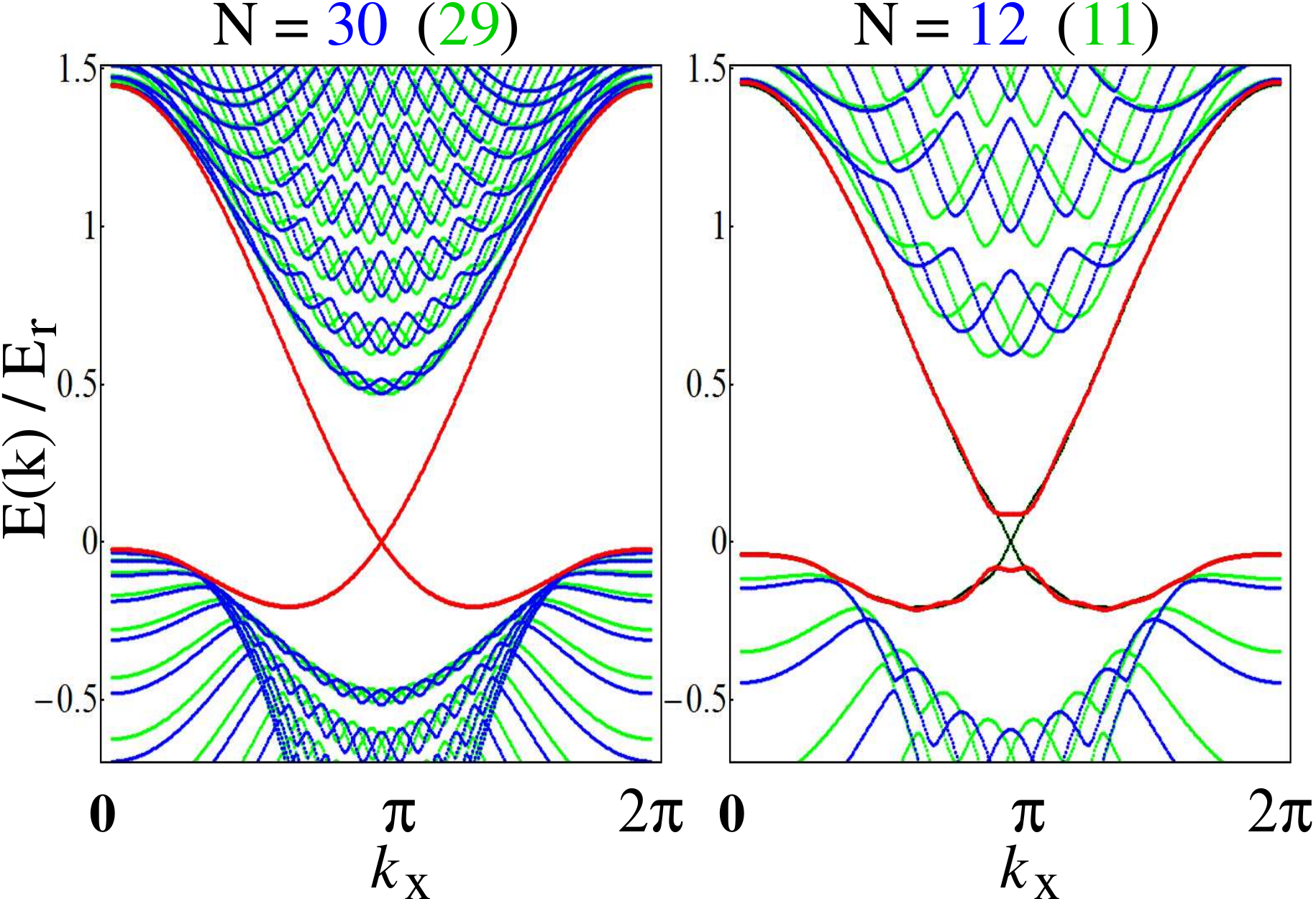}
\end{center}
\caption{(Color online) Opening of a gap in the edge state spectrum due to finite size effects. The system is a thin stripe with the same geometry and hopping parameters as in Fig. \ref{Fig24}. The size of the gap is determined by the overlap of edge states localized on opposite boundaries and, consequently, varies roughly exponentially with the width of the stripe and inherits the oscillatory behavior of the edge states. For large $N$ the edge states are gapless and independent of parity. For thin films, a finite gap opens in systems with even number of layers, while systems with odd number of layers remain gapless. Note that the "bulk" spectrum varies dramatically with the system size, in contrast with the edge modes that are practically unaffected if the system size is much larger than their characteristic length scale.
 \vspace{-3mm}}
\label{Fig25}
\end{figure}

 Knowing the precise asymptotic behavior of the edge states for a given system is important for predicting the dependence of the finite size induced gap on the system size. As mentioned above, the size of the gap depends on the overlap between edge states propagating along opposite edges. More precisely, let us consider an infinitely wide stripe described by a certain Hamiltonian and two Kramers degenerate edge states $\psi_{k_x^0}^{(1)}(\delta y)$ and $\psi_{k_x^0}^{(2)}(\delta y)$, where $\delta y$ is the distance from the boundaries along which the states propagate. Next, consider a relatively thin stripe of width $W$ described by the same Hamiltonian. The relevant overlap can be written as
\begin{equation}
s_{12}(w) =  \int_0^w dy~ \psi_{k_x^0}^{(1)}(y) \psi_{k_x^0}^{(1)}(w-y).
\end{equation}
If $w\gg \xi$, where $\xi$ is the characteristic length scale for the exponential decay of the edge states, the overlap is negligible, but as the width of the stripe is reduced we expect $s_{12}(w)$ to grow exponentially. However, this exponential dependence is not monotonic, due to the extra oscillatory components of the wave functions.

The dependence of the spectrum on the width of the stripe is shown in Fig. \ref{Fig25}. The model parameters and the geometry of the stripe are the same as in Fig. \ref{Fig24}. As the width of the system is varied, the bulk spectrum changes dramatically, in contrast to the edge modes that are practically unaffected if the number of layers exceed $N=30$. For thiner stripes, $s_{12}(w)$ becomes finite if the number of layers is even and a gap opens at the degeneracy point. The even-odd effect is a consequence of the oscillatory asymptotic behavior of the edge states at $k_x=\pi$.
\begin{figure}[tbp]
\begin{center}
\includegraphics[width=0.47\textwidth]{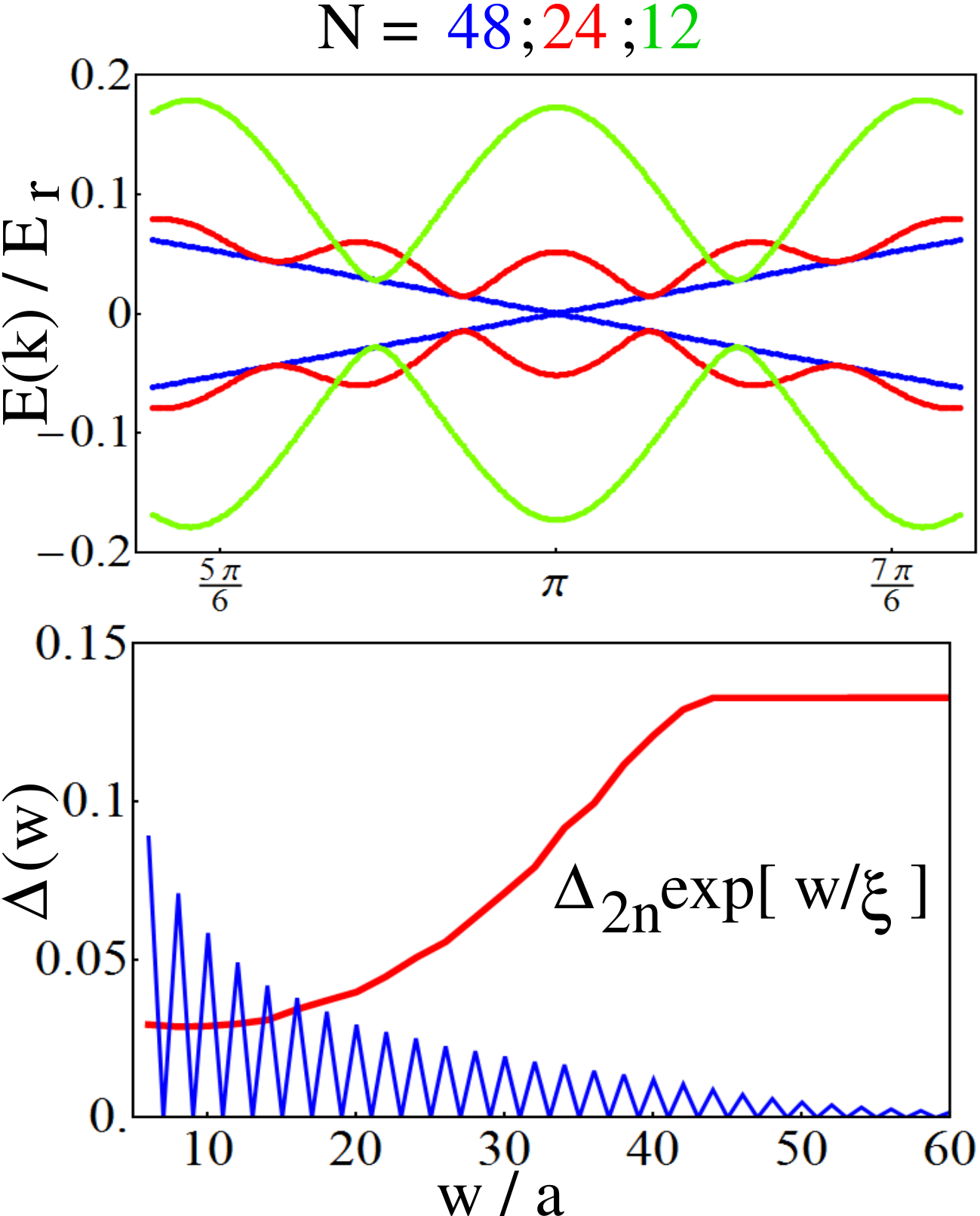}
\end{center}
\caption{(Color online) Gap dependence on the stripe thickness ($w$) for a system with $t_1 = (1+0.05i)E_r$, $t_2=-0.05E_r$, and $t_2^{\prime}=0.05E_r$ and the symmetry axes oriented along the nearest neighbor directions. Upper panel: Edge mode dispersions for stripes with $w=(N-1)a$, where $N$ is the number of layers. The gap at $k_x=\pi$ is almost zero for $N=48$, $0.1E_r$ for $N=24$, and $0.36E_r$ for $N=12$. Edge states with $k_x\neq \pi$ have extra oscillatory components (similar to that shown in the lower panel of Fig. \ref{Fig24}) that may generate a vanishing overlap between edge states located on opposite edges. Consequently, for $N<44$ the gap minima shift away from  $k_x\neq \pi$. (Lower panel) Dependence of the minimum gap amplitude on the stripe thickness. Note the even-odd oscillations. The long period extra oscillatory components of the edge states (see Fig. \ref{Fig24}) determine a deviation from an exponential dependence of the gap amplitude on the stripe thickness for  $N<44$ in stripes with even number of layers (red/smooth line).
 \vspace{-3mm}}
\label{Fig26}
\end{figure}

To investigate further the consequences of the oscillatory asymptotic behavior on the finite size induced gap we consider a system described by the square super-lattice model given by Eq. (\ref{modelH}) with the parameters $t_1 = (1+0.05i)E_r$, $t_2=-0.05E_r$, and $t_2^{\prime}=0.05E_r$. The system has a stripe geometry with edges along one of the nearest neighbor directions. The asymptotic behavior of the edge states is qualitatively similar to that shown in Fig. \ref{Fig24}, but the characteristic length scales $\xi$ are significantly larger due to the smaller value of the bulk gap. The dispersion of the edge modes for thin stripes with three different numbers of layers is shown in the upper panel of Fig. \ref{Fig26}. For $N>44$ layers, a small gap opens at the degeneracy point $k_x=\pi$ due to the overlap between the edge states propagating on the two edges. The wavefunctions at $k_x=\pi$ are characterized by oscillations with a period $2a$, in addition to the exponential decay. This  leads to the vanishing of the overlap, and implicitly of the induced gap, in stripes with odd number of layers, as shown in the lower panel of Fig. \ref{Fig26}. By reducing the system size, the gap induced at $k_x=\pi$ increases exponentially if the number of layers is even. However, edge states with $k_x\neq \pi$ are characterized by extra oscillatory components, similar to the situation shown in the lower panel of Fig. \ref{Fig24}, and the overlap of these states may vanish for certain stripe widths even if the number of layers is even. Consequently, the minimum gap shifts away from $k_x=\pi$ when $N<44$ and the dependence of the (minimum) gap amplitude on the system size is no longer exponential (see Fig. \ref{Fig26}).
\begin{figure}[tbp]
\begin{center}
\includegraphics[width=0.47\textwidth]{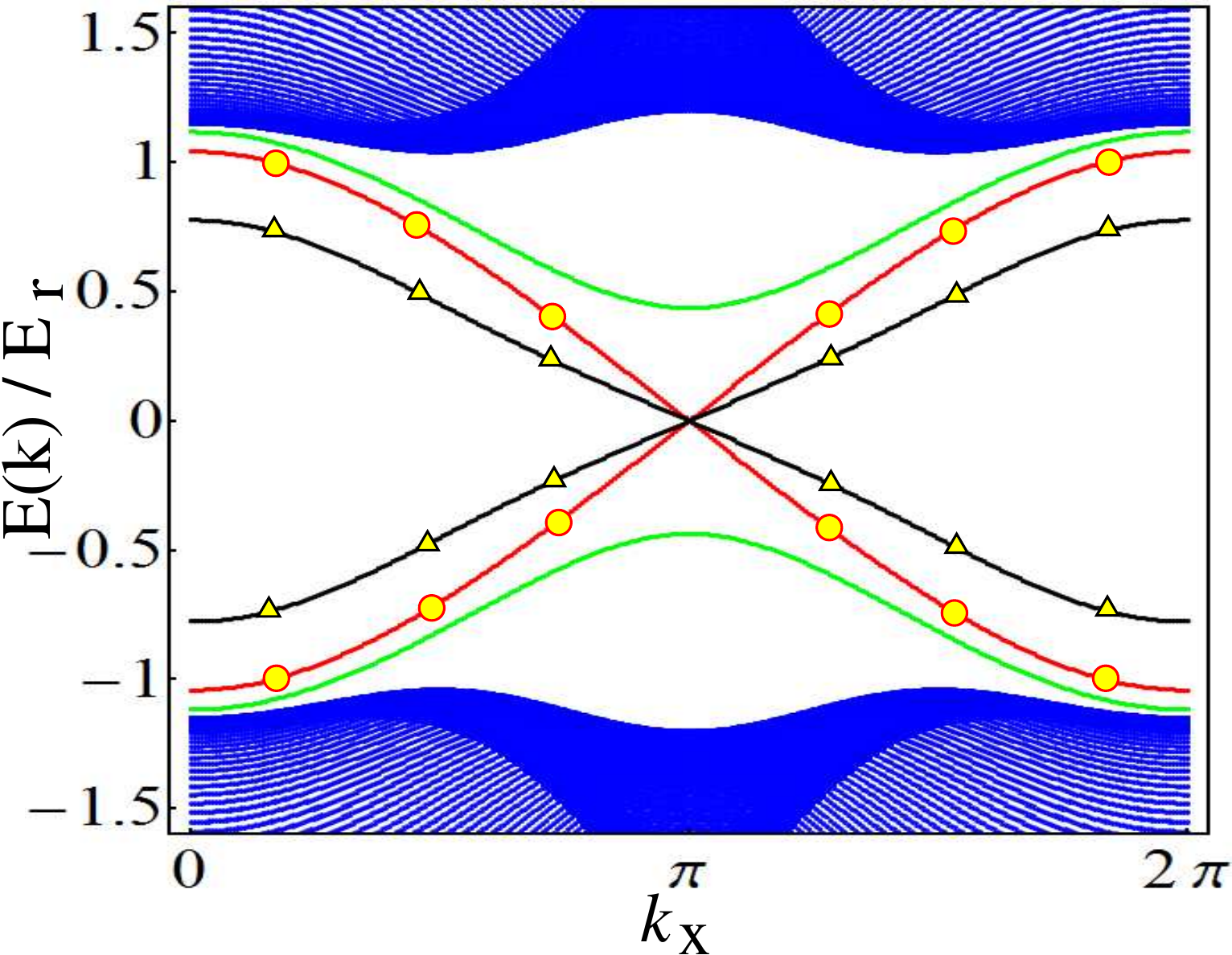}
\end{center}
\caption{(Color online) In-gap states localized along a line defect in a system with no boundaries described by Eq. (\ref{modelH}) with $t_1 = (1+0.3i)E_r$, $t_2=-0.3E_r$, and $t_2^{\prime}=0.3E_r$. The one-dimensional defect is oriented along the nearest neighbor direction and is characterized by (see main text): $\delta t = 0.5 E_r$ and no lattice mismatch (green/light gray lines), $\delta t = 0.5 E_r$ and a lattice mismatch (red lines and circles), $\delta t = \mbox{Re}[t_1] = E_r$ and a lattice mismatch (black lines and triangles). In thin stripes the edge states can overlap with states localized along extended defects and increase the size of the finite size induced gap.
\vspace{-3mm}}
\label{Fig27}
\end{figure}

We have shown that the amplitude of the edge states gap induced by finite size effects depends on the asymptotic behavior of the edge states. Of course, the presence of perturbations, such as impurities, will also impact the size of this gap. For example, the exact vanishing of the gap in stripes with an odd number of layers does not happen in the presence of impurities. More generally, in the presence of disorder one expects the amplitude of the induced oscillatory effects to decrease. Another interesting question concerns the effect of extended defects, such as a one-dimensional lattice mismatch, on the edge states and on the finite size induced gap. Although in cold atom systems the relevance of these type of defects is not clear, this type of problem is highly relevant for solid-state topological insulators. For example, scanning tunneling spectroscopy (STS) measurements on bulk crystals of $Bi_2Se_3$ and $Bi_2Te_3$ have revealed mechanical instabilities of the surface due to the strongly layered structure of these materials, which causes microcracking between the layers~\cite{Ura1,Ura2}. We will not address this issue in detail, but note that the presence of these extended perturbations induces states inside the bulk gap that are spatially localized in the vicinity of the defect. As an example, we show in Fig. \ref{Fig27} the spectrum of a two-dimensional square superlattice model with  $t_1 = (1+0.3i)E_r$, $t_2=-0.3E_r$ and $t_2^{\prime}=0.3E_r$. The system has no boundaries but has a one-dimensional defect along one of the nearest-neighbor directions. The defect is modeled as a pair of lattice lines coupled by nearest neighbor hoppings $\delta t$. A lattice mismatch corresponds to the case when the nearest-neighbor hopping couples sites of the same sublattice. Note that the limit $\delta t \rightarrow 0$ corresponds to the stripe geometry. As shown in Fig. \ref{Fig27}, in the absence of a lattice mismatch the dispersion of the in-gap modes is similar to that of topologically trivial edge states, while in the presence of the lattice mismatch the dispersion is similar to that of topological edge modes. The main difference between edge states and these extended defect states is that the defect modes are not connected to the bulk bands. In thin stripes these extended defect states can overlap with the edge states modifying significantly the size of the induced gap. Also, if such extended defect states are present close to the edge/surface of a topological insulator, the transport properties  of the edge states can be significantly affected.
\begin{figure}[tbp]
\begin{center}
\includegraphics[width=0.47\textwidth]{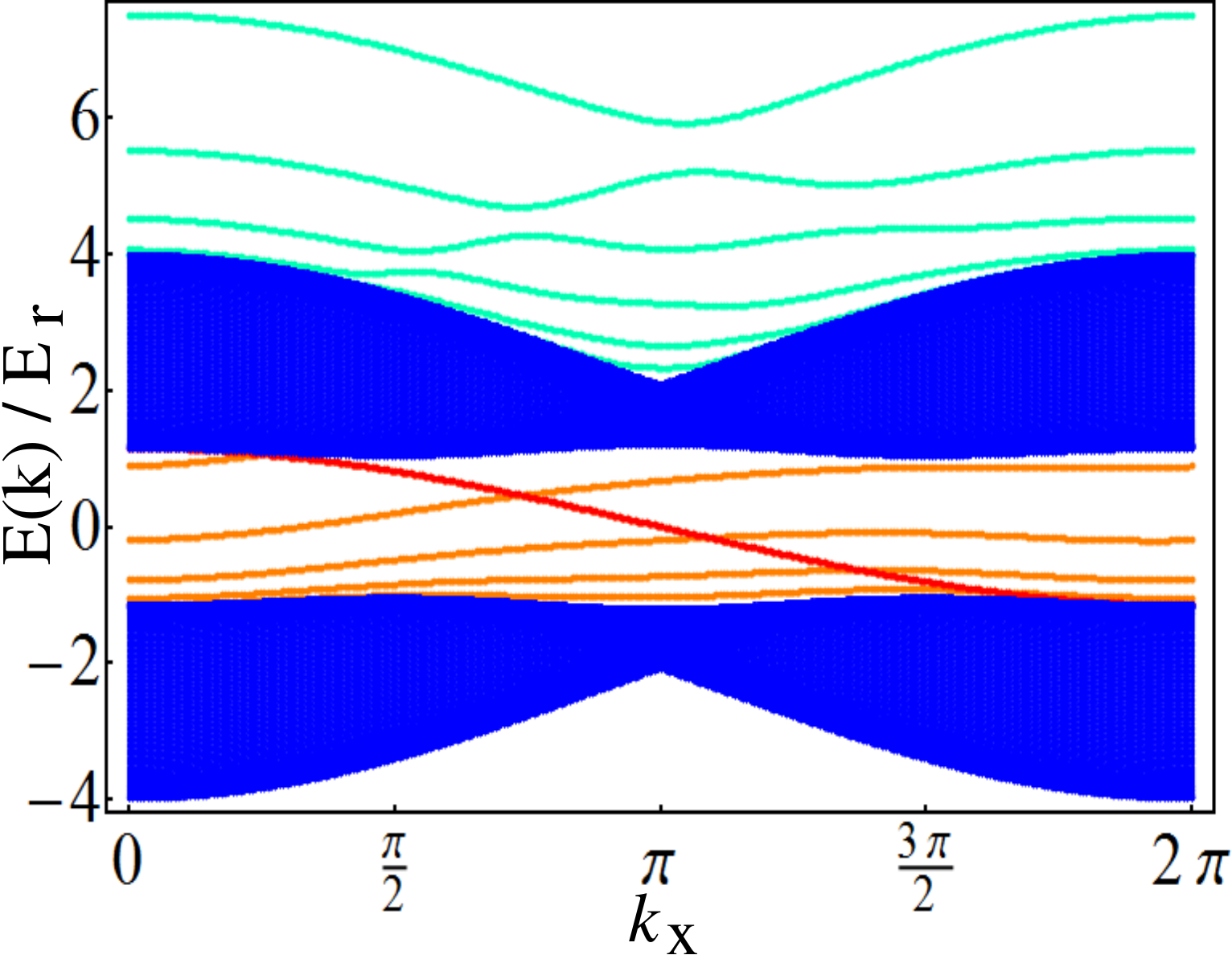}
\end{center}
\caption{(Color online) Spectrum for a system with a soft boundary described by Eq. (\ref{modelH}) with $t_1 = (1+0.3i)E_r$, $t_2=-0.3E_r$, and $t_2^{\prime}=0.3E_r$ in the stripe geometry. The stripe is oriented along the nearest-neighbor directions and is finite along the $y$ direction. The confinement at $y\approx 0$ is given by the exponential potential $V_c(y) = 5.0 \mbox{exp}[-y/(2a)]$, while the confinement at the opposite boundary is given by an infinite potential wall. The dispersion of the edge mode near the soft boundary (orange/light gray lines inside the gap) differs significantly from the dispersion of the hard boundary edge mode (red/dark gray line inside the gap) but retains the fundamental property of a topological insulator edge mode, i.e., it connects the two bulk bands (note that $k_x=2\pi$ has to be identified with $k_x=0$). Notice the topologically trivial edge modes that are generated at  high energies (light green or light gray lines).
 \vspace{-3mm}}
\label{Fig28}
\end{figure}

\subsection{Effects of the confining potential}\label{confining}

The properties of the edge states are strongly dependent on the boundary conditions, as we have shown in the previous subsection. In cold-atom systems these boundary conditions are determined by the extra confining potential that supplements the optical lattice confinement, i.e., the term $V_c({\bf r})$ in Eq. (\ref{H1}). So far, in all the calculations we have used hard wall boundary conditions, which for the stripe geometry are equivalent with having a confining potential
\begin{equation}
V_c(y) = \left\{\begin{array}{l}
0~~~~~\mbox{if}~~ 0\leq y\leq W, \\
~\\
\infty~~~~\mbox{if}~~ y<0,~~\mbox{or}~~y>W,
\end{array}\right.
\end{equation}
The question that we want to address next is how are the edge states properties modified if we relax the hard wall boundary conditions. In general, a given confining potential sharply defines the boundary of a system if it is characterized by values much smaller than the topological insulator gap  over a large area (volume), the "bulk" of the system, and increases to values larger than the bandwidth within a length scale much smaller that the linear size of the "bulk." The region defined by this length scale is the "boundary region." A harmonic confining potential, as typically used in cold-atom experiments, does not contain any intrinsic length scale that could define a boundary. The "boundary" of a system of atoms in a harmonic trap potential $V_c(r)$ has a characteristic length given  by  $d=L_{\Delta}-L_W$, where $L_{\Delta}$ is the length scale at which the confining potential becomes comparable to the bulk gap, $V_c(L_{\Delta}) = \Delta$, and $L_W$ is  the length scale at which the confining potential becomes comparable to the bandwidth, $V_c(L_W) = W~$\cite{StanSarma}.  For a very smooth potential, the "boundary region" can represent a significant fraction of the bulk. The system confined by such a potential is an inhomogeneous system with no insulating properties, at least in the absence of inter particle interactions, which represents an inhomogeneous  topological  metal~\cite{StanSarma}. Before discussing this case, it is instructive to study the effect of a confining potential that has an intrinsic length scale on the properties of the edge states. The particular question that we address is how the edge states depend on the characteristic  length scale of an exponential confining potential.
\begin{figure}[tbp]
\begin{center}
\includegraphics[width=0.47\textwidth]{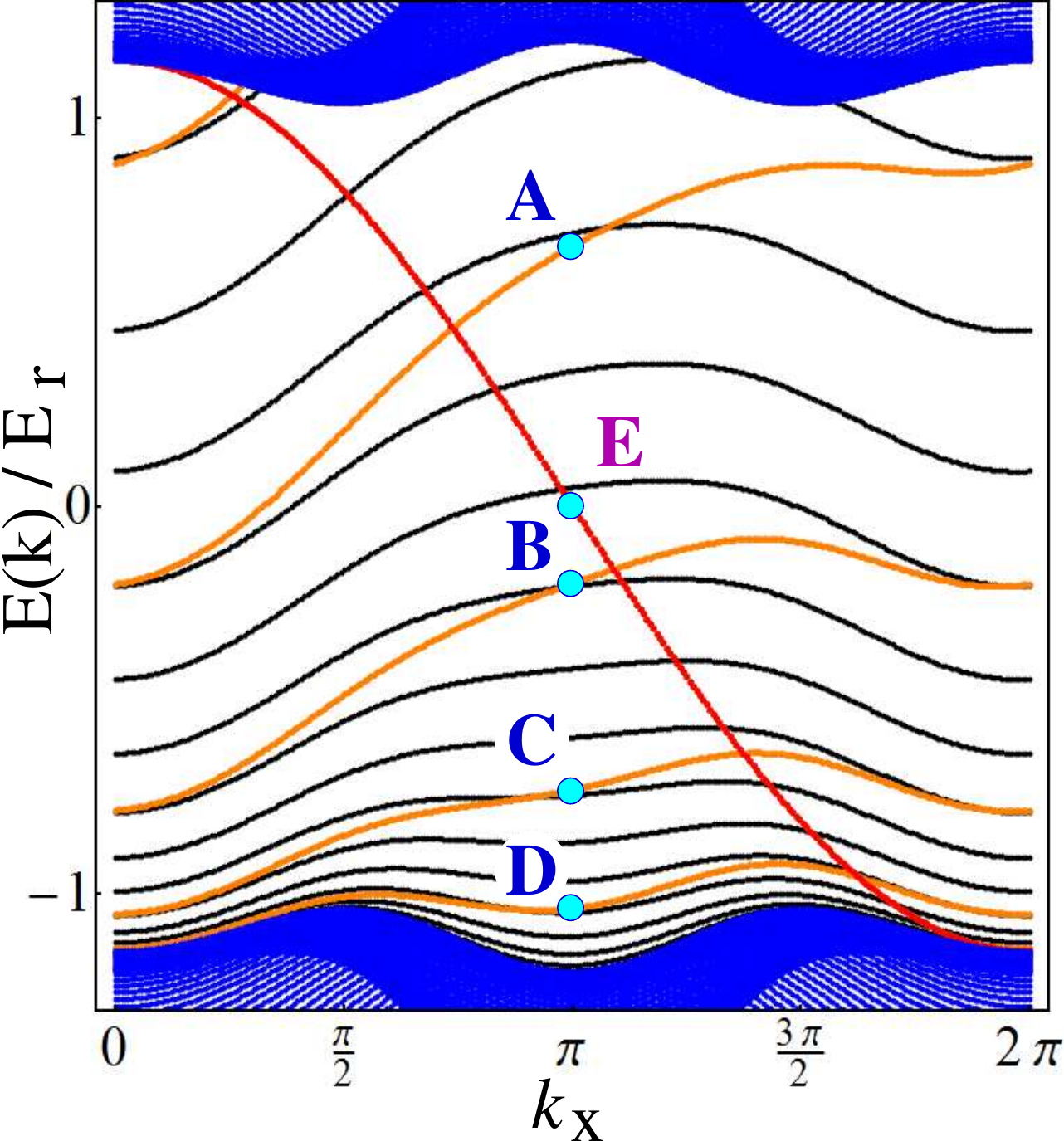}
\end{center}
\caption{(Color online) Detail of the band structure shown in Fig. \ref{Fig28}. For comparison the dispersion of an edge mode corresponding to a softer boundary, $V_c(y) = 5.0 \mbox{exp}[-y/(6.5a)]$, is also shown (black lines).
 \vspace{-3mm}}
\label{Fig29}
\end{figure}

We restrict our analysis to the stripe geometry and consider the simple square super lattice model described by Eq. (\ref{modelH}). To detect easily the changes in the edge mode dispersion we choose a set of model parameters that generates a large bulk gap: $t_1 = (1+0.3i)E_r$, $t_2=-0.3E_r$, and $t_2^{\prime}=0.3E_r$. One edge of the stripe, $y=W$, is defined by a hard wall boundary condition, while the opposite "edge" is soft and generated by an exponential confining potential. Explicitly we have:
\begin{equation}
V_c(y) = \left\{\begin{array}{l}
5E_r e^{-\frac{y}{2a}}~~~~~~~~\mbox{if} ~~ y\leq W,\\
~\\
\infty~~~~~~~~~~~~~~~~~\mbox{if} ~~ y> W.
\end{array}\right.
\end{equation}
Notice that  the characteristic length scale for the soft boundary, $\xi_c = 2a$, is larger than (but comparable to) the typical decay length of the edge states that are localized near the hard wall, $\xi = 0.8a$.  The corresponding spectrum is shown in Fig. \ref{Fig28}. The edge mode represented by the red line propagates along the hard boundary, while the orange lines represent the edge states near the soft boundary. Note that the soft mode retains the fundamental property of a topological edge mode to continuously cross the gap and connect the two bulk bands.

Increasing the characteristic length scale of the boundary region will further soften the edge mode. For comparison, a detail of the gap region from Fig. \ref{Fig28} together with the edge mode corresponding  to a confining potential with $\xi_c = 6.5a$ are shown in Fig. \ref{Fig29}. We conclude that in a clean system and in the absence of interactions, topological edge states exist for any value of $\xi_c$. In a real system, because the edge mode softens with increasing $\xi_c$ and the gaps between different branches become smaller, the spectrum of the edge modes will be significantly renormalized by perturbations such as disorder or interactions. However, due to the topological nature of the edge states, these perturbations will not  open a gap in the spectrum.

\begin{figure}[tbp]
\begin{center}
\includegraphics[width=0.47\textwidth]{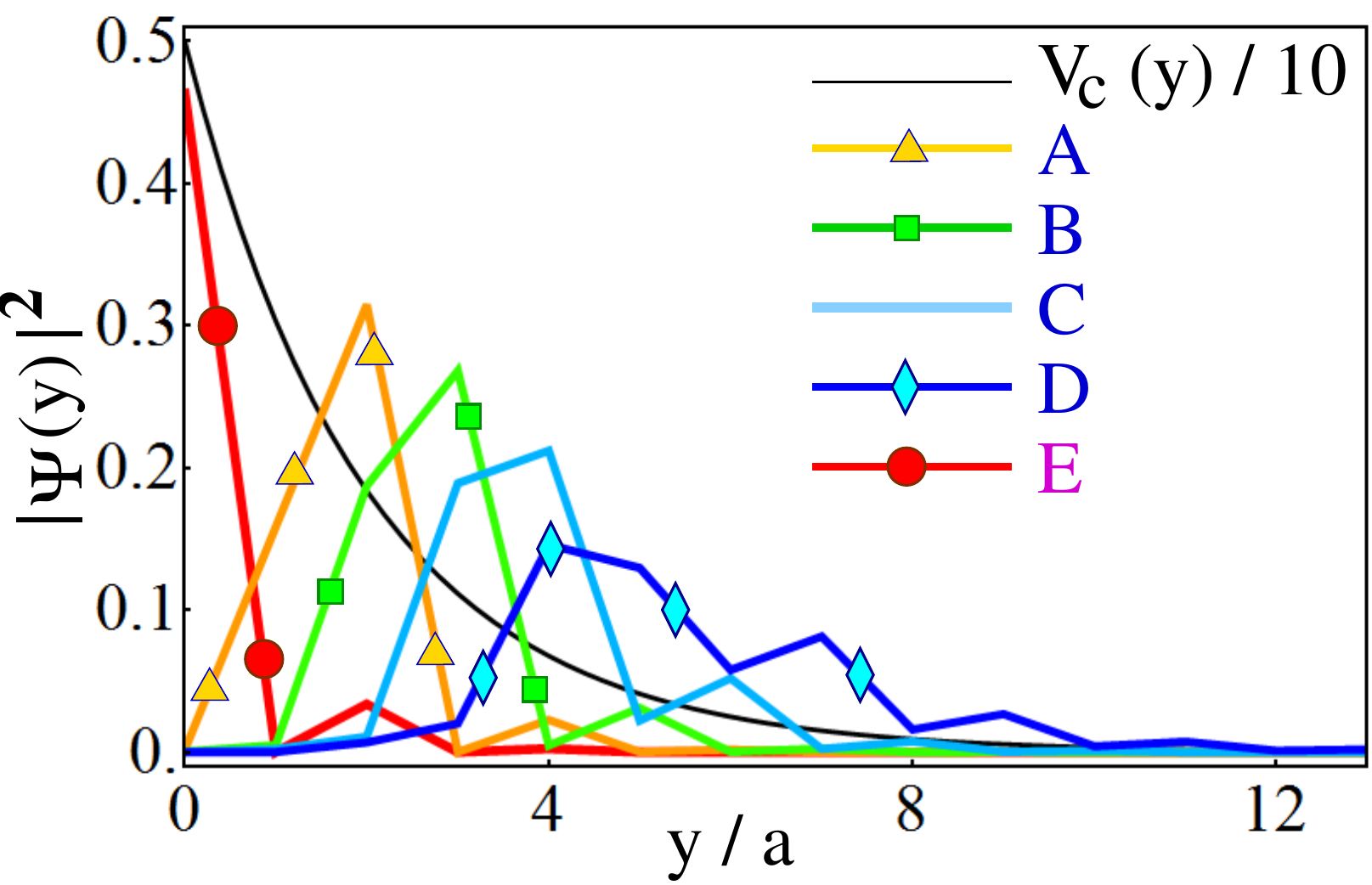}
\end{center}
\caption{(Color online) Amplitudes for the edge states localized near a soft boundary marked in Fig. \ref{Fig29}. The corresponding confining potential, $V_c(y)$ is also shown. For comparison we also show $\vert\Psi_E(W-y)\vert^2$ for the state E which is localized near the hard boundary.
 \vspace{-3mm}}
\label{Fig30}
\end{figure}

The next question that we want to clarify concerns the spatial profile of the edge states in the vicinity of the soft boundary. The amplitude of the states marked A-E in Fig. \ref{Fig29} are shown in Fig. \ref{Fig30}. Near the soft boundary, four different states, A-D, are characterized by the same wave vector, $k_x=\pi/a$. Each state has a maximum amplitude inside the boundary region and decay exponentially in the bulk. Note that the highest energy state, $\psi_A$, has the maximum near the point $y_{\Delta}$ where the confining potential equals the bulk gap, $V_c(y_{\Delta}) = \Delta$. No topological edge states exist in the region where the confining potential exceeds the gap. The edge states with lower energies have the maxima at points with decreasing values of the confining potential and are progressively less confined as the edge mode eventually merges with the bulk.

\begin{figure}[tbp]
\begin{center}
\includegraphics[width=0.47\textwidth]{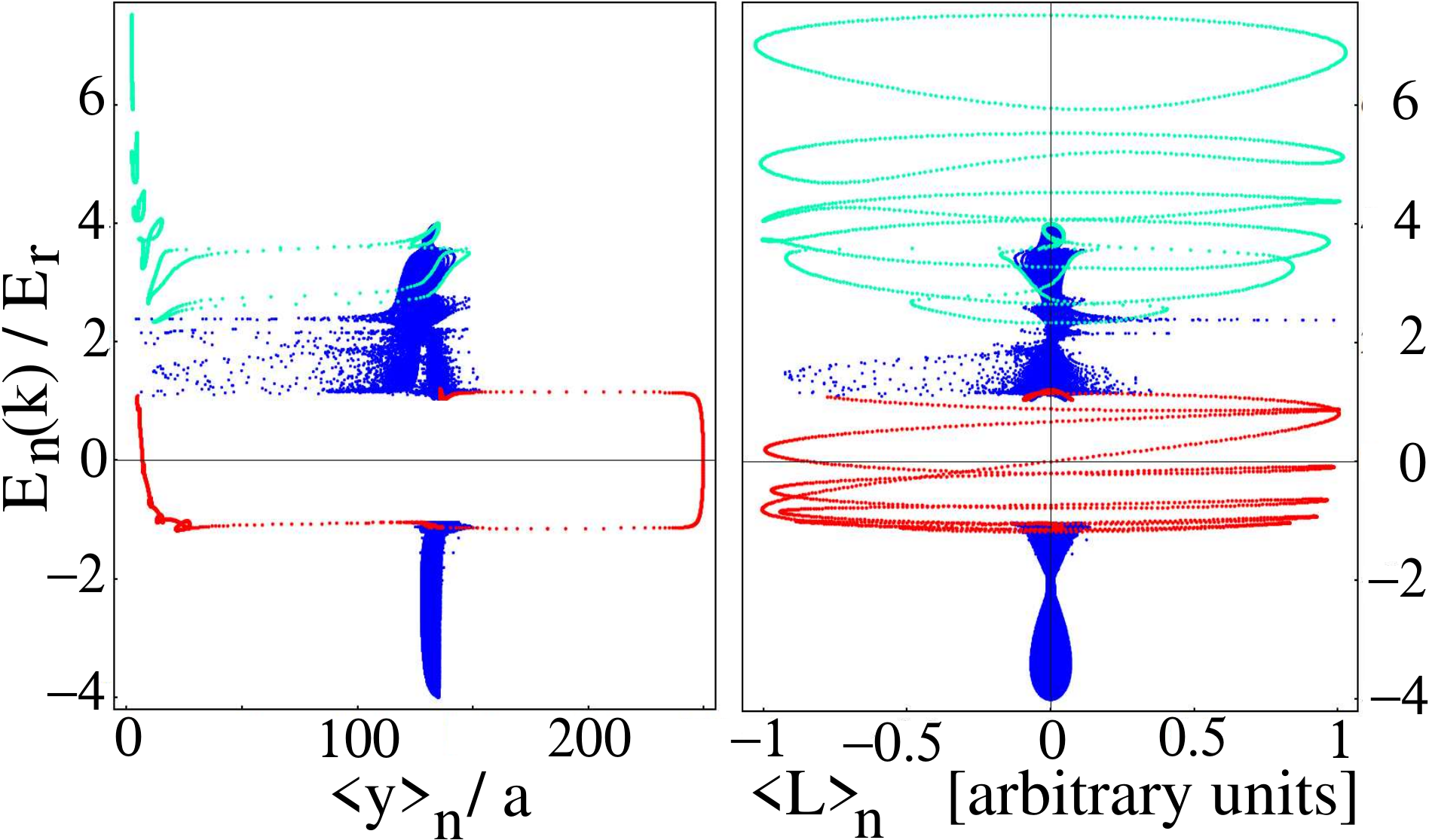}
\end{center}
\caption{(Color online) Average position (left panel) and average angular momentum (right pane) for a stripe with a soft boundary and same parameters as in Fig. \ref{Fig28}.}
\label{Fig31}
\end{figure}

To have a global characterization of the spatial amplitude distribution for all the single particle states it is useful to represent the energy of a given state as function of the average position. The resulting diagram is shown in Fig. \ref{Fig31} (left panel) together with diagram showing the energy as function of the average angular momentum (right panel). This type of diagram is particularly useful in analyzing systems with geometries that have no
translation invariance, as, for example, the disk geometry discussed above. In the left panel of  Fig. \ref{Fig31} one can clearly see the difference between the soft boundary near $y=0$ and the hard wall at $y=250a$. Notice that the topological edge states are not the only edge states localized in the vicinity of the soft boundary, as topologically trivial edge states exist at higher energies. The softening of the edge mode produces a spiral-like shape of energy versus angular-momentum curve. Note that for certain energies the edge states propagating along opposite edges do not necessarily counter-propagate, as one would naively expect given their chiral nature. This is generally the case for systems with inequivalent edges. The consequences of this observation for the transport properties of topological insulators with asymmetric boundaries remain an interesting open question.

\begin{figure}[tbp]
\begin{center}
\includegraphics[width=0.47\textwidth]{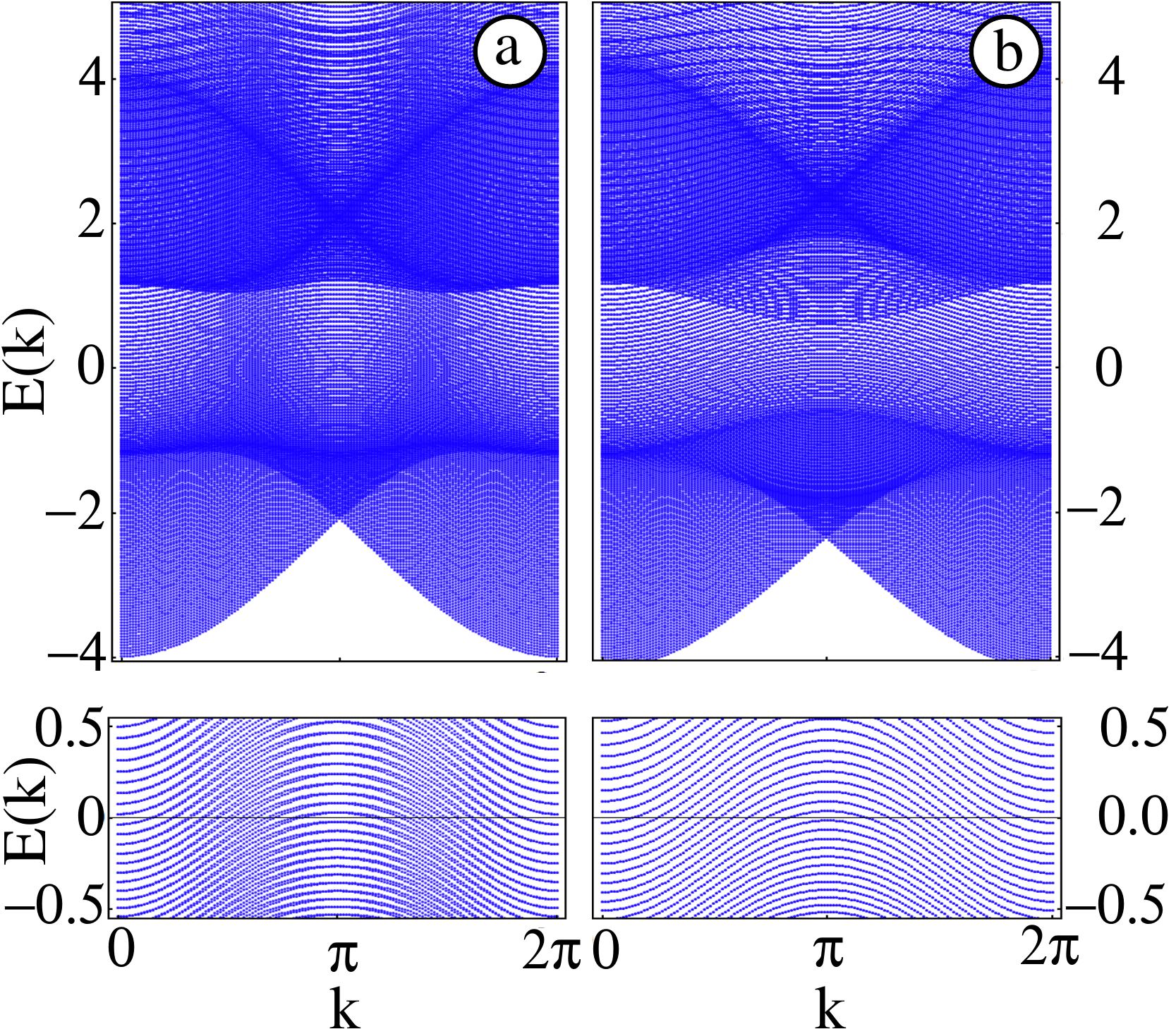}
\end{center}
\caption{(Color online) Spectrum of a system with harmonic confining potential. Panel (a) corresponds to a square superlattice  topological "insulator" model with the same parameters as in Fig \ref{Fig28}, while panel (b) represents a standard insulator model on the square superlattice with $t_1 = E_r$, $t_2=-0.3E_r$,  $t_2^{\prime}=0.3E_r$ and a  staggered on-site potential $\pm\Gamma$ with $\Gamma= E_r$.  The lower panels show the low energy spectra that characterize the inhomogeneous topological (left) and normal (right) metals. Note that the topological metal has a spectrum consisting of two modes that are qualitatively the same as the soft mode shown in Fig. \ref{Fig29} (in the present case both boundaries are soft).  By contrast, the spectrum of the normal metal  contains many disconnected modes.} 
\label{Fig32}
\end{figure}

We return now to the case of smooth confining potentials and in particular to harmonic potentials. As mentioned above, in this case one cannot talk, strictly speaking, about a boundary, but rather a boundary region that can cover, in principle, a significant area. This boundary region is occupied by an inhomogeneous  topological metal~\cite{StanSarma}. The spectrum that characterizes this inhomogeneous metal can be understood qualitatively from Fig. \ref{Fig29}. In the limit of a smooth confining potential, the dispersion of the "edge" mode becomes weaker and the gaps between different branches become smaller, leading to a quasicontinuous spectrum.  Explicitly, we consider a system in the stripe geometry characterized by the same parameters as in Fig. \ref{Fig28} and confined by a harmonic potential $V_c(y) = \frac{1}{2}\Delta y^2/y_0^2$, where $\Delta$ is the bulk gap of the uniform system and $y_0=60a$. The corresponding spectrum is shown in Fig. \ref{Fig32}. For comparison we also show the spectrum of a model with trivial topological properties (panel b), i.e., a square superlattice model with staggered on-site potential, which in the absence of a  confining potential would describe a standard insulator at half filling. Note that from the overall features of the spectra no distinction can be made between the two systems, unlike the hard boundary case when the presence of the characteristic edge modes can clearly identify the topological insulator. Nonetheless, a detailed analysis of the low energy spectrum reveals a qualitative difference between the two systems. Indeed, the topological metal has a spectrum consisting of two modes that are qualitatively the same as the soft mode shown in Fig. \ref{Fig29}, while the spectrum of the normal metal  contains many disconnected modes. In terms of average orbital momentum, this difference would correspond to the difference shown in Fig. \ref{Fig31}between the in-gap "spiral" and the disconnected high-energy "rings." Perturbations can open a gap in the normal metal spectrum [Fig. \ref{Fig32}(b)] but not in the topological metal  [Fig. \ref{Fig32}(a)].

\section{\bf Detection of topological edge states}\label{SecDetect}

Probing quantum Hall states in a condensed matter system typically involves transport measurements. However, this is a rather difficult task in a cold-atom system. Alternatively, as the nontrivial topological properties represent a feature of the single particle Hamiltonian best revealed by the presence of chiral edge states, it was proposed~\cite{StanSarma} to directly map out these edge states using bosons. The procedure involves loading bosons into the edge states and then  imaging the atoms. Note that this technique does not involve the realization of an equilibrium topological insulating state but rather a real-space analysis of the properties of the single particle states. In some sense, it is an effective way of "seeing" a topological phase, something one cannot easily realize in the condensed matter context. First, the optical lattice is loaded with atoms and cooled so that the bosons occupy the lowest-energy single-particle states. Then, a sequence of staged resonance excitation processes is used to promote atoms into states of increasing angular momentum, for example, via a sequence of the two-photon-stimulated Raman transitions\cite{Andersen}. These  intermediate states provide the overlap needed to make resonant Raman transitions to the edge states possible. Finally, 
the atoms loaded into the edge states are imaged  using a direct {\it in situ}
imaging technique, such as the method developed by the Greiner group~\cite{Nelson,gillen2008hst}.

\begin{figure}[tbp]
\begin{center}
\includegraphics[width=0.47\textwidth]{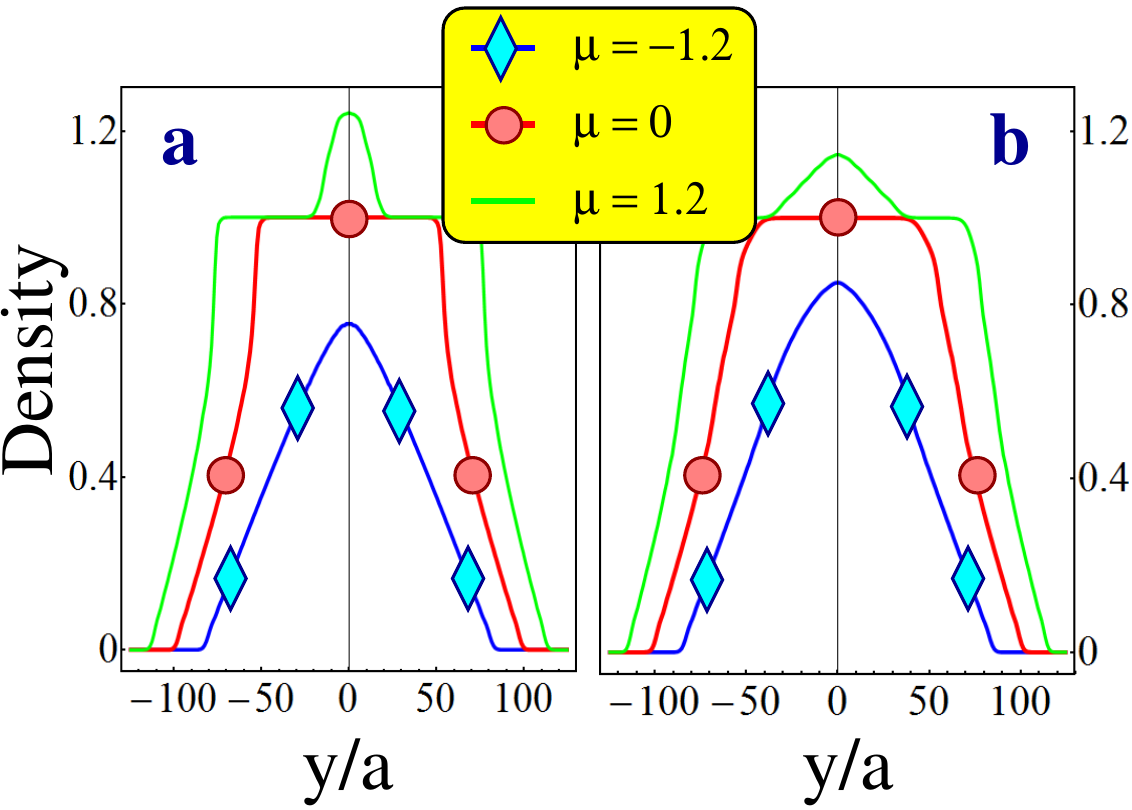}
\end{center}
\caption{(Color online) Density profiles for a fermionic system with harmonic confinement and different values of the chemical potential. Panel (a) corresponds to a topological insulator model, while panel (b) shows the density of a standard insulator. The parameters for the models are the same as in Fig. \ref{Fig32}. Note the plateaus characterized by $\rho=1$ (particles per lattice site), which correspond to the insulating phase.  No qualitative difference can be observed between the two types of models.} 
\label{Fig33}
\end{figure}

Another possibility is to perform density profile measurements on fermionic atomic systems~\cite{Wang,Umuca}. We emphasize that such a measurement can probe the existence of an incompressible insulator, but cannot distinguish between a TI and a trivial insulator, unless supplemented by another probe. To illustrate this point, we calculate explicitly the density profile for a system with harmonic confinement. Note that, in the presence of a smooth confining potential the spectral analysis does not offer much information about the system (see Fig. \ref{Fig32}). In particular, there is no sharp criterion for distinguishing the  "bulk" states from the "edge" states. However, a clear signature for the existence of an incompressible "insulating" bulk surrounded by a compressible metal can be seen in the particle density. Assuming that the system contains fermions with a chemical potential $\mu$, the particle density is defined as
\begin{equation}
\rho({\mathbf r}) = \sum_n^{\epsilon_n\leq\mu}~|\psi_n({\mathbf r})|^2,
\end{equation}
where $\psi_n({\mathbf r})$ are single particle wave functions with energies $\epsilon_n$. In the absence of interactions, a small region of the system around ${\mathbf r}$ characterized by an effective "chemical potential" $\mu -V_c({\mathbf r})$ that falls within the gap corresponding to a uniform infinite system should be an insulator with one particle per site, $\rho({\mathbf r})=1$. Figure \ref{Fig33} shows the results for the topological and standard insulator models with the same parameters as in Fig. \ref{Fig32}. For $\mu=-1.2E_r$ the effective local chemical potential will reside in the lower band or below for any given position and the system is entirely metallic. If  $\mu=0$, the effective chemical potential will reside inside the bulk gap for $-y_0\leq y\leq y_0$. This region, characterized by $\rho=1$, represents the insulating incompressible "bulk" and is surrounded by the compressible metal. Increasing the chemical potential further will determine the occupation of bulk states from the second band and the appearance of a compressible island near the bottom of the trap. In conclusion measuring the density profile represents a possible way to probe the existence of an insulating phase. However, such a measurement cannot distinguish between a normal insulator and a topological insulator. Complete evidence should be obtained by probing the characteristic boundary states. How to probe them remains an important open question in this field. We note here that there are two different regimes in which these states can be investigated: (i) In systems with sharp boundaries there are a relatively small number of well-defined chiral edge states with energies within the bulk gap and (ii) in systems with shallow trapping the insulating "bulk" is surrounded by a boundary region populated with a large number of boundary states which, when occupied, form a topological metal. The two regimes are adiabatically connected. A possible advantage of the shallow trapping is that the ratio between the number of  "boundary" and "bulk"  states can become significant. We suggest that the chiral nature of the boundary states of a topological insulator with broken time reversal symmetry  could help distinguishing them from the boundary states of a standard insulator. Perturbations with opposite angular orientations should determine different responses from the topological metal, in contrast to the normal metal. 

\begin{figure}[tbp]
\begin{center}
\includegraphics[width=0.47\textwidth]{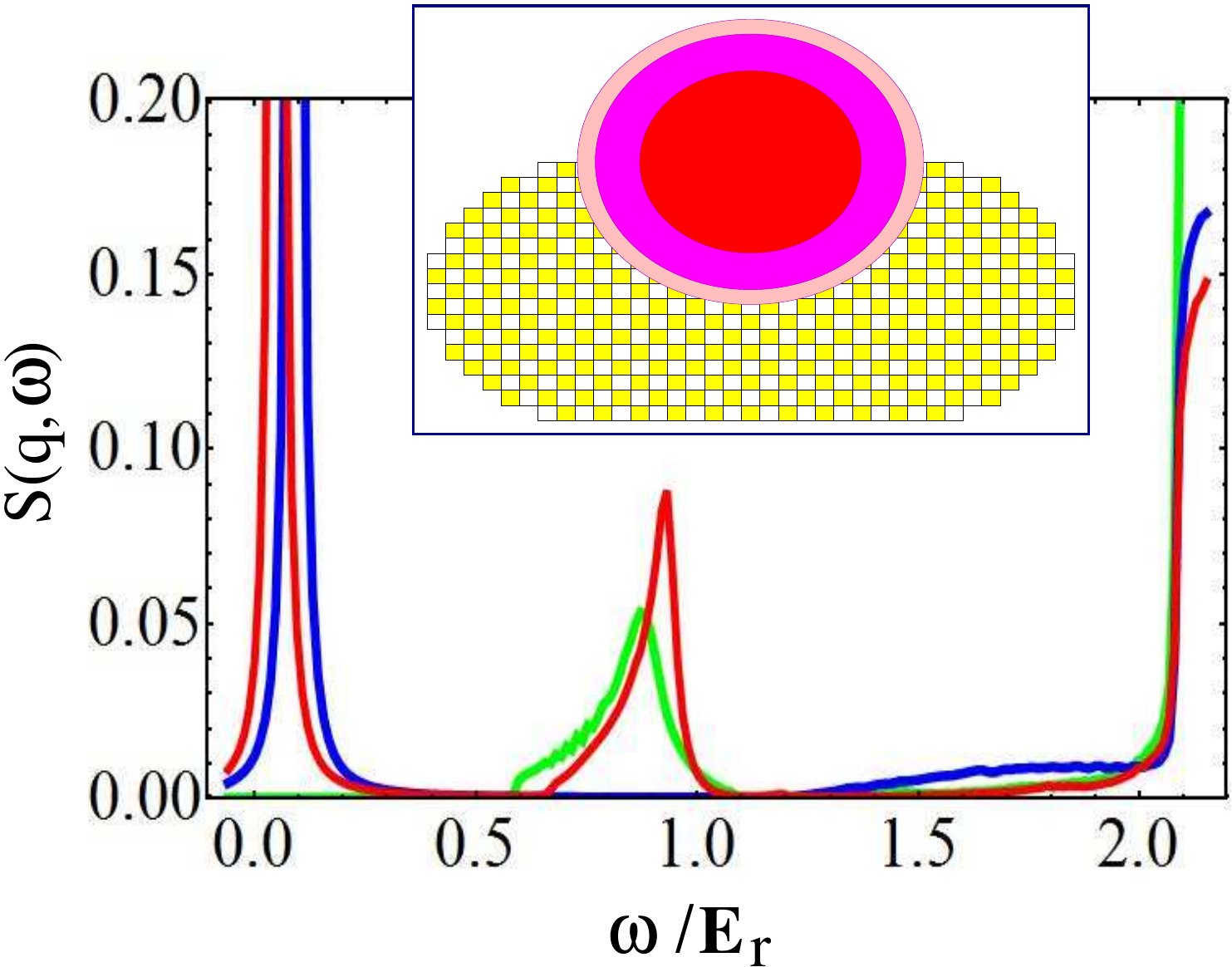}
\end{center}
\caption{(Color online) Dynamical structure factor for a system described by Eq. (\ref{modelH}) with $t_1 = (1+0.3i)E_r$, $t_2=-0.3E_r$ and $t_2^{\prime}=0.3E_r$. The blue (dark gray) line corresponds to a sharp boundary, while the red (gray) line is for a soft boundary with the same profile as in Fig. \ref{Fig28}. The extra in-gap contribution represents transitions between different branches of the soft edge mode. The green (light gray) line corresponds to the same parameters as the red curve and opposite scattering wave vector $-{\mathbf q}$. Note the absence of the low-energy peak. The inset is a schematic illustration of the intersecting laser beams illuminating a portion of the system.} 
\label{Fig34}
\end{figure}

The third type of probe that we propose for detecting topological phases is optical Bragg spectroscopy~\cite{Stamper1999,Liu2010}. As in the case of the density profile measurement, using this probe requires the realization of an equilibrium insulating state. This is obtained by loading fermions into the optical lattice so that the chemical potential lie within the bulk insulating gap. The atomic system is illuminated with a pair of lasers with wave vectors ${\mathbf k}_1$ and ${\mathbf k}_2$, respectively, and a frequency difference $\omega=\omega_1-\omega_2$ much smaller than their detuning from an atomic resonance. The two beams create a traveling intensity modulation $I_{\rm mod}({\mathbf r}, t) = I\cos({\mathbf q}\cdot{\mathbf r} -\omega t)$, where ${\mathbf q}= {\mathbf k}_1-{\mathbf k}_2$. Consequently, the atoms experience a potential proportional to $I_{\rm mod}$ due to the ac Stark effect and may scatter. The optimal geometry for detecting the edge states involves shining the lasers on a certain part of the atomic system that contains a portion of the boundary (see Fig. \ref{Fig34}). Therefore, we will take into account the laser beam profile, which we assume to be Gaussian. The term in the Hamiltonian describing the light-atom interaction  can be expressed using the second-quantized notation as
\begin{equation}
H_{\rm int} = \Omega \int d{\mathbf r}~ \left[e^{-\frac{2 r^2}{w^2}}e^{-i{\mathbf q}\cdot {\mathbf r} - i\omega t} \hat{\psi}({\mathbf r})^{\dagger} \hat{\psi}({\mathbf r}) + h.c.\right], 
\end{equation}
where $\Omega$ is the effective two-photon Rabi frequency, $w$ is the beam width, and $\hat{\psi}({\mathbf r})$ is the atom field operator.  The response of the many body system to this perturbation can be evaluated using Fermi's golden rule. If we neglect the beam profile, light Bragg scattering  measures the dynamical structure factor $S({\mathbf q}, \omega)$, i.e., the density correlations. In general, we have
\begin{eqnarray}
\tilde{S}({\mathbf q}, \omega) &=& \sum_{\nu_i, \nu_f}\sum_{{k}_i,{k}_f} \left[1-f(\epsilon_f)\right]f(\epsilon_i) \label{Sqw}\\
&\times& \vert\langle\Phi_{\nu_f {k}_f}\vert H_{\rm int}\vert \Phi_{\nu_i {k}_i}\rangle\vert^2\delta[\hbar\omega-\epsilon_f+\epsilon_i], \nonumber
\end{eqnarray} 
where $\vert \Phi_{\nu_i {k}_i}\rangle$ and $\vert \Phi_{\nu_f {k}_f}\rangle$ are the initial atomic state and the state after scattering, respectively, and $f(\epsilon)$ is the Fermi distribution function. We assume that the wave vector ${\mathbf q}$ is oriented along the $x$ axis and that the laser beam illuminates a sufficiently small region near the boundary with local properties similar to those of a stripe (Fig. \ref{Fig34}). Taking into account the beam profile, the matrix elements in Eq. (\ref{Sqw}) become $\langle\Phi_{\nu_f {k}_f}\vert H_{\rm int}\vert \Phi_{\nu_i {k}_i}\rangle \propto \exp\left\{-\frac{w^2}{8}(q-k_f-k_i)^2\right\}\Lambda_{\nu_f\nu_i}^{k_f k_i}$, where we approximate the wave functions $\Phi_{\nu k}({\mathbf r})$ with those of a system with stripe geometry that has similar local properties, and 
\begin{equation}
\Lambda_{\nu_f\nu_i}^{k_f k_i} = \sum_{\alpha, n}\Psi_{\nu_f k_f}^*(y_n, \alpha) 
\Psi_{\nu_i k_i}(y_n, \alpha) e^{-\frac{2y_n^2}{w^2}}e^{i(k_f-k_i)\delta_{\alpha}}.
\end{equation}    
For a stripe oriented along the nearest-neighbor direction which is finite in the $y$ direction one has $\Phi_{\nu k}({\mathbf r}) = \sum_{m, n, \alpha}\Psi_{\nu k}(y_n, \alpha)e^{i k x_m} \psi(x-x_m-\delta_{\alpha}, y-y_n)$,
where $\psi(x, y)$ are orbitals localized in the vicinity of the lattice nodes.  The eigenstates $\Psi_{\nu k}(y_n, \alpha)$ are indexed by the wave vector $k$ (oriented along the $x$ direction) and the discrete quantum number $\nu$ and depend on the position $y_n$ and the sublattice index $\alpha =1, 2$. The two sublattices are shifted by one lattice spacing, i.e., $\delta_1=0$ and $\delta_2=a$.

First we consider the TI model on a square super-lattice described by Eq. (\ref{modelH}) and hard wall boundary conditions. We assume that the system has a portion of the boundary oriented along the nearest-neighbor direction that is longer than the width of the intersecting laser beams (Fig. \ref{Fig34}) and focus the lasers on that edge. Note that the light field will cover a significant part of the system, but no other edge. The dynamical structure factor (blue/dark gray line in Fig. \ref{Fig34}) is characterized by a gap corresponding to transitions between bulk states and a low-energy peak at $\omega \approx q v_0$, where $v_0$ is the velocity of the edge mode in the vicinity of the chemical potential, associated with the edge states. In addition, edge-bulk transitions generate a small in-gap contribution. 

Next we consider a soft boundary. The edge mode softens and develops multiple branches (see Fig. \ref{Fig28}). The low-energy peak moves to lower frequencies due to a smaller value of $v_0$ and, in addition, an extra peak develops inside the gap (see the red/gray curve in Fig. \ref{Fig34}). This peak corresponds to transitions between different branches of the edge mode. The chiral character of the edge states can be probed by reversing the scattering wave vector, ${\mathbf q} \rightarrow -{\mathbf q}$, or by inverting the vector potential that generates the Peierls phases, i.e., $t_1 \rightarrow t_1^*$. As a result, $q v_0 <0$ and there are no low-energy transitions. Consequently, the characteristic low energy peak in the structure factor is absent (green/light gray line in Fig. \ref{Fig34}). Note that the secondary in-gap peak corresponding to interbranch transitions is weakly modified. In a chiral metal generated by very soft confining (see Fig. \ref{Fig32}) inter-branch contributions that  occur at low energy become dominant and  eventually destroy this signature of the chiral character of the edge states.  

\section{Summary}\label{summary}

In this article, we have presented a comprehensive analysis of  two-dimensional optical lattice Hamiltonians that give rise to topological insulating states with broken time-reversal symmetry. We have extracted the main ingredients responsible for the appearance of such states and shown that there are an infinite number of lattice models that possess the non-trivial topological structure. It is suggested that the choice of a model is to be dictated by experimental convenience and as such, square superlattices may be an optimal choice from the experimental point of view. We also note here that 
using the setups to create spin-orbit-coupled systems proposed in Refs.~\cite{SZG,SAG,Spiel}, one can engineer in a similar way time-reversal topological insulator systems. However, in cold-atom settings these systems would represent a higher degree of experimental complexity (in contrast to solid-state systems) and therefore the intimately related and simpler  lattice quantum Hall states studied here represent a natural starting point for initial  experiments in this line of research.

One particularly important question that remains open is how to experimentally probe the nontrivial topological properties of a given optical lattice model. We note that our analysis here, as well as most other existing analyses of noninteracting topological insulator models, represent essentially mathematical studies of a complicated {\em single-particle} Hamiltonian, which is not specific to the types of particles that would occupy the lattice sites in a physical model. In fact, our calculations in the finite-size case are basically solving for the spectrum of a large and finite matrix, either associated with the sites of a lattice in real space or those in dual space, which are related to each other via discrete Fourier transform. The resulting spectrum exists in and of itself, and how to probe this spectrum and in particular its topological states within the band gap is a completely separate issue of great physical importance. In solids, the choice of particles to occupy the bands is limited to electron excitations or fermionic quasiparticles arising in various mean-field-like treatments of interacting models (e.g., Bogoliubov excitations in superconductors). The experimental signatures there include charge or spin transport dominated by the edge modes or direct probes of the gapless spectrum at the boundary, which is especially relevant in three-dimensional 
topological insulators. However, cold atoms are drastically different. First, because there are many possible choices of atoms that can be loaded in the optical lattice, which could be either bosons or fermions. Second, these particles are necessarily electrically neutral and therefore transport like probes, while not impossible in principle, are probably too difficult to realize in practice especially if quantized transport is the goal of such a measurement. Hence, other approaches need to be developed and we point out here that cold-atoms bosons, rather than fermions, may become the first line of choice to visualize the topological properties of the spectrum. 

While fermions can indeed be used to fill up the band up to the band gap, so the Fermi level crosses the topological modes, there are little observable consequences for e.g. time-of-flight measurement, which will generally be dominated by the less interesting bulk contributions.  One possible solution to increase the boundary contribution is to use shallow trapping potentials. Alternatively, one can use non-interacting or weakly-interacting bosons which occupy or condense into the lowest-energy states in the spectrum, while in thermodynamic equilibrium at low temperatures. Such low-energy states are not topological. However as pointed out in Ref.~\cite{StanSarma}, using  two-photon-stimulated Raman transitions (such as that used in Ref.~\cite{Andersen}), one can transfer a macroscopic number of bosons from the condensate specifically into the topological edge states and then use time-of-flight measurement to observe a vortex, associated with a few-lattice-constant-thick chiral topological modes. We emphasize here that while such a non-equilibrium measurement would not represent a thermodynamic topological phase,  it would, however, lead to an impressive and explicit manifestation of the non-trivial topology of the underlying exotic spectrum.

Another particularly interesting avenue is to use interacting bosons, e.g. bosons with strong on-site repulsion. Since bosons with hard-core repulsion are not equivalent to free fermions in two dimensions, the noninteracting spectrum will necessarily be modified and it is an interesting open problem as to what states may arise out such a system. However, it is conceivable that  at the fillings corresponding to a Mott-insulating phase, the chiral hopping terms will constraint the bosons into a topological Mott-insulating state.

Finally, we reiterate that measuring a system with nontrivial topological properties can be realized following two possible avenues. Any successful experiment that would be capable of probing well defined  topological edge states will need to address the problem of creating "sharp-enough" boundaries of the trapping potential. We have shown in this article and Ref.~\cite{StanSarma} that a standard quadratic trap is not sufficient for this purpose, as it leads to an inhomogeneous metallic phase. Adding any term to the confining potential capable of introducing a boundary length scale would solve this problem and should give rise to detectable edge states. Another possibility  would be to create a different type of inhomogeneity in the bulk of the system, e.g., by adding a strongly repulsive potential in the center of a regular trap or by strongly altering the lattice hopping terms along a certain line of links of the optical lattice. Topological edge modes are bound to appear not only at the external boundaries of the system, but in all such cases, as long as the perturbation is larger than the relevant bandwidth. The second avenue toward detecting topological quantum states is  to probe the inhomogeneous topological metal that forms in weakly confined systems. The significantly increased fraction of boundary states could lead to observable consequences in a time-of-flight  experiment. Regardless of the boundary sharpness, the existence of an incompressible insulating "bulk" can be revealed by a density profile measurement. Nonetheless, we emphasize that, while the presence of a density plateau proves the existence of an insulator, it does not determine its topological nature. Ultimately, distinguishing between a topological insulator and a standard insulator requires probing the gapless states that form at their boundaries. 

This work is supported by US-ARO and JQI-NSF-PFC.

\bibliography{refsTopIns}

\begin{thebibliography}{81}
\expandafter\ifx\csname natexlab\endcsname\relax\def\natexlab#1{#1}\fi
\expandafter\ifx\csname bibnamefont\endcsname\relax
  \def\bibnamefont#1{#1}\fi
\expandafter\ifx\csname bibfnamefont\endcsname\relax
  \def\bibfnamefont#1{#1}\fi
\expandafter\ifx\csname citenamefont\endcsname\relax
  \def\citenamefont#1{#1}\fi
\expandafter\ifx\csname url\endcsname\relax
  \def\url#1{\texttt{#1}}\fi
\expandafter\ifx\csname urlprefix\endcsname\relax\def\urlprefix{URL }\fi
\providecommand{\bibinfo}[2]{#2}
\providecommand{\eprint}[2][]{\url{#2}}

\bibitem[{\citenamefont{Kitaev}()}]{PTTI}
\bibinfo{author}{\bibfnamefont{A.}~\bibnamefont{Kitaev}},
  \bibinfo{howpublished}{e-print arXiv:0901.2686 (2009)}.

\bibitem[{\citenamefont{Ryu et~al.}()\citenamefont{Ryu, Schnyder, Furusaki, and
  Ludwig}}]{Ludwig2}
\bibinfo{author}{\bibfnamefont{S.}~\bibnamefont{Ryu}},
  \bibinfo{author}{\bibfnamefont{A.}~\bibnamefont{Schnyder}},
  \bibinfo{author}{\bibfnamefont{A.}~\bibnamefont{Furusaki}}, \bibnamefont{and}
  \bibinfo{author}{\bibfnamefont{A.}~\bibnamefont{Ludwig}},
  \bibinfo{howpublished}{New J. Phys. {\bf 12}, 065010 (2010)}.

\bibitem[{\citenamefont{Volovik}({\natexlab{a}})}]{Volovik2009}
\bibinfo{author}{\bibfnamefont{G.}~\bibnamefont{Volovik}},
  \bibinfo{howpublished}{JETP Lett. {\bf 91}, 55 (2010)}.

\bibitem[{\citenamefont{Volovik and Yakovenko}()}]{Volovik1989}
\bibinfo{author}{\bibfnamefont{G.}~\bibnamefont{Volovik}} \bibnamefont{and}
  \bibinfo{author}{\bibfnamefont{V.}~\bibnamefont{Yakovenko}},
  \bibinfo{howpublished}{J. Phys.: Cond. Matter {\bf N 31}, 5263 (1989)}.

\bibitem[{\citenamefont{Volovik}({\natexlab{b}})}]{Volovik2009a}
\bibinfo{author}{\bibfnamefont{G.}~\bibnamefont{Volovik}},
  \bibinfo{howpublished}{JETP Lett. {\bf 90}, 587 (2009)}.

\bibitem[{\citenamefont{Klitzing et~al.}(1980)\citenamefont{Klitzing, Dorda,
  and Pepper}}]{QHE1}
\bibinfo{author}{\bibfnamefont{K.~v.} \bibnamefont{Klitzing}},
  \bibinfo{author}{\bibfnamefont{G.}~\bibnamefont{Dorda}}, \bibnamefont{and}
  \bibinfo{author}{\bibfnamefont{M.}~\bibnamefont{Pepper}},
  \bibinfo{journal}{Phys. Rev. Lett.} \textbf{\bibinfo{volume}{45}},
  \bibinfo{pages}{494} (\bibinfo{year}{1980}).

\bibitem[{\citenamefont{Tsui et~al.}(1982)\citenamefont{Tsui, Stormer, and
  Gossard}}]{QHE2}
\bibinfo{author}{\bibfnamefont{D.~C.} \bibnamefont{Tsui}},
  \bibinfo{author}{\bibfnamefont{H.~L.} \bibnamefont{Stormer}},
  \bibnamefont{and} \bibinfo{author}{\bibfnamefont{A.~C.}
  \bibnamefont{Gossard}}, \bibinfo{journal}{Phys. Rev. Lett.}
  \textbf{\bibinfo{volume}{48}}, \bibinfo{pages}{1559} (\bibinfo{year}{1982}).

\bibitem[{\citenamefont{Landau}(1937)}]{Landau}
\bibinfo{author}{\bibfnamefont{L.~D.} \bibnamefont{Landau}},
  \bibinfo{journal}{Phys. Z. Sowjetunion} \textbf{\bibinfo{volume}{11}},
  \bibinfo{pages}{26} (\bibinfo{year}{1937}).

\bibitem[{\citenamefont{Wen}(1991)}]{Wen}
\bibinfo{author}{\bibfnamefont{X.~G.} \bibnamefont{Wen}},
  \bibinfo{journal}{\prb} \textbf{\bibinfo{volume}{44}}, \bibinfo{pages}{2664}
  (\bibinfo{year}{1991}).

\bibitem[{\citenamefont{Korenman and Drew}(1987)}]{Korenman1986}
\bibinfo{author}{\bibfnamefont{V.}~\bibnamefont{Korenman}} \bibnamefont{and}
  \bibinfo{author}{\bibfnamefont{H.~D.} \bibnamefont{Drew}},
  \bibinfo{journal}{Phys. Rev. B} \textbf{\bibinfo{volume}{35}},
  \bibinfo{pages}{6446} (\bibinfo{year}{1987}).

\bibitem[{\citenamefont{Agassi and Korenman}(1988)}]{Korenman1987}
\bibinfo{author}{\bibfnamefont{D.}~\bibnamefont{Agassi}} \bibnamefont{and}
  \bibinfo{author}{\bibfnamefont{V.}~\bibnamefont{Korenman}},
  \bibinfo{journal}{Phys. Rev. B} \textbf{\bibinfo{volume}{37}},
  \bibinfo{pages}{10095} (\bibinfo{year}{1988}).

\bibitem[{\citenamefont{Moessner and Sondhi}(2001)}]{Moessner}
\bibinfo{author}{\bibfnamefont{R.}~\bibnamefont{Moessner}} \bibnamefont{and}
  \bibinfo{author}{\bibfnamefont{S.~L.} \bibnamefont{Sondhi}},
  \bibinfo{journal}{Phys. Rev. Lett.} \textbf{\bibinfo{volume}{86}},
  \bibinfo{pages}{1881} (\bibinfo{year}{2001}).

\bibitem[{\citenamefont{Balents et~al.}(2002)\citenamefont{Balents, Fisher, and
  Girvin}}]{Balents}
\bibinfo{author}{\bibfnamefont{L.}~\bibnamefont{Balents}},
  \bibinfo{author}{\bibfnamefont{M.~P.~A.} \bibnamefont{Fisher}},
  \bibnamefont{and} \bibinfo{author}{\bibfnamefont{S.~M.}
  \bibnamefont{Girvin}}, \bibinfo{journal}{Phys. Rev. B}
  \textbf{\bibinfo{volume}{65}}, \bibinfo{pages}{224412}
  (\bibinfo{year}{2002}).

\bibitem[{\citenamefont{Misguich et~al.}(2002)\citenamefont{Misguich, Serban,
  and Pasquier}}]{Misguich}
\bibinfo{author}{\bibfnamefont{G.}~\bibnamefont{Misguich}},
  \bibinfo{author}{\bibfnamefont{D.}~\bibnamefont{Serban}}, \bibnamefont{and}
  \bibinfo{author}{\bibfnamefont{V.}~\bibnamefont{Pasquier}},
  \bibinfo{journal}{Phys. Rev. Lett.} \textbf{\bibinfo{volume}{89}},
  \bibinfo{pages}{137202} (\bibinfo{year}{2002}).

\bibitem[{\citenamefont{Kitaev}(2003)}]{Kitaev}
\bibinfo{author}{\bibfnamefont{A.~Y.} \bibnamefont{Kitaev}},
  \bibinfo{journal}{Ann. Phys. (N.Y.)} \textbf{\bibinfo{volume}{303}},
  \bibinfo{pages}{2} (\bibinfo{year}{2003}).

\bibitem[{\citenamefont{Wen}(2003)}]{Wen2003}
\bibinfo{author}{\bibfnamefont{X.-G.} \bibnamefont{Wen}},
  \bibinfo{journal}{Phys. Rev. Lett.} \textbf{\bibinfo{volume}{90}},
  \bibinfo{pages}{016803} (\bibinfo{year}{2003}).

\bibitem[{\citenamefont{Kane and Mele}(2005{\natexlab{a}})}]{KaneMele}
\bibinfo{author}{\bibfnamefont{C.~L.} \bibnamefont{Kane}} \bibnamefont{and}
  \bibinfo{author}{\bibfnamefont{E.~J.} \bibnamefont{Mele}},
  \bibinfo{journal}{\prl} \textbf{\bibinfo{volume}{95}},
  \bibinfo{pages}{226801} (\bibinfo{year}{2005}{\natexlab{a}}).

\bibitem[{\citenamefont{Kane and Mele}(2005{\natexlab{b}})}]{KaneMele2005}
\bibinfo{author}{\bibfnamefont{C.~L.} \bibnamefont{Kane}} \bibnamefont{and}
  \bibinfo{author}{\bibfnamefont{E.~J.} \bibnamefont{Mele}},
  \bibinfo{journal}{Phys. Rev. Lett.} \textbf{\bibinfo{volume}{95}},
  \bibinfo{pages}{146802} (\bibinfo{year}{2005}{\natexlab{b}}).

\bibitem[{\citenamefont{Fu and Kane}(2006)}]{FuKane}
\bibinfo{author}{\bibfnamefont{L.}~\bibnamefont{Fu}} \bibnamefont{and}
  \bibinfo{author}{\bibfnamefont{C.~L.} \bibnamefont{Kane}},
  \bibinfo{journal}{Physical Review B (Condensed Matter and Materials Physics)}
  \textbf{\bibinfo{volume}{74}}, \bibinfo{pages}{195312}
  (\bibinfo{year}{2006}).

\bibitem[{\citenamefont{Bernevig and Zhang}(2006)}]{BernevigZ}
\bibinfo{author}{\bibfnamefont{B.~A.} \bibnamefont{Bernevig}} \bibnamefont{and}
  \bibinfo{author}{\bibfnamefont{S.-C.} \bibnamefont{Zhang}},
  \bibinfo{journal}{Physical Review Letters} \textbf{\bibinfo{volume}{96}},
  \bibinfo{pages}{106802} (\bibinfo{year}{2006}).

\bibitem[{\citenamefont{Murakami}(2006)}]{Murakami}
\bibinfo{author}{\bibfnamefont{S.}~\bibnamefont{Murakami}},
  \bibinfo{journal}{Physical Review Letters} \textbf{\bibinfo{volume}{97}},
  \bibinfo{pages}{236805} (\bibinfo{year}{2006}).

\bibitem[{\citenamefont{Wu et~al.}(2006)\citenamefont{Wu, Bernevig, and
  Zhang}}]{Wu}
\bibinfo{author}{\bibfnamefont{C.}~\bibnamefont{Wu}},
  \bibinfo{author}{\bibfnamefont{B.~A.} \bibnamefont{Bernevig}},
  \bibnamefont{and} \bibinfo{author}{\bibfnamefont{S.-C.} \bibnamefont{Zhang}},
  \bibinfo{journal}{Physical Review Letters} \textbf{\bibinfo{volume}{96}},
  \bibinfo{pages}{106401} (\bibinfo{year}{2006}).

\bibitem[{\citenamefont{Moore and Balents}(2007)}]{Moore}
\bibinfo{author}{\bibfnamefont{J.~E.} \bibnamefont{Moore}} \bibnamefont{and}
  \bibinfo{author}{\bibfnamefont{L.}~\bibnamefont{Balents}},
  \bibinfo{journal}{Physical Review B} \textbf{\bibinfo{volume}{75}},
  \bibinfo{pages}{121306} (\bibinfo{year}{2007}).

\bibitem[{\citenamefont{Fu et~al.}()\citenamefont{Fu, Kane, and
  Mele}}]{FuKaneMele}
\bibinfo{author}{\bibfnamefont{L.}~\bibnamefont{Fu}},
  \bibinfo{author}{\bibfnamefont{C.~L.} \bibnamefont{Kane}}, \bibnamefont{and}
  \bibinfo{author}{\bibfnamefont{E.}~\bibnamefont{Mele}},
  \bibinfo{howpublished}{Phys. Rev. Lett. {\bf 98}, 106803 (2007); L. Fu and C.
  L. Kane, Phys. Rev. B 76, 045302 (2007)}.

\bibitem[{\citenamefont{Konig et~al.}(2007)\citenamefont{Konig, Wiedmann,
  Brune, Roth, Buhmann, Molenkamp, Qi, and Zhang}}]{Konig}
\bibinfo{author}{\bibfnamefont{M.}~\bibnamefont{Konig}},
  \bibinfo{author}{\bibfnamefont{S.}~\bibnamefont{Wiedmann}},
  \bibinfo{author}{\bibfnamefont{C.}~\bibnamefont{Brune}},
  \bibinfo{author}{\bibfnamefont{A.}~\bibnamefont{Roth}},
  \bibinfo{author}{\bibfnamefont{H.}~\bibnamefont{Buhmann}},
  \bibinfo{author}{\bibfnamefont{L.~W.} \bibnamefont{Molenkamp}},
  \bibinfo{author}{\bibfnamefont{X.-L.} \bibnamefont{Qi}}, \bibnamefont{and}
  \bibinfo{author}{\bibfnamefont{S.-C.} \bibnamefont{Zhang}},
  \bibinfo{journal}{Science} \textbf{\bibinfo{volume}{318}},
  \bibinfo{pages}{766} (\bibinfo{year}{2007}).

\bibitem[{\citenamefont{Hsieh et~al.}(2008)\citenamefont{Hsieh, Qian, Wray,
  Xia, Hor, Cava, and Hasan}}]{Hsieh2008}
\bibinfo{author}{\bibfnamefont{D.}~\bibnamefont{Hsieh}},
  \bibinfo{author}{\bibfnamefont{D.}~\bibnamefont{Qian}},
  \bibinfo{author}{\bibfnamefont{L.}~\bibnamefont{Wray}},
  \bibinfo{author}{\bibfnamefont{Y.}~\bibnamefont{Xia}},
  \bibinfo{author}{\bibfnamefont{Y.~S.} \bibnamefont{Hor}},
  \bibinfo{author}{\bibfnamefont{R.~J.} \bibnamefont{Cava}}, \bibnamefont{and}
  \bibinfo{author}{\bibfnamefont{M.~Z.} \bibnamefont{Hasan}},
  \bibinfo{journal}{Nature} \textbf{\bibinfo{volume}{452}},
  \bibinfo{pages}{970} (\bibinfo{year}{2008}).

\bibitem[{\citenamefont{Hsieh et~al.}(2009{\natexlab{a}})\citenamefont{Hsieh,
  Xia, Wray, Qian, Pal, Dil, Osterwalder, Meier, Bihlmayer, Kane
  et~al.}}]{Hsieh2009}
\bibinfo{author}{\bibfnamefont{D.}~\bibnamefont{Hsieh}},
  \bibinfo{author}{\bibfnamefont{Y.}~\bibnamefont{Xia}},
  \bibinfo{author}{\bibfnamefont{L.}~\bibnamefont{Wray}},
  \bibinfo{author}{\bibfnamefont{D.}~\bibnamefont{Qian}},
  \bibinfo{author}{\bibfnamefont{A.}~\bibnamefont{Pal}},
  \bibinfo{author}{\bibfnamefont{J.~H.} \bibnamefont{Dil}},
  \bibinfo{author}{\bibfnamefont{J.}~\bibnamefont{Osterwalder}},
  \bibinfo{author}{\bibfnamefont{F.}~\bibnamefont{Meier}},
  \bibinfo{author}{\bibfnamefont{G.}~\bibnamefont{Bihlmayer}},
  \bibinfo{author}{\bibfnamefont{C.~L.} \bibnamefont{Kane}},
  \bibnamefont{et~al.}, \bibinfo{journal}{Science}
  \textbf{\bibinfo{volume}{323}}, \bibinfo{pages}{919}
  (\bibinfo{year}{2009}{\natexlab{a}}).

\bibitem[{\citenamefont{Hsieh et~al.}(2009{\natexlab{b}})\citenamefont{Hsieh,
  Xia, Qian, Wray, Dil, Meier, Osterwalder, Patthey, Checkelsky, Ong
  et~al.}}]{Hsieh2009a}
\bibinfo{author}{\bibfnamefont{D.}~\bibnamefont{Hsieh}},
  \bibinfo{author}{\bibfnamefont{Y.}~\bibnamefont{Xia}},
  \bibinfo{author}{\bibfnamefont{D.}~\bibnamefont{Qian}},
  \bibinfo{author}{\bibfnamefont{L.}~\bibnamefont{Wray}},
  \bibinfo{author}{\bibfnamefont{J.~H.} \bibnamefont{Dil}},
  \bibinfo{author}{\bibfnamefont{F.}~\bibnamefont{Meier}},
  \bibinfo{author}{\bibfnamefont{J.}~\bibnamefont{Osterwalder}},
  \bibinfo{author}{\bibfnamefont{L.}~\bibnamefont{Patthey}},
  \bibinfo{author}{\bibfnamefont{J.~G.} \bibnamefont{Checkelsky}},
  \bibinfo{author}{\bibfnamefont{N.~P.} \bibnamefont{Ong}},
  \bibnamefont{et~al.}, \bibinfo{journal}{Nature}
  \textbf{\bibinfo{volume}{460}}, \bibinfo{pages}{1101}
  (\bibinfo{year}{2009}{\natexlab{b}}).

\bibitem[{\citenamefont{Roushan et~al.}(2009)\citenamefont{Roushan, Seo,
  Parker, Hor, Hsieh, Qian, Richardella, Hasan, Cava, and Yazdani}}]{Roushan}
\bibinfo{author}{\bibfnamefont{P.}~\bibnamefont{Roushan}},
  \bibinfo{author}{\bibfnamefont{J.}~\bibnamefont{Seo}},
  \bibinfo{author}{\bibfnamefont{C.~V.} \bibnamefont{Parker}},
  \bibinfo{author}{\bibfnamefont{Y.~S.} \bibnamefont{Hor}},
  \bibinfo{author}{\bibfnamefont{D.}~\bibnamefont{Hsieh}},
  \bibinfo{author}{\bibfnamefont{D.}~\bibnamefont{Qian}},
  \bibinfo{author}{\bibfnamefont{A.}~\bibnamefont{Richardella}},
  \bibinfo{author}{\bibfnamefont{M.~Z.} \bibnamefont{Hasan}},
  \bibinfo{author}{\bibfnamefont{R.~J.} \bibnamefont{Cava}}, \bibnamefont{and}
  \bibinfo{author}{\bibfnamefont{A.}~\bibnamefont{Yazdani}},
  \bibinfo{journal}{Nature} \textbf{\bibinfo{volume}{460}},
  \bibinfo{pages}{1106} (\bibinfo{year}{2009}).

\bibitem[{\citenamefont{Wu}(2008)}]{CWu2008}
\bibinfo{author}{\bibfnamefont{C.}~\bibnamefont{Wu}}, \bibinfo{journal}{Phys.
  Rev. Lett.} \textbf{\bibinfo{volume}{101}}, \bibinfo{pages}{186807}
  (\bibinfo{year}{2008}).

\bibitem[{\citenamefont{Shao et~al.}(2008)\citenamefont{Shao, Zhu, Sheng, Xing,
  and Wang}}]{Wang}
\bibinfo{author}{\bibfnamefont{L.~B.} \bibnamefont{Shao}},
  \bibinfo{author}{\bibfnamefont{S.-L.} \bibnamefont{Zhu}},
  \bibinfo{author}{\bibfnamefont{L.}~\bibnamefont{Sheng}},
  \bibinfo{author}{\bibfnamefont{D.}~\bibnamefont{Xing}}, \bibnamefont{and}
  \bibinfo{author}{\bibfnamefont{Z.~D.} \bibnamefont{Wang}},
  \bibinfo{journal}{\prl} \textbf{\bibinfo{volume}{101}},
  \bibinfo{pages}{246810} (\bibinfo{year}{2008}).

\bibitem[{\citenamefont{Stanescu et~al.}(2009)\citenamefont{Stanescu, Galitski,
  Vaishnav, Clark, and Sarma}}]{StanSarma}
\bibinfo{author}{\bibfnamefont{T.~D.} \bibnamefont{Stanescu}},
  \bibinfo{author}{\bibfnamefont{V.}~\bibnamefont{Galitski}},
  \bibinfo{author}{\bibfnamefont{J.~Y.} \bibnamefont{Vaishnav}},
  \bibinfo{author}{\bibfnamefont{C.~W.} \bibnamefont{Clark}}, \bibnamefont{and}
  \bibinfo{author}{\bibfnamefont{S.~D.} \bibnamefont{Sarma}},
  \bibinfo{journal}{Physical Review A} \textbf{\bibinfo{volume}{79}},
  \bibinfo{pages}{053639} (\bibinfo{year}{2009}).

\bibitem[{\citenamefont{Haldane}(1988)}]{Haldane}
\bibinfo{author}{\bibfnamefont{F.~D.~M.} \bibnamefont{Haldane}},
  \bibinfo{journal}{\prl} \textbf{\bibinfo{volume}{61}}, \bibinfo{pages}{2015}
  (\bibinfo{year}{1988}).

\bibitem[{\citenamefont{Stanescu et~al.}({\natexlab{a}})\citenamefont{Stanescu,
  Zhang, and Galitski}}]{SZG}
\bibinfo{author}{\bibfnamefont{T.~D.} \bibnamefont{Stanescu}},
  \bibinfo{author}{\bibfnamefont{C.}~\bibnamefont{Zhang}}, \bibnamefont{and}
  \bibinfo{author}{\bibfnamefont{V.~M.} \bibnamefont{Galitski}},
  \bibinfo{howpublished}{Phys. Rev. Lett. {\bf 99}, 110403 (2007)}.

\bibitem[{\citenamefont{Dum and Olshanii}(1996)}]{Dum}
\bibinfo{author}{\bibfnamefont{R.}~\bibnamefont{Dum}} \bibnamefont{and}
  \bibinfo{author}{\bibfnamefont{M.}~\bibnamefont{Olshanii}},
  \bibinfo{journal}{\prl} \textbf{\bibinfo{volume}{76}}, \bibinfo{pages}{1788}
  (\bibinfo{year}{1996}).

\bibitem[{\citenamefont{Dutta et~al.}(1999)\citenamefont{Dutta, Teo, and
  Raithel}}]{Dutta}
\bibinfo{author}{\bibfnamefont{S.~K.} \bibnamefont{Dutta}},
  \bibinfo{author}{\bibfnamefont{B.~K.} \bibnamefont{Teo}}, \bibnamefont{and}
  \bibinfo{author}{\bibfnamefont{G.}~\bibnamefont{Raithel}},
  \bibinfo{journal}{\prl} \textbf{\bibinfo{volume}{83}}, \bibinfo{pages}{1934}
  (\bibinfo{year}{1999}).

\bibitem[{\citenamefont{Jaksch and Zoller}(2003)}]{Jaksch}
\bibinfo{author}{\bibfnamefont{D.}~\bibnamefont{Jaksch}} \bibnamefont{and}
  \bibinfo{author}{\bibfnamefont{P.}~\bibnamefont{Zoller}},
  \bibinfo{journal}{New J. Phys.} \textbf{\bibinfo{volume}{5}},
  \bibinfo{pages}{56} (\bibinfo{year}{2003}).

\bibitem[{\citenamefont{Juzeliunas and {O}hberg}(2004)}]{oberg1}
\bibinfo{author}{\bibfnamefont{G.}~\bibnamefont{Juzeliunas}} \bibnamefont{and}
  \bibinfo{author}{\bibfnamefont{P.}~\bibnamefont{{O}hberg}},
  \bibinfo{journal}{\prl} \textbf{\bibinfo{volume}{93}},
  \bibinfo{pages}{033602} (\bibinfo{year}{2004}).

\bibitem[{\citenamefont{Juzeliunas et~al.}(2005)\citenamefont{Juzeliunas,
  {O}hberg, Ruseckas, and Klein}}]{oberg2}
\bibinfo{author}{\bibfnamefont{G.}~\bibnamefont{Juzeliunas}},
  \bibinfo{author}{\bibfnamefont{P.}~\bibnamefont{{O}hberg}},
  \bibinfo{author}{\bibfnamefont{J.}~\bibnamefont{Ruseckas}}, \bibnamefont{and}
  \bibinfo{author}{\bibfnamefont{A.}~\bibnamefont{Klein}},
  \bibinfo{journal}{\pra} \textbf{\bibinfo{volume}{71}},
  \bibinfo{pages}{053614} (\bibinfo{year}{2005}).

\bibitem[{\citenamefont{Juzeliunas and {O}hberg}()}]{obergRusec}
\bibinfo{author}{\bibfnamefont{G.}~\bibnamefont{Juzeliunas}} \bibnamefont{and}
  \bibinfo{author}{\bibfnamefont{P.}~\bibnamefont{{O}hberg}},
  \bibinfo{howpublished}{Phys. Rev. Lett. {\bf 93}, 033602 (2004); J. Ruseckas,
  G. Juzeliunas, P. {O}hberg and M. Fleischhauer, Phys. Rev. Lett. {\bf 95},
  010404 (2005)}.

\bibitem[{\citenamefont{Sorensen et~al.}(2005)\citenamefont{Sorensen, Demler,
  and Lukin}}]{Soren}
\bibinfo{author}{\bibfnamefont{A.}~\bibnamefont{Sorensen}},
  \bibinfo{author}{\bibfnamefont{E.}~\bibnamefont{Demler}}, \bibnamefont{and}
  \bibinfo{author}{\bibfnamefont{M.}~\bibnamefont{Lukin}},
  \bibinfo{journal}{\prl} \textbf{\bibinfo{volume}{94}},
  \bibinfo{pages}{086803} (\bibinfo{year}{2005}).

\bibitem[{\citenamefont{Zhu et~al.}(2006)\citenamefont{Zhu, Fu, Wu, Zhang, and
  Duan}}]{Zhu}
\bibinfo{author}{\bibfnamefont{S.}~\bibnamefont{Zhu}},
  \bibinfo{author}{\bibfnamefont{H.}~\bibnamefont{Fu}},
  \bibinfo{author}{\bibfnamefont{C.}~\bibnamefont{Wu}},
  \bibinfo{author}{\bibfnamefont{S.}~\bibnamefont{Zhang}}, \bibnamefont{and}
  \bibinfo{author}{\bibfnamefont{L.}~\bibnamefont{Duan}},
  \bibinfo{journal}{\prl} \textbf{\bibinfo{volume}{97}},
  \bibinfo{pages}{240401} (\bibinfo{year}{2006}).

\bibitem[{\citenamefont{Osterloh et~al.}(2005)\citenamefont{Osterloh, Baig,
  Santos, Zoller, and Lewenstein}}]{Osterloh}
\bibinfo{author}{\bibfnamefont{K.}~\bibnamefont{Osterloh}},
  \bibinfo{author}{\bibfnamefont{M.}~\bibnamefont{Baig}},
  \bibinfo{author}{\bibfnamefont{L.}~\bibnamefont{Santos}},
  \bibinfo{author}{\bibfnamefont{P.}~\bibnamefont{Zoller}}, \bibnamefont{and}
  \bibinfo{author}{\bibfnamefont{M.}~\bibnamefont{Lewenstein}},
  \bibinfo{journal}{\prl} \textbf{\bibinfo{volume}{95}},
  \bibinfo{pages}{010403} (\bibinfo{year}{2005}).

\bibitem[{\citenamefont{Lin et~al.}()\citenamefont{Lin, Compton, Perry,
  Phillips, Porto, and Spielman}}]{Spiel}
\bibinfo{author}{\bibfnamefont{Y.-J.} \bibnamefont{Lin}},
  \bibinfo{author}{\bibfnamefont{R.~L.} \bibnamefont{Compton}},
  \bibinfo{author}{\bibfnamefont{A.~R.} \bibnamefont{Perry}},
  \bibinfo{author}{\bibfnamefont{W.~D.} \bibnamefont{Phillips}},
  \bibinfo{author}{\bibfnamefont{J.~V.} \bibnamefont{Porto}}, \bibnamefont{and}
  \bibinfo{author}{\bibfnamefont{I.~B.} \bibnamefont{Spielman}},
  \bibinfo{howpublished}{Phys.Rev.Lett. {\bf 102}, 130401 (2009)}.

\bibitem[{\citenamefont{Goldman et~al.}()\citenamefont{Goldman, Satija,
  Nikolic, Bermudez, Martin-Delgado, Lewenstein, and Spielman}}]{SpielII}
\bibinfo{author}{\bibfnamefont{N.}~\bibnamefont{Goldman}},
  \bibinfo{author}{\bibfnamefont{I.}~\bibnamefont{Satija}},
  \bibinfo{author}{\bibfnamefont{P.}~\bibnamefont{Nikolic}},
  \bibinfo{author}{\bibfnamefont{A.}~\bibnamefont{Bermudez}},
  \bibinfo{author}{\bibfnamefont{M.}~\bibnamefont{Martin-Delgado}},
  \bibinfo{author}{\bibfnamefont{M.}~\bibnamefont{Lewenstein}},
  \bibnamefont{and} \bibinfo{author}{\bibfnamefont{I.~B.}
  \bibnamefont{Spielman}}, \bibinfo{howpublished}{e-print arXiv:1002.0219}.

\bibitem[{\citenamefont{Ruseckas et~al.}(2005)\citenamefont{Ruseckas,
  Juzeliunas, {O}hberg, and Fleischhauer}}]{Rusec}
\bibinfo{author}{\bibfnamefont{J.}~\bibnamefont{Ruseckas}},
  \bibinfo{author}{\bibfnamefont{G.}~\bibnamefont{Juzeliunas}},
  \bibinfo{author}{\bibfnamefont{P.}~\bibnamefont{{O}hberg}}, \bibnamefont{and}
  \bibinfo{author}{\bibfnamefont{M.}~\bibnamefont{Fleischhauer}},
  \bibinfo{journal}{\prl} \textbf{\bibinfo{volume}{95}},
  \bibinfo{pages}{010404} (\bibinfo{year}{2005}).

\bibitem[{\citenamefont{Raghu et~al.}(2008)\citenamefont{Raghu, Qi, Honerkamp,
  and Zhang}}]{TopoMott1}
\bibinfo{author}{\bibfnamefont{S.}~\bibnamefont{Raghu}},
  \bibinfo{author}{\bibfnamefont{X.-L.} \bibnamefont{Qi}},
  \bibinfo{author}{\bibfnamefont{C.}~\bibnamefont{Honerkamp}},
  \bibnamefont{and} \bibinfo{author}{\bibfnamefont{S.-C.} \bibnamefont{Zhang}},
  \bibinfo{journal}{Phys. Rev. Lett.} \textbf{\bibinfo{volume}{100}},
  \bibinfo{pages}{156401} (\bibinfo{year}{2008}).

\bibitem[{\citenamefont{Rachel and Hur}()}]{TopoMott2}
\bibinfo{author}{\bibfnamefont{S.}~\bibnamefont{Rachel}} \bibnamefont{and}
  \bibinfo{author}{\bibfnamefont{K.~L.} \bibnamefont{Hur}},
  \bibinfo{howpublished}{e-print arXiv:1003.2238 (2010)}.

\bibitem[{\citenamefont{Nelson et~al.}(2007)\citenamefont{Nelson, Li, and
  Weiss}}]{Nelson}
\bibinfo{author}{\bibfnamefont{K.~D.} \bibnamefont{Nelson}},
  \bibinfo{author}{\bibfnamefont{X.}~\bibnamefont{Li}}, \bibnamefont{and}
  \bibinfo{author}{\bibfnamefont{D.~S.} \bibnamefont{Weiss}},
  \bibinfo{journal}{Nature Physics} \textbf{\bibinfo{volume}{3}},
  \bibinfo{pages}{556} (\bibinfo{year}{2007}).

\bibitem[{\citenamefont{Bakr et~al.}(2009)\citenamefont{Bakr, Gillen, Peng,
  Folling, and Greiner}}]{gillen2008hst}
\bibinfo{author}{\bibfnamefont{W.~S.} \bibnamefont{Bakr}},
  \bibinfo{author}{\bibfnamefont{J.~I.} \bibnamefont{Gillen}},
  \bibinfo{author}{\bibfnamefont{A.}~\bibnamefont{Peng}},
  \bibinfo{author}{\bibfnamefont{S.}~\bibnamefont{Folling}}, \bibnamefont{and}
  \bibinfo{author}{\bibfnamefont{M.}~\bibnamefont{Greiner}},
  \bibinfo{journal}{Nature} \textbf{\bibinfo{volume}{462}}, \bibinfo{pages}{74}
  (\bibinfo{year}{2009}).

\bibitem[{\citenamefont{Thouless et~al.}(1982)\citenamefont{Thouless, Kohmoto,
  Nightingale, and den Nijs}}]{Thouless}
\bibinfo{author}{\bibfnamefont{D.~J.} \bibnamefont{Thouless}},
  \bibinfo{author}{\bibfnamefont{M.}~\bibnamefont{Kohmoto}},
  \bibinfo{author}{\bibfnamefont{M.~P.} \bibnamefont{Nightingale}},
  \bibnamefont{and} \bibinfo{author}{\bibfnamefont{M.}~\bibnamefont{den Nijs}},
  \bibinfo{journal}{Phys. Rev. Lett.} \textbf{\bibinfo{volume}{49}},
  \bibinfo{pages}{405} (\bibinfo{year}{1982}).

\bibitem[{\citenamefont{Halperin}(1982)}]{Halperin}
\bibinfo{author}{\bibfnamefont{B.~I.} \bibnamefont{Halperin}},
  \bibinfo{journal}{Phys. Rev. B} \textbf{\bibinfo{volume}{25}},
  \bibinfo{pages}{2185} (\bibinfo{year}{1982}).

\bibitem[{\citenamefont{Laughlin}(1981)}]{Laughlin}
\bibinfo{author}{\bibfnamefont{R.~B.} \bibnamefont{Laughlin}},
  \bibinfo{journal}{Phys. Rev. B} \textbf{\bibinfo{volume}{23}},
  \bibinfo{pages}{5632} (\bibinfo{year}{1981}).

\bibitem[{\citenamefont{Yakovenko}(1990)}]{VY1}
\bibinfo{author}{\bibfnamefont{V.~M.} \bibnamefont{Yakovenko}},
  \bibinfo{journal}{Phys. Rev. Lett.} \textbf{\bibinfo{volume}{65}},
  \bibinfo{pages}{251} (\bibinfo{year}{1990}).

\bibitem[{\citenamefont{Tewari et~al.}(2008)\citenamefont{Tewari, Zhang,
  Yakovenko, and Sarma}}]{VY2}
\bibinfo{author}{\bibfnamefont{S.}~\bibnamefont{Tewari}},
  \bibinfo{author}{\bibfnamefont{C.}~\bibnamefont{Zhang}},
  \bibinfo{author}{\bibfnamefont{V.~M.} \bibnamefont{Yakovenko}},
  \bibnamefont{and} \bibinfo{author}{\bibfnamefont{S.~D.} \bibnamefont{Sarma}},
  \bibinfo{journal}{Phys. Rev. Lett.} \textbf{\bibinfo{volume}{100}},
  \bibinfo{pages}{217004} (\bibinfo{year}{2008}).

\bibitem[{\citenamefont{Zhang et~al.}(2008)\citenamefont{Zhang, Tewari,
  Yakovenko, and Sarma}}]{VY3}
\bibinfo{author}{\bibfnamefont{C.}~\bibnamefont{Zhang}},
  \bibinfo{author}{\bibfnamefont{S.}~\bibnamefont{Tewari}},
  \bibinfo{author}{\bibfnamefont{V.~M.} \bibnamefont{Yakovenko}},
  \bibnamefont{and} \bibinfo{author}{\bibfnamefont{S.~D.} \bibnamefont{Sarma}},
  \bibinfo{journal}{Phys. Rev. B} \textbf{\bibinfo{volume}{78}},
  \bibinfo{pages}{174508} (\bibinfo{year}{2008}).

\bibitem[{\citenamefont{Schnyder et~al.}(2008)\citenamefont{Schnyder, Ryu,
  Furusaki, and Ludwig}}]{Ludwig}
\bibinfo{author}{\bibfnamefont{A.~P.} \bibnamefont{Schnyder}},
  \bibinfo{author}{\bibfnamefont{S.}~\bibnamefont{Ryu}},
  \bibinfo{author}{\bibfnamefont{A.}~\bibnamefont{Furusaki}}, \bibnamefont{and}
  \bibinfo{author}{\bibfnamefont{A.~W.~W.} \bibnamefont{Ludwig}},
  \bibinfo{journal}{Physical Review B (Condensed Matter and Materials Physics)}
  \textbf{\bibinfo{volume}{78}}, \bibinfo{pages}{195125}
  (\bibinfo{year}{2008}).

\bibitem[{\citenamefont{Konig et~al.}(2008)\citenamefont{Konig, Buhmann,
  Molenkamp, Hughes, Liu, Qi, and Zhang}}]{Konig2008}
\bibinfo{author}{\bibfnamefont{M.}~\bibnamefont{Konig}},
  \bibinfo{author}{\bibfnamefont{H.}~\bibnamefont{Buhmann}},
  \bibinfo{author}{\bibfnamefont{L.~W.} \bibnamefont{Molenkamp}},
  \bibinfo{author}{\bibfnamefont{T.}~\bibnamefont{Hughes}},
  \bibinfo{author}{\bibfnamefont{C.-X.} \bibnamefont{Liu}},
  \bibinfo{author}{\bibfnamefont{X.-L.} \bibnamefont{Qi}}, \bibnamefont{and}
  \bibinfo{author}{\bibfnamefont{S.-C.} \bibnamefont{Zhang}},
  \bibinfo{journal}{J. Phys. Soc. Jpn.} \textbf{\bibinfo{volume}{77}},
  \bibinfo{pages}{031007} (\bibinfo{year}{2008}).

\bibitem[{\citenamefont{Xu and Moore}(2006)}]{XuMoore}
\bibinfo{author}{\bibfnamefont{C.}~\bibnamefont{Xu}} \bibnamefont{and}
  \bibinfo{author}{\bibfnamefont{J.~E.} \bibnamefont{Moore}},
  \bibinfo{journal}{Physical Review B} \textbf{\bibinfo{volume}{73}},
  \bibinfo{pages}{045322} (\bibinfo{year}{2006}).

\bibitem[{\citenamefont{Ostrovsky et~al.}(2007)\citenamefont{Ostrovsky, Gornyi,
  and Mirlin}}]{Ostrovsky}
\bibinfo{author}{\bibfnamefont{P.~M.} \bibnamefont{Ostrovsky}},
  \bibinfo{author}{\bibfnamefont{I.~V.} \bibnamefont{Gornyi}},
  \bibnamefont{and} \bibinfo{author}{\bibfnamefont{A.~D.}
  \bibnamefont{Mirlin}}, \bibinfo{journal}{Physical Review Letters}
  \textbf{\bibinfo{volume}{98}}, \bibinfo{pages}{256801}
  (\bibinfo{year}{2007}).

\bibitem[{\citenamefont{Obuse et~al.}(2007)\citenamefont{Obuse, Furusaki, Ryu,
  and Mudry}}]{Obuse}
\bibinfo{author}{\bibfnamefont{H.}~\bibnamefont{Obuse}},
  \bibinfo{author}{\bibfnamefont{A.}~\bibnamefont{Furusaki}},
  \bibinfo{author}{\bibfnamefont{S.}~\bibnamefont{Ryu}}, \bibnamefont{and}
  \bibinfo{author}{\bibfnamefont{C.}~\bibnamefont{Mudry}},
  \bibinfo{journal}{Physical Review B} \textbf{\bibinfo{volume}{76}},
  \bibinfo{pages}{075301} (\bibinfo{year}{2007}).

\bibitem[{\citenamefont{Bardarson et~al.}(2007)\citenamefont{Bardarson, o,
  Brouwer, and Beenakker}}]{Bardarson}
\bibinfo{author}{\bibfnamefont{J.~H.} \bibnamefont{Bardarson}},
  \bibinfo{author}{\bibfnamefont{J.~T.} \bibnamefont{o}},
  \bibinfo{author}{\bibfnamefont{P.~W.} \bibnamefont{Brouwer}},
  \bibnamefont{and} \bibinfo{author}{\bibfnamefont{C.~W.~J.}
  \bibnamefont{Beenakker}}, \bibinfo{journal}{Physical Review Letters}
  \textbf{\bibinfo{volume}{99}}, \bibinfo{pages}{106801}
  (\bibinfo{year}{2007}).

\bibitem[{\citenamefont{Satija et~al.}(2006)\citenamefont{Satija, Dakin, and
  Clark}}]{Clark}
\bibinfo{author}{\bibfnamefont{I.~I.} \bibnamefont{Satija}},
  \bibinfo{author}{\bibfnamefont{D.~C.} \bibnamefont{Dakin}}, \bibnamefont{and}
  \bibinfo{author}{\bibfnamefont{C.~W.} \bibnamefont{Clark}},
  \bibinfo{journal}{\prl} \textbf{\bibinfo{volume}{97}},
  \bibinfo{pages}{216401} (\bibinfo{year}{2006}).

\bibitem[{\citenamefont{Grynberg and Robilliard}()}]{Grynberg}
\bibinfo{author}{\bibfnamefont{G.}~\bibnamefont{Grynberg}} \bibnamefont{and}
  \bibinfo{author}{\bibfnamefont{C.}~\bibnamefont{Robilliard}},
  \bibinfo{howpublished}{Phys. Rep. {\bf 355}, 335 (2001)}.

\bibitem[{\citenamefont{Ritt et~al.}(2006)\citenamefont{Ritt, Geckeler, Salger,
  Cennini, and Weitz}}]{Ritt}
\bibinfo{author}{\bibfnamefont{G.}~\bibnamefont{Ritt}},
  \bibinfo{author}{\bibfnamefont{C.}~\bibnamefont{Geckeler}},
  \bibinfo{author}{\bibfnamefont{T.}~\bibnamefont{Salger}},
  \bibinfo{author}{\bibfnamefont{G.}~\bibnamefont{Cennini}}, \bibnamefont{and}
  \bibinfo{author}{\bibfnamefont{M.}~\bibnamefont{Weitz}},
  \bibinfo{journal}{Physical Review A (Atomic, Molecular, and Optical Physics)}
  \textbf{\bibinfo{volume}{74}}, \bibinfo{pages}{063622}
  (\bibinfo{year}{2006}).

\bibitem[{\citenamefont{Rey et~al.}(2007)\citenamefont{Rey, Gritsev, Bloch,
  Demler, and Lukin}}]{Rey}
\bibinfo{author}{\bibfnamefont{A.~M.} \bibnamefont{Rey}},
  \bibinfo{author}{\bibfnamefont{V.}~\bibnamefont{Gritsev}},
  \bibinfo{author}{\bibfnamefont{I.}~\bibnamefont{Bloch}},
  \bibinfo{author}{\bibfnamefont{E.}~\bibnamefont{Demler}}, \bibnamefont{and}
  \bibinfo{author}{\bibfnamefont{M.~D.} \bibnamefont{Lukin}},
  \bibinfo{journal}{Physical Review Letters} \textbf{\bibinfo{volume}{99}},
  \bibinfo{pages}{140601} (\bibinfo{year}{2007}).

\bibitem[{\citenamefont{Trotzky et~al.}()\citenamefont{Trotzky, Cheinet,
  Fölling, Feld, Schnorrberger, Rey, Polkovnikov, Demler, Lukin, and
  Bloch}}]{Trotzky}
\bibinfo{author}{\bibfnamefont{S.}~\bibnamefont{Trotzky}},
  \bibinfo{author}{\bibfnamefont{P.}~\bibnamefont{Cheinet}},
  \bibinfo{author}{\bibfnamefont{S.}~\bibnamefont{Fölling}},
  \bibinfo{author}{\bibfnamefont{M.}~\bibnamefont{Feld}},
  \bibinfo{author}{\bibfnamefont{U.}~\bibnamefont{Schnorrberger}},
  \bibinfo{author}{\bibfnamefont{A.~M.} \bibnamefont{Rey}},
  \bibinfo{author}{\bibfnamefont{A.}~\bibnamefont{Polkovnikov}},
  \bibinfo{author}{\bibfnamefont{E.~A.} \bibnamefont{Demler}},
  \bibinfo{author}{\bibfnamefont{M.~D.} \bibnamefont{Lukin}}, \bibnamefont{and}
  \bibinfo{author}{\bibfnamefont{I.}~\bibnamefont{Bloch}},
  \bibinfo{howpublished}{Science, {\bf 319}, 295 (2008)}.

\bibitem[{\citenamefont{Cheinet et~al.}(2008)\citenamefont{Cheinet, Trotzky,
  Feld, Schnorrberger, Moreno-Cardoner, F\"{o}lling, and Bloch}}]{Cheinet}
\bibinfo{author}{\bibfnamefont{P.}~\bibnamefont{Cheinet}},
  \bibinfo{author}{\bibfnamefont{S.}~\bibnamefont{Trotzky}},
  \bibinfo{author}{\bibfnamefont{M.}~\bibnamefont{Feld}},
  \bibinfo{author}{\bibfnamefont{U.}~\bibnamefont{Schnorrberger}},
  \bibinfo{author}{\bibfnamefont{M.}~\bibnamefont{Moreno-Cardoner}},
  \bibinfo{author}{\bibfnamefont{S.}~\bibnamefont{F\"{o}lling}},
  \bibnamefont{and} \bibinfo{author}{\bibfnamefont{I.}~\bibnamefont{Bloch}},
  \bibinfo{journal}{Physical Review Letters} \textbf{\bibinfo{volume}{101}},
  \bibinfo{pages}{090404} (\bibinfo{year}{2008}).

\bibitem[{\citenamefont{Kohmoto}(1985)}]{Kohmoto}
\bibinfo{author}{\bibfnamefont{M.}~\bibnamefont{Kohmoto}},
  \bibinfo{journal}{Annals of Physics} \textbf{\bibinfo{volume}{160}},
  \bibinfo{pages}{296} (\bibinfo{year}{1985}).

\bibitem[{\citenamefont{Umucalilar et~al.}(2008)\citenamefont{Umucalilar, Zhai,
  and Oktel}}]{Umuca}
\bibinfo{author}{\bibfnamefont{R.~O.} \bibnamefont{Umucalilar}},
  \bibinfo{author}{\bibfnamefont{H.}~\bibnamefont{Zhai}}, \bibnamefont{and}
  \bibinfo{author}{\bibfnamefont{M.~O.} \bibnamefont{Oktel}},
  \bibinfo{journal}{\prl} \textbf{\bibinfo{volume}{100}},
  \bibinfo{pages}{070402} (\bibinfo{year}{2008}).

\bibitem[{\citenamefont{Zhang et~al.}()\citenamefont{Zhang, He, Chang, Song,
  Wang, Chen, Jia, Fang, Dai, Shan et~al.}}]{KeXue2009}
\bibinfo{author}{\bibfnamefont{Y.}~\bibnamefont{Zhang}},
  \bibinfo{author}{\bibfnamefont{K.}~\bibnamefont{He}},
  \bibinfo{author}{\bibfnamefont{C.-Z.} \bibnamefont{Chang}},
  \bibinfo{author}{\bibfnamefont{C.-L.} \bibnamefont{Song}},
  \bibinfo{author}{\bibfnamefont{L.}~\bibnamefont{Wang}},
  \bibinfo{author}{\bibfnamefont{X.}~\bibnamefont{Chen}},
  \bibinfo{author}{\bibfnamefont{J.}~\bibnamefont{Jia}},
  \bibinfo{author}{\bibfnamefont{Z.}~\bibnamefont{Fang}},
  \bibinfo{author}{\bibfnamefont{X.}~\bibnamefont{Dai}},
  \bibinfo{author}{\bibfnamefont{W.-Y.} \bibnamefont{Shan}},
  \bibnamefont{et~al.}, \bibinfo{howpublished}{e-print arXiv:0911.3706 (2009)}.

\bibitem[{\citenamefont{Zhang et~al.}(2009)\citenamefont{Zhang, Qin, Teng, Guo,
  Guo, Dai, Fang, and Wu}}]{ZhangWu}
\bibinfo{author}{\bibfnamefont{G.}~\bibnamefont{Zhang}},
  \bibinfo{author}{\bibfnamefont{H.}~\bibnamefont{Qin}},
  \bibinfo{author}{\bibfnamefont{J.}~\bibnamefont{Teng}},
  \bibinfo{author}{\bibfnamefont{J.}~\bibnamefont{Guo}},
  \bibinfo{author}{\bibfnamefont{Q.}~\bibnamefont{Guo}},
  \bibinfo{author}{\bibfnamefont{X.}~\bibnamefont{Dai}},
  \bibinfo{author}{\bibfnamefont{Z.}~\bibnamefont{Fang}}, \bibnamefont{and}
  \bibinfo{author}{\bibfnamefont{K.}~\bibnamefont{Wu}},
  \bibinfo{journal}{Applied Physics Letters} \textbf{\bibinfo{volume}{95}},
  \bibinfo{pages}{053114} (\bibinfo{year}{2009}).

\bibitem[{\citenamefont{Linder et~al.}(2009)\citenamefont{Linder, Yokoyama, and
  Sudb\o{}}}]{Linder}
\bibinfo{author}{\bibfnamefont{J.}~\bibnamefont{Linder}},
  \bibinfo{author}{\bibfnamefont{T.}~\bibnamefont{Yokoyama}}, \bibnamefont{and}
  \bibinfo{author}{\bibfnamefont{A.}~\bibnamefont{Sudb\o{}}},
  \bibinfo{journal}{Phys. Rev. B} \textbf{\bibinfo{volume}{80}},
  \bibinfo{pages}{205401} (\bibinfo{year}{2009}).

\bibitem[{\citenamefont{Lu et~al.}(2010)\citenamefont{Lu, Shan, Yao, Niu, and
  Shen}}]{LuShen}
\bibinfo{author}{\bibfnamefont{H.-Z.} \bibnamefont{Lu}},
  \bibinfo{author}{\bibfnamefont{W.-Y.} \bibnamefont{Shan}},
  \bibinfo{author}{\bibfnamefont{W.}~\bibnamefont{Yao}},
  \bibinfo{author}{\bibfnamefont{Q.}~\bibnamefont{Niu}}, \bibnamefont{and}
  \bibinfo{author}{\bibfnamefont{S.-Q.} \bibnamefont{Shen}},
  \bibinfo{journal}{Phys. Rev. B} \textbf{\bibinfo{volume}{81}},
  \bibinfo{pages}{115407} (\bibinfo{year}{2010}).

\bibitem[{\citenamefont{Liu et~al.}(2010{\natexlab{a}})\citenamefont{Liu,
  Zhang, Yan, Qi, Frauenheim, Dai, Fang, and Zhang}}]{LiuZhang}
\bibinfo{author}{\bibfnamefont{C.-X.} \bibnamefont{Liu}},
  \bibinfo{author}{\bibfnamefont{H.}~\bibnamefont{Zhang}},
  \bibinfo{author}{\bibfnamefont{B.}~\bibnamefont{Yan}},
  \bibinfo{author}{\bibfnamefont{X.-L.} \bibnamefont{Qi}},
  \bibinfo{author}{\bibfnamefont{T.}~\bibnamefont{Frauenheim}},
  \bibinfo{author}{\bibfnamefont{X.}~\bibnamefont{Dai}},
  \bibinfo{author}{\bibfnamefont{Z.}~\bibnamefont{Fang}}, \bibnamefont{and}
  \bibinfo{author}{\bibfnamefont{S.-C.} \bibnamefont{Zhang}},
  \bibinfo{journal}{Phys. Rev. B} \textbf{\bibinfo{volume}{81}},
  \bibinfo{pages}{041307} (\bibinfo{year}{2010}{\natexlab{a}}).

\bibitem[{\citenamefont{Urazhdin et~al.}({\natexlab{a}})\citenamefont{Urazhdin,
  Bilc, Tessmer, Mahanti, Kyratsi, and Kanatzidis}}]{Ura1}
\bibinfo{author}{\bibfnamefont{S.}~\bibnamefont{Urazhdin}},
  \bibinfo{author}{\bibfnamefont{D.}~\bibnamefont{Bilc}},
  \bibinfo{author}{\bibfnamefont{S.}~\bibnamefont{Tessmer}},
  \bibinfo{author}{\bibfnamefont{S.}~\bibnamefont{Mahanti}},
  \bibinfo{author}{\bibfnamefont{T.}~\bibnamefont{Kyratsi}}, \bibnamefont{and}
  \bibinfo{author}{\bibfnamefont{M.}~\bibnamefont{Kanatzidis}},
  \bibinfo{howpublished}{Phys. Rev. B 66, 161306(R) (2002)}.

\bibitem[{\citenamefont{Urazhdin et~al.}({\natexlab{b}})\citenamefont{Urazhdin,
  Bilc, Tessmer, Mahanti, Kyratsi, and Kanatzidis}}]{Ura2}
\bibinfo{author}{\bibfnamefont{S.}~\bibnamefont{Urazhdin}},
  \bibinfo{author}{\bibfnamefont{D.}~\bibnamefont{Bilc}},
  \bibinfo{author}{\bibfnamefont{S.}~\bibnamefont{Tessmer}},
  \bibinfo{author}{\bibfnamefont{S.}~\bibnamefont{Mahanti}},
  \bibinfo{author}{\bibfnamefont{T.}~\bibnamefont{Kyratsi}}, \bibnamefont{and}
  \bibinfo{author}{\bibfnamefont{M.}~\bibnamefont{Kanatzidis}},
  \bibinfo{howpublished}{Phys. Rev. B 69, 085313 (2004)}.

\bibitem[{\citenamefont{Andersen et~al.}(2006)\citenamefont{Andersen, Ryu,
  Clade, Natarajan, Vaziri, Helmerson, and Phillips}}]{Andersen}
\bibinfo{author}{\bibfnamefont{M.~F.} \bibnamefont{Andersen}},
  \bibinfo{author}{\bibfnamefont{C.}~\bibnamefont{Ryu}},
  \bibinfo{author}{\bibfnamefont{P.}~\bibnamefont{Clade}},
  \bibinfo{author}{\bibfnamefont{V.}~\bibnamefont{Natarajan}},
  \bibinfo{author}{\bibfnamefont{A.}~\bibnamefont{Vaziri}},
  \bibinfo{author}{\bibfnamefont{K.}~\bibnamefont{Helmerson}},
  \bibnamefont{and} \bibinfo{author}{\bibfnamefont{W.~D.}
  \bibnamefont{Phillips}}, \bibinfo{journal}{\prl}
  \textbf{\bibinfo{volume}{97}}, \bibinfo{pages}{170406}
  (\bibinfo{year}{2006}).

\bibitem[{\citenamefont{Stamper-Kurn et~al.}(1999)\citenamefont{Stamper-Kurn,
  Chikkatur, G\"orlitz, Inouye, Gupta, Pritchard, and Ketterle}}]{Stamper1999}
\bibinfo{author}{\bibfnamefont{D.~M.} \bibnamefont{Stamper-Kurn}},
  \bibinfo{author}{\bibfnamefont{A.~P.} \bibnamefont{Chikkatur}},
  \bibinfo{author}{\bibfnamefont{A.}~\bibnamefont{G\"orlitz}},
  \bibinfo{author}{\bibfnamefont{S.}~\bibnamefont{Inouye}},
  \bibinfo{author}{\bibfnamefont{S.}~\bibnamefont{Gupta}},
  \bibinfo{author}{\bibfnamefont{D.~E.} \bibnamefont{Pritchard}},
  \bibnamefont{and} \bibinfo{author}{\bibfnamefont{W.}~\bibnamefont{Ketterle}},
  \bibinfo{journal}{Phys. Rev. Lett.} \textbf{\bibinfo{volume}{83}},
  \bibinfo{pages}{2876} (\bibinfo{year}{1999}).

\bibitem[{\citenamefont{Liu et~al.}(2010{\natexlab{b}})\citenamefont{Liu, Liu,
  Wu, and Sinova}}]{Liu2010}
\bibinfo{author}{\bibfnamefont{X.-J.} \bibnamefont{Liu}},
  \bibinfo{author}{\bibfnamefont{X.}~\bibnamefont{Liu}},
  \bibinfo{author}{\bibfnamefont{C.}~\bibnamefont{Wu}}, \bibnamefont{and}
  \bibinfo{author}{\bibfnamefont{J.}~\bibnamefont{Sinova}},
  \bibinfo{journal}{Phys. Rev. A} \textbf{\bibinfo{volume}{81}},
  \bibinfo{pages}{033622} (\bibinfo{year}{2010}{\natexlab{b}}).

\bibitem[{\citenamefont{Stanescu et~al.}({\natexlab{b}})\citenamefont{Stanescu,
  Anderson, and Galitski}}]{SAG}
\bibinfo{author}{\bibfnamefont{T.~D.} \bibnamefont{Stanescu}},
  \bibinfo{author}{\bibfnamefont{B.}~\bibnamefont{Anderson}}, \bibnamefont{and}
  \bibinfo{author}{\bibfnamefont{V.~M.} \bibnamefont{Galitski}},
  \bibinfo{howpublished}{Phys. Rev. A {\bf 78} , 023616 (2008)}.

\end{thebibliography}

\end{document}